\documentclass[twocolumn]{aastex63}

\usepackage{afterpage}

\usepackage{booktabs} 
\let\tablenum\relax 
\usepackage{siunitx} 
\usepackage{amsmath,amstext,color}
\usepackage{gensymb}
\usepackage{natbib}
\usepackage{hyperref}
\usepackage{longtable}
\usepackage{threeparttablex}
\usepackage{booktabs}

\usepackage{color}
\usepackage{graphicx}
\usepackage{amsmath}
\usepackage{amssymb}

\usepackage[]{appendix}

\begin{document}
\title{Uniform Modeling of Observed Kilonovae: Implications for Diversity and the Progenitors of Merger-Driven Long Gamma-Ray Bursts}
\correspondingauthor{Jillian Rastinejad}
\email{jillianrastinejad2024@u.northwestern.edu}

\shorttitle{Uniform Modeling of Kilonovae}
\shortauthors{Rastinejad et al.}
\newcommand{\NU}{\affiliation{Center for Interdisciplinary Exploration and Research in Astrophysics (CIERA) and Department of Physics and Astronomy, Northwestern University, Evanston, IL 60208, USA}}

\newcommand{\Radboud}{\affiliation{Department of Astrophysics/IMAPP, Radboud University, 6525 AJ Nijmegen, The Netherlands}}

\newcommand{\Leicester}{\affiliation{School of Physics and Astronomy, University of Leicester, University Road, Leicester, LE1 7RH, UK}}
\author[0000-0002-9267-6213]{J. C.~Rastinejad}
\NU

\author[0000-0002-7374-935X]{W.~Fong}
\NU

\author[0000-0002-5740-7747]{C.~D.~Kilpatrick}
\NU

\author[0000-0002-2555-3192]{M.~Nicholl}
\affil{Astrophysics Research Centre, School of Mathematics and Physics, Queen’s University Belfast, BT7 1NN, UK}

\author[0000-0002-4670-7509]{B. D. Metzger}
\affil{Department of Physics and Columbia Astrophysics Laboratory, Columbia University, Pupin Hall, New York, NY 10027, USA}
\affil{Center for Computational Astrophysics, Flatiron Institute, 162 5th Ave, New York, NY 10010, USA}

\begin{abstract}
We present uniform modeling of eight kilonovae, five following short gamma-ray bursts (GRBs; including GRB\,170817A) and three following long GRBs. We model their broadband afterglows to determine the relative contributions of afterglow and kilonova emission. We fit the kilonovae using a three-component model in \texttt{MOSFiT}, and report population median ejecta masses for the total, blue ($\kappa_{\rm B} = 0.5$ cm$^2$ g$^{-1}$), purple ($\kappa_{\rm P} = 3$ cm$^2$ g$^{-1}$), and red ($\kappa_{\rm R} = 10$ cm$^2$ g$^{-1}$) components. The kilonova of GW\,170817 is near the sample median in most derived properties. We investigate trends between the ejecta masses and the isotropic-equivalent and beaming-corrected $\gamma$-ray energies ($E_{\gamma, {\rm iso}}$, $E_{\gamma}$), as well as rest-frame durations ($T_{\rm 90, rest}$). We find long GRB kilonovae have higher median red ejecta masses ($M_{\rm ej, R} \gtrsim 0.05 M_{\odot}$) compared to on-axis short GRB kilonovae ($M_{\rm ej, R} \lesssim 0.02 M_{\odot}$). We also observe a weak scaling between the total and red ejecta masses with $E_{\gamma, {\rm iso}}$ and $E_{\gamma}$, though a larger sample is needed to establish a significant correlation. These findings imply a connection between merger-driven long GRBs and larger tidal dynamical ejecta masses, which may indicate that their progenitors are asymmetric compact object binaries. We produce representative kilonova light curves and find that the planned depths and cadences of the Rubin and Roman Observatory surveys will be sufficient for order-of-magnitude constraints on $M_{\rm ej, B}$ (and, for Roman, $M_{\rm ej, P}$ and $M_{\rm ej, R}$) of future kilonovae at $z \lesssim 0.1$.
\end{abstract}


\section{Introduction}

On August 17, 2017, the nearly coincident detection of a binary neutron star (BNS) merger through gravitational waves (GWs, GW\,170817; \citealt{gw170817mma, gw170817}) and a short gamma-ray burst (GRB\,170817A; \citealt{Goldstein+2017,Savchenko+17}) confirmed the long-theorized connection between these ``multi-messenger'' signals. Adding to this long-awaited breakthrough, an optical-near-IR counterpart (``AT\,2017gfo'') to the GW and GRB signals was observed \citep{Arcavi+17,Coulter+17,Lipunov+17,Tanvir+17,Soares-Santos+17,Valenti+17} and strongly resembled theoretical predictions for a kilonova, a thermal transient powered by the radioactive decay of elements beyond the iron-peak formed through rapid neutron capture nucleosynthesis (``$r$-process''; e.g.,  \citealt{Burbidge+57,Cameron57,LattimerSchramm74,Li&Paczynski98,Metzger+10}). Indeed, the counterpart's rapid temporal evolution pointed to a relatively low ejecta mass ($\lesssim 0.1 M_{\odot}$), expected for neutron star mergers, while its reddening on $\sim$day timescales indicated a high (compared to that of iron-group elements) ejecta opacity produced by newly-created heavy elements (e.g., \citealt{barneskasen13,tanaka+13,kasen+17}). Modeling of AT\,2017gfo demonstrated that the light curve was well-explained with a two-component model, in which each component is parameterized by a different ejecta mass, velocity, and opacity (e.g., \citealt{Cowperthwaite+17,Drout+17,kasen+17,kilpatrick+17,Tanvir+17,Troja+17,villar+17}), providing evidence for multiple emission sources from the merger.

The first definitive kilonova opened a host of new questions and predictions that may only be studied with a wider population of observed events. Rapid progress on the theoretical end of kilonova studies has resulted in models that incorporate additional physics, including updated neutrino schemes and viewing angle dependencies (e.g., \citealt{Bulla19,Nicholl+21,Wollaeger+21,Chase+22,Curtis+23}), and a mapping of progenitors, remnants, and novel emission mechanisms to predicted light curve properties (e.g., \citealt{Arcavi18,Gottlieb+18,Metzger+18,Kawaguchi+20b,Gompertz+23_NSBH}). In parallel, GW observations of diverse BNS and NSBH progenitors (e.g., \citealt{LVC_GW190425,LVC_GW190814,LVK_230529}) and recent kilonovae discoveries following LGRBs \citep{Rastinejad+22,Troja+22,Yang+22,Gillanders+23,Levan+24,Yang+24} further motivate expectations for observed kilonova diversity. Looking forward, constraints on light curve diversity are critical for kilonova search strategies with next-generation facilities such as the Rubin Observatory and the \textit{Roman Space Telescope} (Roman; \citealt{Margutti+18_LSSTTOO,Andreoni_romankilonova}).

Despite progress on the theoretical end, constraints on kilonova ejecta parameters from observations other than AT\,2017gfo remain limited due to the low rates of detected neutron star mergers (e.g., \citealt{MandelBroekgaarden22,GWTC-3}). Kilonovae are also relatively faint and fleeting transients, rendering them difficult to observe at distances beyond $z\approx 0.1$ without rapid response and highly-sensitive telescopes.
Several kilonovae following GRBs have been discovered contemporaneously and in archival data (e.g., \citealt{berger+13,Tanvir+13,yang+15,Gompertz+18}) and, when combined with deep upper limits, reveal a span of $\approx$100 in optical luminosity (e.g., \citealt{Gompertz+18,Rossi+20,Rastinejad+21}). However, the majority of previous fits to observed light curves have been with by multiple codes with varying assumptions (e.g., \citealt{lamb+19,troja+19,O'Connor+21}). Thus, direct comparisons of derived parameters (e.g., mass, velocity, composition) between individual events are inadvisable. 

Previous uniform modeling of GRB kilonovae used a single component model \citep{Ascenzi+19}, prohibiting a search for diversity among emission components. Further, a uniform analysis has not been performed since 2019 \citep{Ascenzi+19}, thus excluding three recent events (GRBs\,200522A, 211211A and 230307A; \citealt{Fong+21,O'Connor+21,Rastinejad+22,Troja+22,Levan+24,Yang+24}). Surprisingly, two of these kilonovae followed long-duration GRBs. The progenitors and/or emission mechanisms driving their longer $\gamma$-ray signals remain an unknown, with several observationally untested theories (e.g., \citealt{Yang+22,Gottlieb+23}). Beyond significantly expanding the sample size, these new events motivate an updated, multi-component modeling endeavor to explore kilonova diversity and compare $\gamma$-ray and kilonova properties across the short and long GRB populations.

Here, we perform uniform, multi-component modeling of a sample of eight kilonovae discovered over 2005--2023. We provide compiled multi-wavelength light curves for future fitting efforts with additional models. In Section~\ref{sec:obs} we describe our sample selection and provide details for each event. In Section~\ref{sec:ag_mod} we detail our modeling of the synchrotron afterglow component to extricate the kilonova emission. In Section~\ref{sec:KN_mod} we describe our kilonova modeling procedure and report our results. In Section~\ref{sec:discussion} we discuss the implications of our results for future work. We assume a cosmology of $H_{0}$ = 69.6~km~s$^{-1}$~Mpc$^{-1}$, $\Omega_{M}$ = 0.286, $\Omega_{vac}$ = 0.714 \citep{Bennett+14} and report magnitudes in the AB system throughout this work.

\section{Observations}
\label{sec:obs}

\subsection{Sample Selection}
\label{subsec:sample_sel}

We begin with the list of claimed kilonovae compiled in previous literature and covering the full sample of short GRBs discovered in the post-\textit{Swift} era \citep{Gompertz+18,Rastinejad+21}. These include GRBs\,050709 \citep{jin+16}, 050724A \citep{Gao+17}, 060614 \citep{yang+15}, 070714B \citep{Gao+17}, 070809 \citep{Jin+20}, 130603B \citep{berger+13,Tanvir+13}, 150101B \citep{troja+18}, 160821B \citep{lamb+19, troja+19}, 170817A (GW\,170817) and 200522A \citep{Fong+21,O'Connor+21}. Though GRBs\,080503 and 150424A have been claimed as a kilonova candidate \citep{Perley09,Bucciantini+12,Rossi+22}, we do not include these events as their redshifts, and thus their kilonova luminosities, remain uncertain \citep{Perley09,Fong+22}. In addition, we also consider the more recent kilonova candidates GRB\,211211A \citep{Rastinejad+22,Troja+22}, and GRB\,230307A \citep{Levan+24,Yang+24}, both of which were identified following recent merger-driven long GRBs\footnote{Though we note there are alternate explanations for the red excess following these events (e.g., \citealt{BarnesMetzger23,Waxman+22}.)}.

We limit our sample to events with sufficient observations (typically, multiple X-ray detections, several optical-near-IR detections past 2 days and at least one radio observation) for robust afterglow and multi-component kilonova modeling. To evaluate the availability of X-ray and radio observations, we use the catalog of \citet{fong+15}, supplemented with data from the Gamma-ray Circular Notices (GCNs) and the literature \citep{Fox+05,Mangano+07,Londish+06,berger+13,Tanvir+13,Fong+14,lamb+19,troja+19,Mei+22,Rastinejad+22,Troja+22,Levan+24,Yang+24}. We restrict our sample to events with two or more optical-near-IR observations past 2 days for two reasons. First, this is the timescale on which we roughly expect the kilonova to significantly contribute to the observed flux. Second, multiple observations allow us to assess fading, which in turn constrains the ejecta mass and velocity, and/or color, giving insight to the relative contribution from different components (e.g., \citealt{barneskasen13}). To evaluate available optical and near-IR data for each burst we consider data collected and published in short GRB kilonova compilations \citep{Gompertz+18,Rastinejad+21} for bursts prior to 2021 and the literature for those from 2021-2023. We do not include GRBs\,050724A, 070714B and 150101B in our sample as they each have only a single optical detection past 2~days. We further remove GRB\,070809 from our sample as it was not observed in the radio and no optical data is available past $\approx 1.5$~days.

We are left with eight events that meet our criteria: GRBs\,050709, 060614, 130603B, 160821B, 170817A (GW\,170817), 200522A, 211211A, and 230307A. The redshift range of these events is relatively low for short GRBs ($z = 0.008 - 0.554$; \citealt{Fong+22}), but greater than the expected horizon for GW-detected compact object mergers in O4 and O5 \citep{lvc_2016_LRR}. We describe each burst and the dataset used in our analysis in the following sections. We list the main properties of each burst in Table~\ref{tab:grbprops}, including the $\gamma$-ray durations, GRB fluence ($f_{\gamma}$), isotropic-equivalent $\gamma$-ray energy ($E_{\gamma, {\rm iso}}$; calculated using the method of \citealt{fong+15}), and galaxy type \citep{Nugent+22,Levan+24}. For \textit{Swift} GRBs we collect values for $f_{\gamma}$ (15-350 keV) and the durations over which 50\% and 90\% of the gamma-ray fluence was detected (T$_{50}$ and T$_{90}$) from the \textit{Swift}-BAT catalog\footnote{\url{https://swift.gsfc.nasa.gov/results/batgrbcat/}} \citep{Lien+16}. For GRBs detected with other satellites we collect values from the GCNs and literature \citep{Fox+05,Fermi_160821b,Goldstein+2017,Veres+23,230307a_t90_GBM}. In the following sub-sections, we summarize the burst discovery for each event in our sample and the source of data used in our analysis.

\begin{deluxetable*}{clCcCCCCc}
\savetablenum{1}
\tabletypesize{\small}
\centering
\tablecolumns{10}
\tabcolsep0.07in
\tablecaption{GRB Sample \& Properties
\label{tab:grbprops}}
\tablehead {
\colhead {GRB}		&
\colhead {$z$}		& 
\colhead {A$_{V, {\rm MW}}^{1}$} &
\colhead {$\gamma$-ray Tel.} &
\colhead {$T_{\rm 90}$} &
\colhead {$T_{\rm 50}$} &
\colhead {$f_{\gamma}$} &
\colhead {$E_{\gamma, {\rm iso}, 52}$} &
\colhead {Host Type$^2$}
\\
\colhead {}		&
\colhead {}		&
\colhead {(mag)} &
\colhead {} &
\colhead {(s)} &
\colhead {(s)} &
\colhead {($10^{-6}$ erg cm$^{-2}$)} &
\colhead {($10^{52}$ erg)} &
\colhead {}		
		}
\startdata
050709 & 0.161 & 0.030 & \textit{HETE-II} & 0.07 & - & 1.0 & 0.031 & SF \\ 
060614 & 0.125 & 0.059 & \textit{Swift} & 109 \pm 3 & 43.2 \pm 0.8 & 28.0 \pm 4 & 0.51 \pm 0.07 & SF \\
130603B & 0.356 & 0.063 & \textit{Swift} & 0.18 \pm 0.02 & 0.06 \pm 0.004 &  1.8 \pm 0.1 & 0.29 \pm 0.02 & SF \\ 
160821B & 0.162 & 0.123 & \textit{Fermi} & $\approx$1  & - & 1.7 \pm 0.2 & 0.011 \pm 0.001 & SF \\ 
&&& \textit{Swift} & 0.48 \pm 0.07 & 0.28 \pm 0.05 & 0.17 \pm 0.03 & 0.005 \pm 0.001 & - \\ 
170817A & 0.008 & 0.338 & \textit{Fermi}$^*$ & 2.0 \pm 0.5$^*$ & - &  28 \pm 2^{*} & 0.0006 \pm 0.00004$^*$ & Q \\ 
200522A & 0.554 & 0.071 & \textit{Swift} &  0.62 \pm 0.08 & 0.38 \pm 0.05 &  0.20 \pm 0.04 & 0.081 \pm 0.02 & SF \\ 
211211A & 0.076 & 0.048 & \textit{Fermi} & 34.3 \pm 0.6 & - & 507 \pm 10 & 0.68 \pm 0.01 & SF \\
&&& \textit{Swift} & 50.7 \pm 0.9 & 21.2 \pm 0.2 &  255 \pm 4 & 1.72 \pm 0.03 & - \\ 
230307A & 0.065 & 0.239 & \textit{Fermi} & $\approx$35 & - & 2950 & 2.9 & SF 
\enddata
\tablecomments{
$^1$Milky Way extinction values taken from \citet{SchlaflyFinkbeiner11}.\\
$^2$Host galaxy type (star-forming, SF, or quiescent, Q) taken from the uniformly modeled sample of \citet{Nugent+22} with the exception of GRB\,2303037A for which we use the analysis of \citet{Levan+24}. \\ 
$^*$We include this information for context on GRB\,170817A. However, as the event's $\gamma$-ray emission source is debated and likely distinct from typical on-axis events (e.g., \citealt{Lazzati+17,Gottlieb+18,Ioka+19,MarguttiChornock20}) we do use this value in our analysis. \\ }
\end{deluxetable*}

\subsection{GRB\,050709}
\label{subsec:050709}

GRB\,050709 was detected by the High Energy Transient Explorer II (HETE-II; \citealt{HETE}) at 22:36:37 UT on 2005 July 9, with a T$_{90}=0.07$~s \citep{Hjorth+05}. Follow-up observations with the \textit{Chandra X-ray Observatory} (\textit{Chandra}) revealed a new X-ray source within the HETE-II localization \citep{Fox+05}. Follow-up observations by the \textit{Swift} X-Ray Telescope (XRT; \citealt{Swift-XRT}) and \textit{Chandra} confirmed this counterpart as fading and thus, likely related to GRB\,050709 \citep{Fox+05}. At the position of the X-ray source, an optical counterpart was detected, embedded in a star-forming galaxy at $z=0.161$ \citep{Fox+05,Hjorth+05}. The Very Large Array (VLA) observed the position of the X-ray counterpart over four epochs, but did not detect any significant emission \citep{Fox+05}.

Imaging with the \textit{Hubble Space Telescope} ($HST$) revealed long-lived emission in the F814W band, which has been attributed to a kilonova by numerous groups in the literature \citep{jin+16,Gompertz+18,Ascenzi+19}. We combine the X-ray, radio \citep{Fox+05} and optical-near-IR \citep{Fox+05,Hjorth+05,Covino+06} datasets for our analysis. We present all observations used in our analysis in Table~\ref{tab:obs}.

\subsection{GRB\,060614}
\label{subsec:060614}

GRB\,060614 was discovered by the \textit{Swift} Burst Alert Telescope (BAT; \citealt{Swift_BAT}) on 2006 June 14 at 12:43:48.5 UT with a T$_{90} = 109 \pm 3$~s \citep{Barthelmy_GCN5256}. Prompt follow-up by \textit{Swift}-XRT and the \textit{Swift} Ultra-Violet Optical Telescope (UVOT; \citealt{Swift-UVOT}) revealed bright counterparts to the burst \citep{Mangano+07}. The early ultraviolet through optical afterglow spectral energy distribution (SED) provides evidence for low line-of-sight extinction and a $z < 1.3$ origin \citep{Gehrels+06}. Subsequent follow-up by ground-based observatories revealed an optical counterpart with a spectroscopic redshift of $z=0.125$ \citep{Price+06,Fugazza+06_GCN} on the outskirts of a star-forming galaxy at the same distance. The optical counterpart was monitored to late times with $HST$ \citep{yang+15}.

Numerous deep imaging observations and spectra, extending out to $\approx 65$~days, failed to reveal the expected counterpart to a long-duration GRB, a SN Ic-BL \citep{DellaValle+06,Fynbo+06,GalYam+06}. This fact, combined with its intriguing gamma-ray properties, motivated a later analysis showing that the optical light curve reddened at later times, suggesting instead the presence of a kilonova \citep{yang+15}. 

For our analysis, we use the XRT and UVOT light curves (though we do not use UVOT detections past 10 days as the exposures are on $\gtrsim$ day timescales; \citealt{Mangano+07}), ATCA radio upper limits \citep{Londish+06} and combine the optical datasets in the literature \citep{Cobb+06_GCN,DellaValle+06,Fynbo+06,Schmidt+06_GCN,Xu+09,yang+15}.

\subsection{GRB\,130603B}
\label{subsec:130603b}

GRB\,130603B was detected by \textit{Swift}-BAT and the Konus-\textit{Wind} Observatory \citep{bat130603b_Gcn,konus_130603b} on 2013 June 13 at 15:49:14 UT with T$_{90} = 0.18 \pm 0.02$~s \citep{berger+13}. Prompt \textit{Swift}-XRT observations revealed an X-ray counterpart \citep{bat130603b_Gcn}. Rapid optical follow-up revealed the optical afterglow within the X-ray localization \citep{levan+13_gcn}. Spectroscopy of the afterglow identified a GRB origin of $z=0.356$ \citep{foley+13_gcn,thone+13_gcn}. Additional multi-wavelength follow-up detected a radio counterpart and a well-sampled optical and X-ray afterglow (e.g., \citealt{Fong+14}). Later observations with $HST$ revealed that the optical counterpart had significantly reddened, resulting in a bright near-IR detection and deep limits in the optical bands. This represented the first bona-fide claim of an $r$-process-enriched kilonova \citep{berger+13,Tanvir+13}.

The kilonova detection of GRB\,130603B has been modeled by numerous groups in the literature, though with only a single detection, precise ejecta parameters remain uncertain (e.g., \citealt{berger+13,Tanvir+13,Hotokezaka+13,barnes+16}). We employ the multi-wavelength dataset compiled in \citet{Fong+14} and combine the optical through near-IR datasets published in the literature \citep{berger+13,Cucchiara+13,deUP+13,Tanvir+13,Fong+14}.

\subsection{GRB\,160821B}
\label{subsec:160821b}

GRB\,160821B was detected by the \textit{Fermi Space Telescope} Gamma-ray Burst Monitor (GBM; \citealt{Meegan+09}) and the \textit{Swift}-BAT on 2016 August 21 at 22:29:13 UT. It was classified as a short GRB with T$_{90} = 0.048 \pm 0.07$~s (\textit{Swift}; \citealt{lamb+19}) and T$_{90} \approx 1$~s (\textit{Fermi}; \citealt{Fermi_160821b}). \textit{Swift}-XRT quickly localized a counterpart \citep{Evans+16_GCN}. UVOT observed the location but did not detect a counterpart \citep{Breeveld+16_GCN}. Prompt follow-up identified optical and radio counterparts on the outskirts of a bright galaxy at $z=0.1616$ \citep{Fong+16_GCN,Levan+16_GCN,Xu+16_GCN}, motivating further multi-color follow-up by $HST$ and ground observatories (e.g., \citealt{Kasliwal+17,lamb+19,troja+19}).

The afterglow and kilonova of GRB\,160821B have been analyzed by multiple groups in the literature (e.g., \citealt{Kasliwal+17,lamb+19,troja+19}), and show evidence for an early reverse shock (e.g., \citealt{lamb+19}) and a bluer kilonova relative to AT\,2017gfo (e.g., \citealt{Rastinejad+21}). For our analysis, we collect \textit{Swift} and \textit{XMM-Newton} \citep{troja+19} and VLA radio observations \citep{lamb+19,Fong+21} from the literature. We combine the optical-NIR datasets \citep{Kasliwal+17,lamb+19,troja+19} and report where we draw specific measurements from in Table~\ref{tab:obs}.

\subsection{GRB\,170817A/GW\,170817}
\label{subsec:170817}

GRB\,170817A was detected by the \textit{Fermi}-GBM and INTernational Gamma-ray Astrophysics Laboratory (INTEGRAL; \citealt{INTEGRAL+03}) on 2017 August 17 at 12:41:06 UT, 1.7~s after the LVK-detected BNS merger GW\,170817 \citep{gw170817,gw170817mma,Goldstein+2017,Savchenko+17}. The gamma-ray duration was T$_{90} = 2.0 \pm 0.5$~s (50–300 keV; \textit{Fermi}; \citealt{Goldstein+2017}). An optical counterpart (AT\,2017gfo) was first found $\delta t =  0.452$~days following the event discovery \citep{Arcavi+17,Coulter+17,Lipunov+17,Tanvir+17,Soares-Santos+17,Valenti+17}. The community observed the transient in exquisite temporal and color detail, revealing a fast-fading, reddening source with spectroscopic evidence of the radioactive decay of lanthanide elements (e.g., \citealt{Chornock+17,Gillanders+22,kasen+17,kilpatrick+17,Nicholl+17,Pian+17,Smartt+17,Watson+19,Hotokezaka+23}).

We utilize the optical-NIR dataset compiled in \citealt{villar+17} with observations from the literature \citep{Andreoni+17,Arcavi+17,Coulter+17,Cowperthwaite+17,Diaz+17,Drout+17,Evans+17,Hu+17,Kasliwal+17,Lipunov+17, Pian+17, Pozanenko+17, Shappee+17, Smartt+17, Tanvir+17,Troja+17, Utsumi+17,Valenti+17} extending out to $\approx 25$ days. Specifically, we employ the set of observations used in their analysis, as there are known inconsistencies amongst the full dataset \citep{villar+17}. We do not collect multi-wavelength data to model this component as it is well-established that the off-axis afterglow did not contaminate observations on the timescales of the kilonova ($\delta t \lesssim 30$~days; e.g., \citealt{Lyman+18,Fong+19,MarguttiChornock20,Kilpatrick+21}).

\subsection{GRB\,200522A}
\label{subsec:200522a}

GRB\,200522A was discovered on 2020 May 22 at 11:41:34 UT by \textit{Swift}-BAT with T$_{90} = 0.62 \pm 0.08$~s \citep{200522a_Swift}. A prompt XRT counterpart location was reported \citep{200522a_Swift}, within which a catalogued galaxy with photometric redshift $z_{\rm phot} = 0.4 \pm 0.1$ was noted \citep{Alam+15,Fong+20_GCN}. Subsequent observations identified a radio counterpart \citep{200522a_VLA,Fong+21} and secured the host redshift to $z = 0.554$ \citep{Fong+21,O'Connor+21}. Follow-up by $HST$ and ground observatories uncovered a fading optical-near-IR counterpart embedded in the host galaxy \citep{LCO_GCN_200522a,Fong+21,O'Connor+21}.

We draw the \textit{Swift}-XRT observations from the United Kingdom Swift Science Data Centre (UKSSDC; \citealt{Evans+07,Evans+09}) and incorporate \textit{Chandra} and VLA observations \citep{Fong+21}. We combine the optical-near-IR data in the literature \citep{Fong+21,O'Connor+21}.

\subsection{GRB\,211211A}
\label{subsec:211211a}

GRB\,211211A was detected on 11 December 2021 at 13:09:59 UT by \textit{Swift}-BAT, \textit{Fermi}-GBM, INTEGRAL, and the CALET Gamma-ray Burst Monitor \citep{grb211211a_gbm,GRB211211a_bat,grb211211a_CALET,grb211211a_integral}. It was reported as a bright, long-duration GRB with T$_{90} =  50.7 \pm 0.9$~s (\textit{Swift}) and T$_{90} = 34.3 \pm 0.6$~s (\textit{Fermi}; \citealt{Veres+23}). \textit{Swift}-XRT and UVOT promptly identified counterparts to the burst \citep{grb211211a_xrt,grb211211a_uvot}. Notably, the detection of the afterglow in the $UVW2$ filter limits the event origin to $z<1.4$ \citep{Rastinejad+22}. An early optical counterpart was discovered proximate to galaxy SDSS J140910.47+275320.8 \citep{GCN_211211A_KAITag}. Later spectroscopy revealed a featureless afterglow \citep{Rastinejad+22} and a galaxy redshift of $z=0.0763$, rendering it one of the most nearby GRBs observed across all durations \citep{Rastinejad+22,Troja+22}. 

Motivated by the low redshift of the putative host galaxy, the counterpart was followed in the optical and near-IR. These observations revealed a fast-fading, red transient with similar luminosities and behavior to AT\,2017gfo \citep{Rastinejad+22}. Later ($\delta t \approx$2-3 weeks), deep optical upper limits revealed no sign of a supernova counterpart to a luminosity lower than that of any known GRB-SN \citep{Rastinejad+22, Troja+22}, further motivating an interpretation of the red excess as a kilonova. The kilonova of GRB\,211211A has been modeled in the literature by numerous groups (e.g., \citealt{Rastinejad+22,Troja+22,Mei+22,Yang+22,Kunert+23}). For our analysis, we use \textit{Swift}-XRT observations from UKSSDC, \textit{XMM-Newton} upper limits \citep{Mei+22} and a VLA radio observation \citep{Rastinejad+22}. We use the \textit{Swift}-UVOT and optical-near-IR datasets from \citealt{Rastinejad+22}, and add early optical data from \citealt{Troja+22}.

\subsection{GRB\,230307A}
\label{subsec:230307a}

GRB\,230307A was detected on 7 March 2023 15:44:06.67 UT by \textit{Fermi}-GBM with a duration of $\approx 35$~s \citep{230307a_GBM,230307a_t90_GBM}. The burst was also observed by GECAM, the InterPlanetary Network (IPN), AGILE, AstroSAT, and GRBAlpha \citep{GECAM_GCN_230307a,IPN_GCN_230307a,AGILE_GCN_230307a,AstroSAT_GCN_230307a,GRBAlpha_GCN_230307a}, and was quickly noted as the second-brightest GRB seen by \textit{Fermi} to date \citep{Burns_GCN_230307a}. The ULTRACAM instrument mounted on the 3.5~m New Technology Telescope (NTT) and \textit{Swift}-XRT undertook wide-field searches for an optical and X-ray counterpart, respectively, discovering coincident candidates at $\delta t = 1.4-1.7$~days \citep{XRT_GCN_230307a,Levan+24}. The counterpart was offset 30.2$''$ (38.9~kpc) from a bright spiral galaxy confirmed at $z=0.0646$ \citep{Levan+24}, and an extensive multi-wavelength follow-up campaign was initiated. 

Despite GRB\,230307A's nominal long duration, its high-energy properties, including its spectral lag and X-ray flux decay, provided evidence for a compact object merger origin similar to GRB\,211211A. Near-IR follow-up from ground-based observatories and \textit{JWST} revealed a late-time red excess, light curve shape, and spectral features expected for a kilonova (e.g., \citealt{Gillanders+23,Gillanders+24,Levan+24,Yang+24}). We combine the multi-wavelength datasets presented in the literature, which includes \textit{Chandra}, \textit{Swift}, and \textit{XMM} X-ray observations, broad coverage by ground observatories, late-time \textit{HST} and \textit{JWST} observations, and ATCA and AMI-LA radio observations \citep{Levan+24,Yang+24} for our analysis. 

\section{Afterglow Modeling}
\label{sec:ag_mod}

In addition to a radioactive decay-powered kilonova, BNS mergers are expected to launch a relativistic jet whose interaction with the surrounding medium produces broadband synchroton emission, or the ``afterglow'' (e.g., \citealt{Sari+98}). The afterglow flux is a potential contaminant for modeling kilonovae, especially for on-axis events in which the afterglow often dominates the total luminosity at early times, motivating us to model the afterglows of each event in our sample, with the exception of GW\,170817 (see Section~\ref{subsec:170817}). Our afterglow model uses the formulae of \citet{GranotSari02} and methods described in \citet{Laskar+14} to describe synchrotron emission from a forward shock (FS), produced by the interaction of the GRB's collimated jet and the surrounding medium, incorporating the effects of Inverse Compton cooling \citep{SariEsin01,Laskar+15}. 

\begin{figure*}
\centering
\includegraphics[angle=0,width=\textwidth]{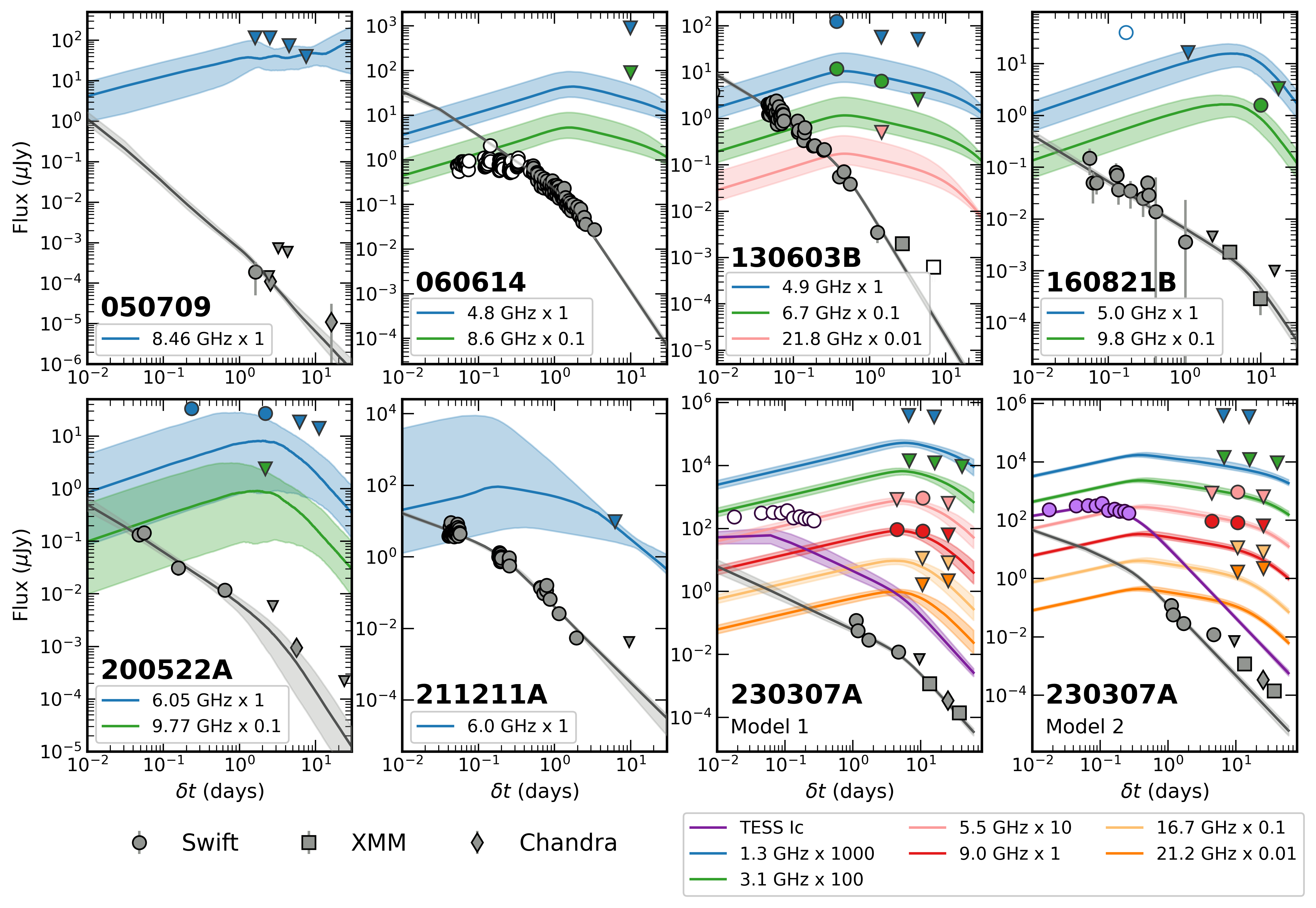}
\caption{The X-ray and radio detections (circles) and 3$\sigma$ upper limits (triangles) of the seven events in our sample with on-axis afterglows. In each panel, we show the model light curves' median and 68\% confidence range (shaded region) along with the X-ray (grey) and radio (colored and labeled) observations. Open symbols denote data that was masked in the afterglow fit (Section~\ref{sec:ag_mod}). We show two models for GRB\,230307A. The first (``Model 1'') is fit to the X-ray and radio data only and provides a poor fit to the early TESS observations. The second (``Model 2'') is fit with the TESS, radio and X-ray observations. It provides a better constraint on the optical contribution but cannot fully account for the late-time X-ray and radio observations.}
\label{fig:ag_data}
\end{figure*}

The parameters fit in this afterglow model are the jet isotropic-equivalent energy ($E_{\rm K,iso}$), the circumburst density of the surrounding medium ($n_0$), the input electron distribution power law index ($p$), the fraction of energy deposited into non-thermal relativistic electrons ($\epsilon_{\rm e}$), the line-of-sight extinction in $B$-band ($A_{\rm B}$; fixed to 0 in our fits except for GRB\,130603B, see below), and the time of the jet break ($t_{\rm jet}$), which is directly related to the jet opening angle \citep{Rhoads99,Sari+99}. For each event we assume a radially-homogeneous ISM-like environment ($k=0$), as this is expected in the local environments for most NS mergers. We fix the energy deposited in magnetic fields ($\epsilon_{\rm B}$) to 0.01, the median for short GRBs \citep{fong+15,Schroeder+24}, as the model fails to converge to a reasonable solution when leaving this parameter free. We fit the afterglow model using the Markov Chain Monte Carlo (MCMC) \texttt{emcee} package \citep{emcee13}, enforcing a minimum $10\%$ uncertainty on all detections to capture realistic measurement errors. We run each fit using 128 walkers for 5000 iterations and discard the first $10\%$ of steps as burn-in. For each event, we employ the redshifts and extinction values listed in Table~\ref{tab:grbprops} and use values from the literature as starting parameters.

As the kilonova and afterglow both contribute flux in the optical-to near-IR wavelengths, disentangling emission between the two can be difficult, especially at early times. In particular, observations of AT\,2017gfo demonstrated that kilonovae may have early ($\delta t \lesssim 1$~day) blue emission due to large quantities of fast-moving lanthanide-poor ejecta or, potentially, additional energy sources such as free neutron decay (e.g., \citealt{Metzger+15}) or shock-heating (e.g., \citealt{Kasliwal+17,villar+17,Arcavi18,Nicholl+21}). Thus, we attempt to remain agnostic to the precise afterglow contribution in the optical and mask data in the range $10^{13} - 10^{16}$~Hz (effectively, fitting only the radio and X-ray observations) in our fits to GRBs\,160821B, 200522A, 211211A and 230307A. For GRB\,230307A, we find this method significantly underestimates the flux of TESS optical observations at $0.01 \lesssim \delta t \lesssim 0.2$~days (Figure~\ref{fig:ag_data}; ``Model 1'') while a fit including the TESS observations provides a worse fit to the X-ray and radio data (``Model 2''). We perform our analysis on the results from both models due to the uncertainty in the emission source. For GRB\,160821B, we exclude the radio detection at $\delta t = 0.17$~days as it is likely the result of an early reverse shock \citep{lamb+19}, and incompatible with the standard FS model. We do not expect the reverse shock to significantly contaminate the optical flux on the timescales of observations we use to model the kilonova ($\delta t > 0.95$~days).

For the remaining three events we include some early optical data in our fit either due to sparse X-ray and radio detections (GRBs\,050709, 060614) or high line-of-sight extinction that will significantly affect our estimation of the afterglow flux contribution (GRB\,130603B; \citealt{Cucchiara+13,Fong+14,fong+15}). While it is possible that some kilonova flux may be contributing in these optical detections, previous works have shown that these points can be explained with an afterglow model only (e.g., \citealt{Fox+05,Fong+14,Xu+09}). Specifically, for GRB\,050709, we include two early ($\delta t < 2.4$~days) $R$-band optical observations. For GRB\,060614, we include early $RIJK$-band data ($\delta t = 0.7 - 2$~days) in our afterglow fit. We exclude X-ray data of GRB\,060614 prior to $\delta t = 0.5$~days as numerous analyses favor an energy injection scenario to explain the afterglow plateau observed at $\delta t \lesssim 0.5$~day that is incompatible with the standard afterglow model (Figure~\ref{fig:ag_data}; e.g., \citealt{Xu+09,Mangano+07}). We do not expect the exclusion of the energy injection episode in GRB\,060614 to affect our kilonova modeling as the FS is expected to dominate on the timescales and in the filters we use to model the kilonovae. Finally, for GRB\,130603B we include early ($\delta t \lesssim $~day) optical-near-IR detections to measure the line-of-sight dust, $A_B$, which we propagate to the output models we use for subtraction. We also exclude the final \textit{XMM-Newton} observation of GRB\,130603B as it is known to be contaminated by an unrelated X-ray source \citep{Rouco+23}. We denote which observations were used in the afterglow fitting in Table~\ref{tab:obs}.

For all events, we visually inspect the optical-near-IR afterglow models to ensure they are not brighter than measured values beyond the uncertainties. In general, we find that our derived afterglow physical parameters are consistent with those in the literature, though in some cases we find inconsistent values (typically, in the degenerate parameters $E_{\rm K,iso}$ and $n_0$). Variations are likely the result of discrepancies between modeling codes or the fact that we primarily use X-ray and radio data only; however the inferred afterglow parameters are not used in any subsequent analysis\footnote{With the exception of the jet break in GRB\,230307A, which is incorporated in Section~\ref{sec:long_short}.}, so the specific values are not important for this work. We create model light curves at the same rest-frame wavelengths as the optical-near-IR observations for 1000 random draws from the full posterior of each event. From these 1000 draws we calculate the median and 68\% credible flux range, using the 68\% flux range as our uncertainty on the afterglow model. We show our fits to the multi-wavelength data, along with their uncertainties, in Figure~\ref{fig:ag_data}. Following interpolation of the afterglow light curves to the $\delta t$ of each observation, we subtract the median afterglow model flux from each observed flux, producing an ``afterglow-subtracted'' light curve. We combine the data uncertainty and model uncertainty at the time of the observation in quadrature. We present our afterglow-subtracted values and errors in Table~\ref{tab:sub_obs}.

\section{\texttt{MOSFiT} Kilonova Modeling}
\label{sec:KN_mod}
\begin{deluxetable}{cccCC}
\savetablenum{2}
\tabletypesize{\small}
\centering
\tablecolumns{5}
\tabcolsep0.07in
\tablecaption{Kilonova Model Parameters \& Priors
\label{tab:prior}}
\tablehead {
\colhead {Parameter$^{\dagger}$}		&
\colhead {Units}		&
\colhead {Prior}		& 
\colhead {Min.} &
\colhead {Max.} 
}
\startdata
$M_{\rm ej}$ (B, R, P) & $M_{\odot}$ & Log-Uniform & 0.001 & 0.5  \\ 
$V_{\rm ej}$ (B, R, P) & $c$ & Uniform & 0.03 & 0.3 \\
$T_{\rm floor}$ (B, R, P) & K & Log-Uniform & 1000$^*$ & 4000 \\
cos($\theta_{\rm open}$) & -- & Uniform & 0.5 & 0.866 \\
$\sigma$ & -- & Log-Uniform & 0.001 & 100 
\enddata
\tablecomments{$^{\dagger}$``B'', ``P'', and ``R'' refer to the blue, purple and red ejecta components, as described in Section~\ref{sec:KN_mod}. Each components' parameters are modeled separately with different $\kappa$ values but have the same priors, minima, and maxima. \\
$^*$For GRB\,230307A only we allow this minimum to extend to $T_{\rm floor, P} = 800$ K and $T_{\rm floor, R} = 440$ K due to the temporal and wavelength coverage of the kilonova (Section~\ref{sec:KN_mod}).
}
\end{deluxetable}

\subsection{Description of Kilonova Model}
\label{sec:KN_mod_description}
We employ the Python-based Modular Open Source Fitter for Transients (\texttt{MOSFiT}) code \citep{Guillochon+17} to fit kilonova models to the afterglow-subtracted data, for which the kilonova contribution has nominally been isolated. From this modeling, we derive physical parameters describing the observed kilonova emission, which we parameterize in terms of ejecta mass ($M_{\rm ej}$), velocity ($V_{\rm ej}$) and temperature cooling floor ($T_{\rm floor}$).  The latter is the temperature below which the photosphere recedes into the ejecta. Hence, $T_{\rm floor}$ represents an effective emission temperature for the optically-thin nebular phase to follow. We elect to use \texttt{MOSFiT} as its modular design affords us the flexibility to build new modules, freeze or adjust parameters and their priors, add constraints, and test several samplers without a high computational cost. In addition, \texttt{MOSFiT} is a well-tested method that has been used to fit or determine upper ejecta mass limits of several past kilonovae (e.g., \citealt{villar+17,Nicholl+21,Rastinejad+22,Coulter+24}). All of the results presented in this section were performed with nested sampling, as implemented in the \texttt{dynesty} fitting routine \citep{Speagle+20}.

We begin with \texttt{MOSFiT}'s existing three-component (blue, purple and red; see below) kilonova model \citep{Villar+17_model,metzger19}, which assumes analytic forms for the radioactive heating rate (\citealt{Korobkin+12}; equation 1 of \citealt{villar+17}) and the thermalization efficiency (\citealt{barnes+16}; equation 2 of \citealt{villar+17}). The model calculates the bolometric luminosity assuming a central energy source and following an updated \citet{Arnett82} formalism (\citealt{Chatzopoulos+12}; equation 3 of \citealt{villar+17}). Here, each of the three components is parameterized by a constant ``grey'' ejecta opacity ($\kappa$), the value of which correlates with lanthanide or electron fractions in portions of the total ejecta. Kilonova models comprised of two or three components were found to provide a better fit to the well-sampled AT\,2017gfo compared to single component models \citep{Cowperthwaite+17,Drout+17,kasen+17,kilpatrick+17,Tanaka+17,villar+17}. They are also physically motivated by simulations that predict multiple ejecta mechanisms with distinct elemental compositions prior to and following the NS merger (e.g., \citealt{metzgerfernandez14,kasen+17}).  

For our three-component model we employ $\kappa_{\rm R} = 10$~cm$^2$~g$^{-1}$, $\kappa_{\rm P} = 3$~cm$^2$~g$^{-1}$, and $\kappa_{\rm B} = 0.5$~cm$^2$~g$^{-1}$ for the ``red'', ``purple'' and ``blue'' components, respectively \citep{Tanaka+17}. We expect these components to roughly map to the red (lanthanide-rich or $Y_{\rm e} \lesssim 0.2$) tidal dynamical ejecta (e.g., \citealt{Rosswog+99}), the purple (moderately lanthanide-rich or $0.2 \lesssim Y_{\rm e} \lesssim 0.3$) disk wind ejecta (e.g., \citealt{metzgerfernandez14,just+15,FernandezMetzger16,Lippuner+17}), and the blue (lanthanide-poor or $Y_{\rm e} \gtrsim 0.3$) dynamical ejecta shocked at the NS contact interface and ejected near the poles (e.g., \citealt{sekiguchi+16}) or ejected in a magnetized wind from the neutron star remnant prior to black hole formation (e.g., \citealt{Metzger+18,Fong+21,Combi&Siegel23,Curtis+24}). For each component in our model, we measure $M_{\rm ej}$, $V_{\rm ej}$ and $T_{\rm floor}$ (Table~\ref{tab:prior}). We pursue a three-component model (as opposed to a two-component model) as this provides a better mapping of ejecta mechanism to opacity, enabling a search for trends tied to physical properties.

\begin{figure*}
\centering
\includegraphics[angle=0,width=0.9\textwidth]{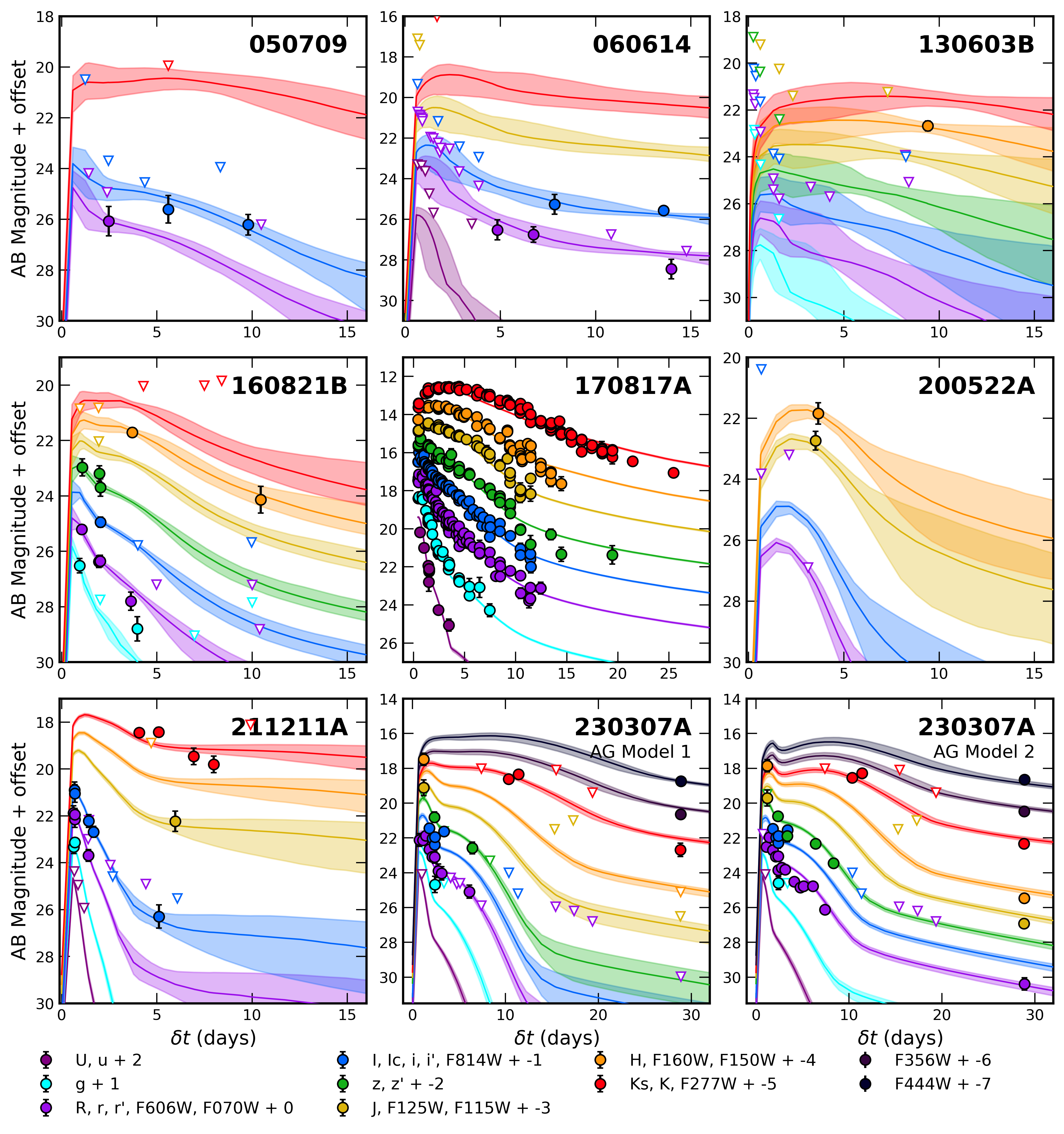}
\caption{Afterglow-subtracted observations of the eight kilonovae in our sample along with best-fit (median and 68\% confidence) model light curves. 3$\sigma$ upper limits are marked with triangles, while detections are marked with circles. We show light curves in the filters where observations (including upper limits) are available for each event. Two models are shown for GRB\,230307A based on the two afterglow models described in Section~\ref{sec:ag_mod} and shown in Figure~\ref{fig:ag_data}.}
\label{fig:kn_models_subdata}
\end{figure*}

The geometry of the mechanism producing each ejecta component and the viewing angle of the observer are known to significantly impact the observed light curve (e.g., \citealt{DarbhaKasen20,Chase+22}), and thus any parameter inference. Of particular relevance to this work, the assumption of an isotropic kilonova will likely introduce a bias in estimating the mass of material ejected along the line-of-sight. Thus, under this assumption, GRB events observed along the jet axis likely have blue and red ejecta components that are overestimated and underestimated, respectively. Instead, here we account for the geometry of the ejecta by modifying the original spherical kilonova model \citep{villar+17} to an aspherical model, wherein a half-opening angle ($\theta_{\rm open}$) defines a conical boundary between red ejecta, confined to the equatorial region, and the blue and purple ejecta, modeled in the direction of the poles. We use the half-opening angle prescription of \citet{DarbhaKasen20}, as implemented in \texttt{MOSFiT} by \citet{Nicholl+21}. The viewing angle ($\theta_{\rm obs}$; defined relative to the axis of the GRB jet) of each event in our sample is well-established, either as relatively pole-on due to the detection of a cosmological GRB or measured through high-precision astrometry in the case of AT\,2017gfo \citep{Mooley+22}. Thus, we fix $\theta_{\rm obs} = 22\degree$ (the central value in the range given) for AT\,2017gfo \citep{Mooley+22} and $\theta_{\rm obs} = 0\degree$ for all other events in our sample. We allow the kilonova ejecta half-opening angle to be a free parameter and include an additional parameter ($\sigma$) to account for white noise in the likelihood function. These, in addition to $M_{\rm ej}$, $V_{\rm ej}$ and $T_{\rm floor}$ for each of the three components, comprise the 11 parameters measured for each kilonova. 

We acknowledge that the number of free parameters exceeds the number of data points for some events in our sample. We perform several test fits fixing each component's $T_{\rm floor}$ and $\theta_{\rm open}$, finding consistent masses compared to runs where these parameters are left free. In the end, we opt to keep these parameters free in order to marginalize over the uncertainties on other parameters when measuring masses. Uniform use of the three-component model is critical to comparisons between kilonovae. In these cases we find very broad posteriors on the masses (e.g., Figure~\ref{fig:med_mass_4panel} and Table~\ref{tab:posterior}), but are still able to constrain them compared to the broad uniform prior.

\renewcommand{\arraystretch}{1.25}
\begin{deluxetable*}{cCCCCCCCCCC}
\savetablenum{3}
\tabletypesize{\small}
\centering
\tablecolumns{11}
\tabcolsep0.05in
\tablecaption{Kilonova Model Posteriors
\label{tab:posterior}}
\tablehead {
\colhead {GRB} &
\colhead {cos($\theta_{\rm open}$)}		&
\colhead {$M_{\rm ej, B}$}		&
\colhead {$M_{\rm ej, P}$}		&
\colhead {$M_{\rm ej, R}$}		&
\colhead {$V_{\rm ej, B}$}		& 
\colhead {$V_{\rm ej, P}$}		&
\colhead {$V_{\rm ej, R}$}		&
\colhead {$T_{\rm floor, B}$ } & 
\colhead {$T_{\rm floor, P}$ } &
\colhead {$T_{\rm floor, R}$ } 
\\
 &
		&
\colhead {($M_{\odot}$)}		&
\colhead {($M_{\odot}$)}		&
\colhead {($M_{\odot}$)}		&
\colhead {($c$)}		& 
\colhead {($c$)}		&
\colhead {($c$)}		&
\colhead {(K) } & 
\colhead {(K) } &
\colhead {(K)} }
\startdata
050709 & 0.68^{+0.13}_{-0.12} & 0.003^{+0.004}_{-0.002} & 0.029^{+0.010}_{-0.009} & 0.015^{+0.072}_{-0.012} & 0.18^{+0.08}_{-0.09} & 0.07^{+0.05}_{-0.03} & 0.16^{+0.09}_{-0.09} & 2040^{+1183}_{-775} & 2807^{+833}_{-1322} & 1904^{+1146}_{-662} \\
060614 & 0.73^{+0.10}_{-0.18} & 0.014^{+0.013}_{-0.010} & 0.029^{+0.034}_{-0.027} & 0.146^{+0.072}_{-0.120} & 0.19^{+0.06}_{-0.07} & 0.15^{+0.09}_{-0.09} & 0.22^{+0.05}_{-0.07} & 2546^{+1059}_{-1157} & 2743^{+675}_{-1280} & 3030^{+513}_{-786} \\
130603B & 0.68^{+0.13}_{-0.12} & 0.006^{+0.017}_{-0.004} & 0.075^{+0.063}_{-0.037} & 0.023^{+0.160}_{-0.020} & 0.14^{+0.11}_{-0.08} & 0.11^{+0.12}_{-0.05} & 0.17^{+0.09}_{-0.09} & 1990^{+1185}_{-737} & 2085^{+1145}_{-793} & 1982^{+1197}_{-711} \\
160821B & 0.63^{+0.14}_{-0.09} & 0.003^{+0.001}_{-0.001} & 0.011^{+0.002}_{-0.002} & 0.011^{+0.021}_{-0.009} & 0.14^{+0.06}_{-0.05} & 0.12^{+0.06}_{-0.03} & 0.15^{+0.09}_{-0.08} & 2038^{+1201}_{-759} & 3792^{+156}_{-1067} & 2069^{+1226}_{-769} \\
170817A & 0.86^{+0.00}_{-0.00} & 0.004^{+0.000}_{-0.000} & 0.019^{+0.001}_{-0.001} & 0.052^{+0.003}_{-0.002} & 0.15^{+0.01}_{-0.01} & 0.15^{+0.01}_{-0.01} & 0.20^{+0.01}_{-0.01} & 1745^{+239}_{-261} & 3152^{+30}_{-29} & 1004^{+6}_{-3} \\
200522A & 0.67^{+0.13}_{-0.12} & 0.046^{+0.009}_{-0.009} & 0.020^{+0.068}_{-0.018} & 0.019^{+0.124}_{-0.016} & 0.23^{+0.04}_{-0.06} & 0.18^{+0.08}_{-0.09} & 0.16^{+0.09}_{-0.09} & 1969^{+1183}_{-725} & 1938^{+1173}_{-694} & 2043^{+1150}_{-756} \\
211211A & 0.85^{+0.01}_{-0.02} & 0.007^{+0.000}_{-0.001} & 0.010^{+0.001}_{-0.002} & 0.130^{+0.051}_{-0.054} & 0.28^{+0.01}_{-0.02} & 0.27^{+0.02}_{-0.04} & 0.27^{+0.02}_{-0.04} & 1800^{+549}_{-427} & 2048^{+463}_{-554} & 1817^{+304}_{-623} \\
230307A$^{\dagger}$ & 0.82^{+0.03}_{-0.10} & 0.012^{+0.001}_{-0.001} & 0.025^{+0.003}_{-0.003} & 0.054^{+0.019}_{-0.022} & 0.20^{+0.02}_{-0.02} & 0.09^{+0.01}_{-0.01} & 0.20^{+0.05}_{-0.06} & 1655^{+229}_{-250} & 830^{+42}_{-22} & 464^{+46}_{-18} \\
230307A$^{\ddagger}$ & 0.54^{+0.09}_{-0.03} & 0.008^{+0.001}_{-0.001} & 0.037^{+0.004}_{-0.005} & 0.003^{+0.009}_{-0.002} & 0.19^{+0.03}_{-0.02} & 0.11^{+0.01}_{-0.01} & 0.19^{+0.07}_{-0.09} & 3718^{+201}_{-357} & 831^{+45}_{-23} & 1626^{+1635}_{-1033} \\
\enddata
\tablecomments{``B'', ``P'', and ``R'' subscripts refer to the blue, purple and red ejecta components, as described in Section~\ref{sec:KN_mod}. \\
$^{\dagger}$Afterglow model 1 fit to X-ray and radio data only.\\
$^{\ddagger}$Afterglow model 2 fit to X-ray, radio and TESS data.
}
\end{deluxetable*}

We list the parameters and priors used in our kilonova model in Table~\ref{tab:prior}. For $M_{\rm ej}$ and $V_{\rm ej}$ we elect to use the widest priors possible that correspond to physical values based on simulations (e.g., \citealt{Radice+18,Nedora+21}). For $T_{\rm floor}$ we use the range 1000-4000 K following the reasoning of \citet{Nicholl+21} based on previous fits to AT\,2017gfo with \texttt{MOSFiT} \citep{Cowperthwaite+17,villar+17}. For GRB\,230307A, \textit{JWST} detections at $\delta t \approx 29$ and 61~days likely occur at epochs when nebular emission is either significantly contributing or dominating the observed flux. As kilonova nebular emission remains a challenge to properly model, we do not include the observations at $\delta t \approx 61$~days in our kilonova fit. To accommodate the observations at $\delta t \approx 29$~days and the mid-IR coverage of the \textit{JWST} detections, we allow the  $T_{\rm floor, P}$ and $T_{\rm floor, R}$ prior range to extend down to 800 and 440 K, respectively. We base our choice of $T_{\rm floor, R} = 440$ K on the blackbody fit to synthetic nebular lanthanide-rich kilonova spectra and \textit{Spitzer} detections of AT\,2017gfo \citep{Kasliwal+22,Hotokezaka+21,BarnesMetzger22}. We choose a median value of $T_{\rm floor, P} = 800$ K as we expect the blackbody temperature of moderately lanthanide-rich ejecta to fall in between that of lanthanide-poor and lanthanide-rich ejecta.

We fit the eight afterglow-subtracted optical-near-IR light curves at $\delta t > 0.5$~days (Table~\ref{tab:sub_obs}) with the three-component kilonova model presented above. We do not include data prior to 0.5~days in our kilonova fits as these timescales may be affected by emission mechanisms beyond radioactive decay, including energy injection in the afterglow (GRB\,060614; \citealt{Mangano+07}), central engine activity (e.g., GRB\,211211A; \citealt{Hamidani+24}), or shock cooling (e.g., \citealt{Rastinejad+22}). For our fits, we include only data points whose combined statistical and systematic (from afterglow fitting; Section~\ref{sec:KN_mod}) errors are $< 0.5$~mag and treat all observations with combined errors $> 0.5$~mag as upper limits. In general, this results in detections prior to $\approx$1 day being treated as upper limits (e.g., Figure~\ref{fig:kn_models_subdata} and Table~\ref{tab:sub_obs}). For GRB\,050709, we employ a threshold of $< 0.6$~mag, as we find the inclusion of the two additional points provides a significantly better fit. We present the median and 68\% confidence range for each parameter measured in our kilonova modeling in Table~\ref{tab:posterior}.

\subsection{Results and Observed $M_{\rm ej}$ Diversity}
\label{sec:results}

In Figure~\ref{fig:kn_models_subdata} we plot median and 68\% confidence range model light curves, constructed from 900 random draws of the full posterior of each GRB (with the exception of GRB\,170817A for which we use 50 random draws due to computational constraints). Overall, we find that the models provide reasonable fits to the data and follow the predicted behavior of kilonovae: rapid decay, especially in bluer bands, and reddening over time. As expected, events with better observational coverage correspond to tighter constraints on the light curves. For AT\,2017gfo, we find that our model provides a good fit to the optical data and the majority of the near-IR observations, but overpredicts the late-time ($\delta t \gtrsim 20$~day) $K$-band observations, a feature also noted in previous \texttt{MOSFiT} fits to this event \citep{villar+17}. We discuss comparisons to previous fits further in Section~\ref{subsec:prev_fits}. 

For GRB\,230307A we find that our choice of afterglow model significantly affects the shape of the model optical light curve at later times, reflected in the disparate best-fit values found for $M_{\rm ej, R}$ and $T_{\rm floor, R}$. The fit to data subtracted with afterglow Model 1 (Figures~\ref{fig:ag_data} and \ref{fig:kn_models_subdata}; ```KN Fit 1'')  follows a typical fading behavior at $\delta t \gtrsim 10$~days. In contrast, the fit to data subtracted with afterglow Model 2 (``KN Fit 2''; in particular the \textit{JWST} F070W detection at $\approx$29~days which is an upper limit in KN Fit 1) produces a flattening in the optical decay past $\delta t \gtrsim 10$~days. To produce this shape, KN Fit 2 requires higher values of $T_{\rm floor, B}$ and $T_{\rm floor, R}$ than KN Fit 1 to explain the emission at later times (Table~\ref{tab:posterior}). We favor KN Fit 1 in our later analysis (though we show both for completeness), for several reasons. First, we prefer the afterglow model that provides a better fit to the X-ray and radio data than the early TESS observations (afterglow Model 1; Figure~\ref{fig:ag_data}) as there are a number of proposed emission mechanisms, including a reverse shock, shock cooling and free-neutron decay \citep{Metzger+15,Gottlieb+18}, that may explain the early TESS excess but could not account for a late-time radio and X-ray excess. Second, the true TESS bandpass is wider than the nominal $I_c$-band reported in the literature (e.g., \citealt{Yang+24}), and may also explain excess emission relative to the model. Third, KN Fit 1 results in more physically realistic values for $T_{\rm floor, R}$, as they are similar to the blackbody temperature that approximates kilonova nebular emission at the wavelengths where the nebular observations occur (e.g., 440 K; \citealt{BarnesMetzger22}). Finally, fits to both afterglow-subtracted datasets without \textit{JWST} observations are more consistent with KN Model 1.

\begin{figure*}
\centering
\includegraphics[angle=0,width=0.95\textwidth]{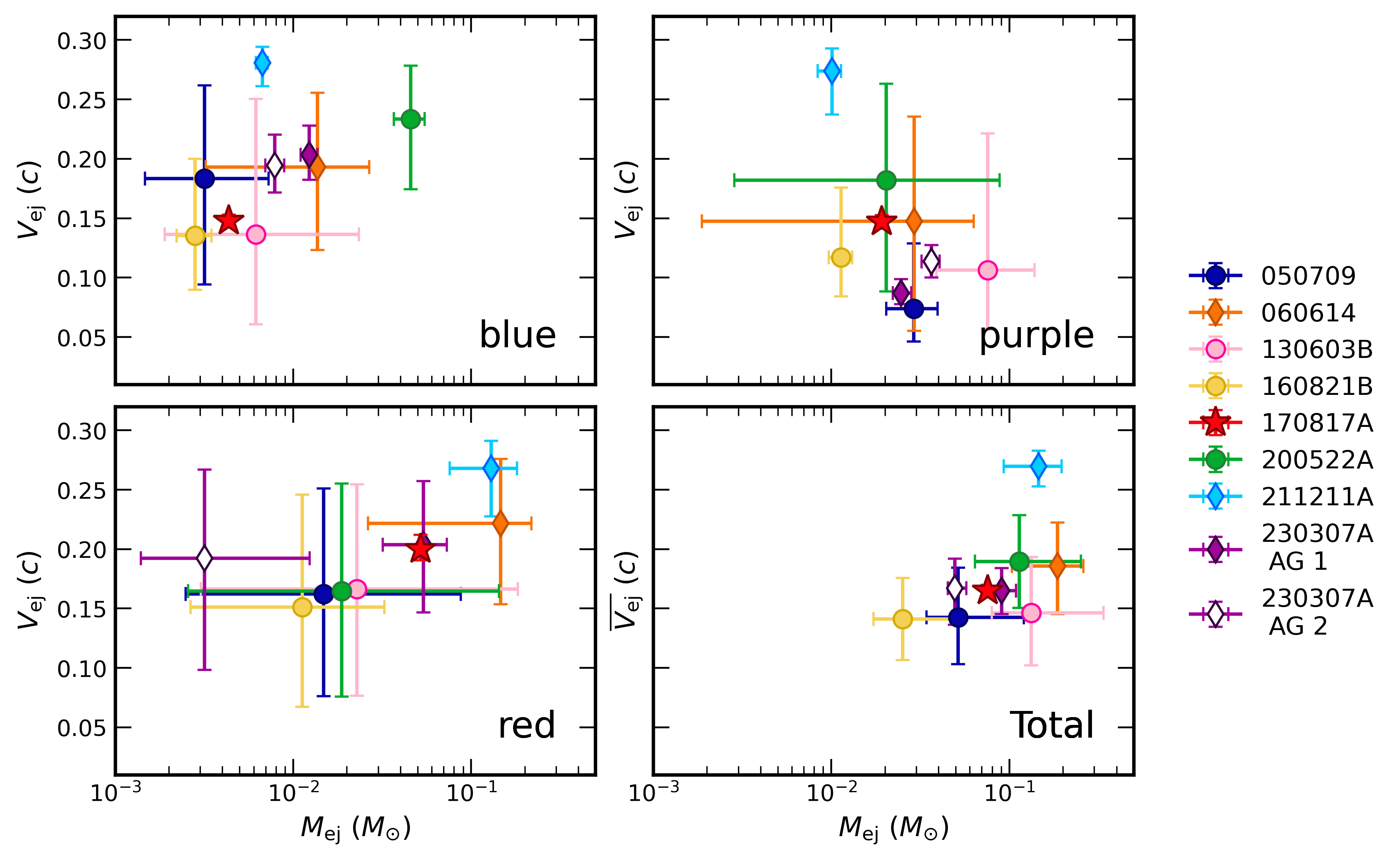}
\caption{Median $M_{\rm ej}$ and $V_{\rm ej}$ (68\% confidence) for the blue (top left), purple (top right), and red (bottom left) components as well as the total ejecta (bottom right). We mark long GRBs with diamonds, GW\,170817/AT\,2017gfo with a star, and the remaining short GRBs with circles. Our analysis highlights that GW\,170817/AT\,2017gfo is a ``typical'' kilonova.}
\label{fig:mej_vej}
\end{figure*}

In Figure~\ref{fig:mej_vej} we show the median and 68\% confidence range for $M_{\rm ej}$ and $V_{\rm ej}$ for each component and the total ejecta. In keeping with previous works (e.g., \citealt{Rastinejad+21}), our fits place the tightest constraints on the blue and purple component parameters and the coarsest constraints on the red component. This is in large part due to the traditional use of more sensitive optical telescopes for afterglow searches, especially for the kilonova candidates detected prior to GW\,170817. Notably, our fits to events with just two or three detections and deep upper limits (GRBs\,050709 and 200522A) place order-of-magnitude or tighter constraints on $M_{\rm ej, B}$ and $M_{\rm ej, P}$. In Section~\ref{sec:long_short} we analyze and compare the kilonova properties of long and short GRBs.

\begin{figure*}
\centering
\includegraphics[angle=0,width=0.8\textwidth]{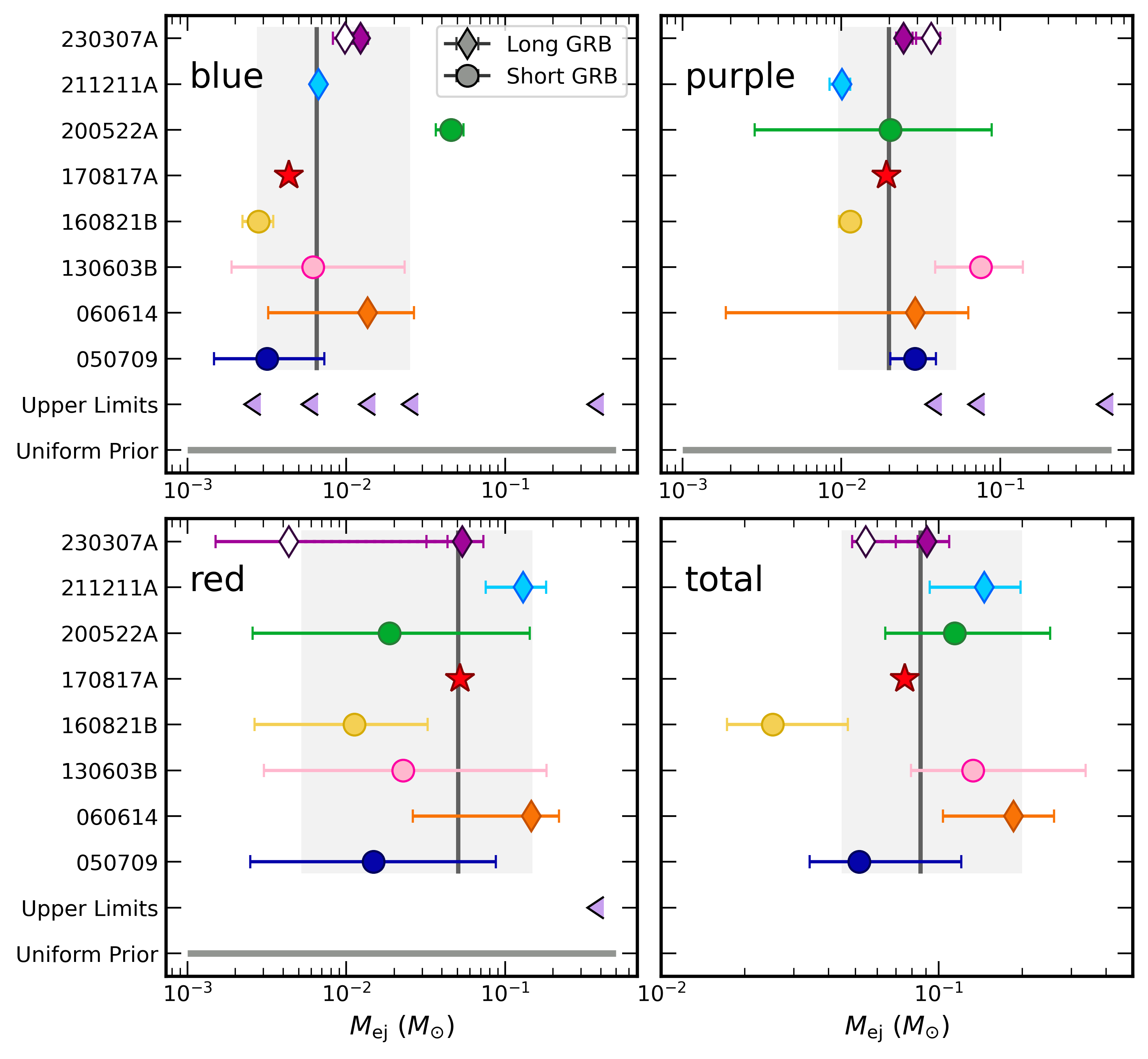}
\caption{The median and 68\% confidence range of the $M_{\rm ej}$ found for each GRB in the blue (top left), purple (top right), and red (bottom left) components as well as the total ejecta mass (bottom left). Long GRBs, short GRBs and AT\,2017gfo are marked with diamonds, circles, and a star, respectively. We show the median and 68\% confidence range across all events in the grey vertical line and shaded region, and mark the log-uniform prior for each component in the grey horizontal line. We also mark upper limits on the component ejecta masses from observations of short GRB afterglows with left-facing light purple triangles (Section~\ref{subsec:uls_from_grbs}). We show our two fits for GRB\,230307A, marking KN Fit 2 with an open symbol as we disfavor this fit in our analysis (Section~\ref{sec:KN_mod}). GW\,170817/AT\,2017gfo falls comfortable within the 68\% confidence range for each ejecta mass.}
\label{fig:med_mass_4panel}
\end{figure*}

Focusing on the blue component, our analysis finds that all except one event prefer $M_{\rm ej, B} \leq 0.01 M_{\odot}$ (Table~\ref{tab:posterior}). We also observe a general trend between increasing $M_{\rm ej, B}$ and $V_{\rm ej, B}$ (Figure~\ref{fig:mej_vej}), though the large error bars on several events prohibit a firm conclusion. Taking 1000 draws from the posterior of each event (and discarding GRB\,230307A KN Fit 2), we calculate a median of $M_{\rm ej, B} = 0.006_{-0.004}^{+0.015} M_{\odot}$ (68\% confidence; Figure~\ref{fig:med_mass_4panel}). 

The blue kilonova emissions may be attributed to either dynamical ejecta heated at the contact surface between the NSs and ejected along the the axis of the jet (e.g., \citealt{Oechslin+07,sekiguchi+16}) or post-merger disk ejecta experiencing neutrino irradiation from a NS remnant, which lowers the lanthanide-richness of any ejecta (e.g., \citealt{metzgerfernandez14,Miller+18}). 
Notably, GRB\,200522A is a significant outlier in $M_{\rm ej, B}$ (Figure~\ref{fig:med_mass_4panel}) with $M_{\rm ej, B} = 0.046 \pm 0.009 M_{\odot}$, consistent with past findings \citep{Fong+21}. We slightly favor a disk-wind source (rather than a shock-heated dynamical source) to explain the majority of GRB\,200522A's larger $M_{\rm ej, B}$ as BNS merger simulations that measure shock-heated dynamical ejecta do not produce $M_{\rm ej, B} \gtrsim 0.01 M_{\odot}$, even when spanning a range in NS masses, mass ratios and two NS equations of state \citep{sekiguchi+16}. In contrast, simulations measuring a disk-wind mass in the case of a long-lived NS remnant produce $M_{\rm ej, B} \approx 0.03 M_{\odot}$ (e.g., \citealt{Lippuner+17}). The extreme luminosity of GRB\,200522A's kilonova has previously been explained with the creation of a magnetar remnant, which may provide an additional blue emission source \citep{Fong+21}. 

Turning to the purple component, we find a population median of $M_{\rm ej, P} = 0.020_{-0.010}^{+0.034} M_{\odot}$ (68\% confidence; Figure~\ref{fig:med_mass_4panel}). This range is broadly consistent with expectations of disk component masses (e.g., \citealt{Lippuner+17}). We find that several events (GRBs\,050709, 130603B and 230307A) prefer lower ejecta velocities compared to those found for the blue or red components. This trend supports the purple component's source as a disk wind which is likely ejected with slower speeds compared to the dynamical red and blue components ($V_{\rm ej} \approx 0.01 - 0.1c$; e.g., \citealt{metzgerfernandez14,Fernandez+15}). We find that GRB\,211211A is an outlier on the higher end in $V_{\rm ej, P}$. As our fit to this event finds high velocities for all three components, we posit this is likely a reflection of the fast-decaying UV-optical emission observed at $\delta t \approx 0.5 - 2$~days, which previous fits have explained with a shock cooling model (e.g., \citealt{Rastinejad+22}).

Finally, for the red component, we find a population median of $M_{\rm ej, R} = 0.051_{-0.045}^{+0.100} M_{\odot}$ (68\% confidence; Figure~\ref{fig:med_mass_4panel}). Amongst the three components, this is highest median, and the widest range in ejecta masses, though we note there are poor constraints on this value for GRBs\,050709, 130603B and 200522A. The red kilonova component can be ascribed to the neutron-rich ejecta tidally stripped from the NS surfaces as the compact objects slowly inspiral (e.g., \citealt{LattimerSchramm74,Li&Paczynski98,Metzger+10,tanaka+13}). Generally, it is expected that red dynamical ejecta mass will increase with larger asymmetry in the progenitor masses (particularly for NSBH events; e.g., \citealt{Foucart+14}) or higher spins (e.g., \citealt{Kyutoku+15,Shibata+19}). Notably, the three highest $M_{\rm ej, R}$ median values are found for the three LGRB events, GRBs\,060614, 211211A and 230307A. We further discuss the implications of this finding and other trends with GRB properties in Section~\ref{sec:long_short}. 

Across the sample, we derive a median total ejecta mass of $M_{\rm ej, tot} = 0.085_{-0.040}^{+0.110} M_{\odot}$ (68\% confidence). In every component and the total, the ejecta mass of GW\,170817/AT\,2017gfo falls comfortably within the 68\% credible range found for all events. This indicates that AT\,2017gfo may be considered a ``representative'' kilonova (Figure~\ref{fig:med_mass_4panel}) compared to the seven kilonovae analyzed here. In contrast, the kilonova of GRB\,160821B falls below the median ejecta masses found for all GRBs, rendering it a critical point in probing kilonova diversity.

\subsection{Constraints on $M_{\rm ej}$ from Additional Short GRB Observations}
\label{subsec:uls_from_grbs}

Next, we briefly explore the possibility that our kilonova sample is observationally biased towards more luminous events. As luminosity roughly scales with a fractional power of $M_{\rm ej}$ (e.g., \citealt{Metzger+18}), a missing population of low-luminosity kilonovae would translate to a population with lower $M_{\rm ej}$ than those reported here.
To evaluate this, we employ observations of seven short GRBs with upper limits and afterglow detections that are less luminous compared to AT\,2017gfo when matched in rest-frame time and band at their known redshifts \citep[see][for details]{Rastinejad+21}. These bursts\footnote{We remove GRB\,060201 from the sample as its host association is inconclusive (e.g., \citealt{Fong+22}).} are GRBs\,050509B \citep{Bloom+06,GCN3409,Hjorth+05,Castro-Tirado+05}, 080905A \citep{Rowlinson+10b,Guelbenzu+12}, 090515 \citep{Rowlinson+10}, 100206 \citep{Guelbenzu+12,Perley+12}, 130822A \cite{GCN15121,Rastinejad+21}, 150120A \citep{Rastinejad+21} and 160624A \citep{O'Connor+21,Rastinejad+21}. For this analysis, we assume all detections are dominated by afterglow flux and treat them as upper limits.

In the observed frame for each event (determined with the redshift catalog of \citealt{Fong+22}), we generate three sets of one-component kilonova light curves, characterized by either the blue, purple, and red fixed opacity value mentioned in Section~\ref{sec:KN_mod_description}. We fix $T_{\rm floor} = 1000$ K and use the corresponding component's median velocity found in Section~\ref{sec:results}. For each component, we produce kilonova models log-spaced in $M_{\rm ej}$ over the range $M_{\rm ej} = 0.001 - 0.5 M_{\odot}$. We compare our GRB observations to the set of kilonova models, and record the highest $M_{\rm ej}$ in each component allowed by the upper limits for each event. In Figure~\ref{fig:med_mass_4panel} we plot constraining ($< 0.5 M_{\odot}$) upper limits on the respective component ejecta masses. We note that our use of one-component models translates to a conservative upper limit as flux from other components is not taken into account in our procedure. 

Similar to previous findings, short GRB observations are most constraining of $M_{\rm ej, B}$ (Figure~\ref{fig:med_mass_4panel}; purple triangles). Past, deep rest-frame near-IR coverage is sparse. Thus, historical short GRB upper limits do not place meaningful constraints on $M_{\rm ej, P}$ and $M_{\rm ej, R}$ (Figure~\ref{fig:med_mass_4panel}). In the blue component, we find one event less massive than AT\,2017gfo (GRB\,080905A; $M_{\rm ej, B} < 0.002 M_{\odot}$) and an additional two events with upper limits below the median value found in Section~\ref{sec:results} (GRBs\,050509B and 130822A; $M_{\rm ej, B} < 0.005 M_{\odot}$). Overall, the existing upper limits span the range of blue ejecta masses. Thus, it is difficult to conclude at present if the population that do not have detected kilonovae also have lower ejecta masses. 

\subsection{Comparison to Previous Kilonova Fits}
\label{subsec:prev_fits}

Comparing our results in Table~\ref{tab:posterior} to those from previous fits, we find our results are generally consistent with those in the literature. We find variation in absolute differences, an expected outcome given the range of kilonova modeling codes used, which we further discuss here. 

For AT\,2017gfo, our fit produces a larger $M_{\rm ej, R}$ and a smaller $M_{\rm ej, B}$ and $M_{\rm ej, P}$ compared to a previous \texttt{MOSFiT} run \citep{villar+17}, though a similar value for the total ejecta mass is reached ($M_{\rm tot} \approx 0.07 M_{\odot}$). We ascribe this difference in the relative component masses to our addition of the geometry prescription (Section~\ref{sec:KN_mod_description}; \citealt{DarbhaKasen20}), which is expected to increase the amount of red mass relative to the blue for viewing angles less than the $\theta_{\rm open}$ (and is observed in \citealt{villar+17} though a larger $\theta_{\rm obs}$ and different asymmetry prescription was used). \citet{Nicholl+21} also model AT\,2017gfo with \texttt{MOSFiT}, incorporating constraints from GW observations, and find comparable $M_{\rm ej, B}$ and $M_{\rm ej, P}$. Their analysis finds a smaller $M_{\rm ej, R}$ ($\approx 0.001 M_{\odot}$), which we attribute to their constraints on the tidal dynamical ejecta based on the BNS mass ratio and chirp mass \citep{Nicholl+21}. Compared to two-component (red and blue components only) fits to AT\,2017gfo, we obtain $M_{\rm ej, B,R}$ values that are within the range but on the upper end of those measured in the literature (e.g., \citealt{Arcavi+17,Chornock+17,kasen+17,Kasliwal+17,McCully+17,Pian+17,Smartt+17,Troja+17,Anand+23_170817}).

Of the remaining kilonovae in our sample, the majority have measured $M_{\rm ej}$ and $V_{\rm ej}$. However, previous fits to these events use a variety of models and methods to measure the kilonova parameters and exact comparisons are not advisable. For GRB\,060614, a previous estimate found $M_{\rm ej, tot} \approx 0.1 M_{\odot}$, on the lower end of our estimate \citep{yang+15}. For GRB\,130603B, previous fits find a wide span in $M_{\rm ej, tot} = 0.01 - 0.1 M_{\odot}$ \citep{berger+13,Tanvir+13,barnes+16}, broadly consistent with our results. For GRB\,160821B, \citet{troja+19} constrain a lanthanide-rich mass to $< 0.006 M_{\odot}$ and a lanthanide-poor mass to $\approx 0.01 M_{\odot}$. For the same burst, \citet{lamb+19} find a dynamical ejecta mass of ($1.0 \pm 0.6$) $\times 10^{-3} M_{\odot}$ and a post-merger ejecta mass of ($1.0 \pm 0.6$) $\times 10^{-2} M_{\odot}$. Within errors, our results are consistent with both findings but are on the upper end of the given ranges. 

GRB\,211211A was fit with a three-component model in \texttt{MOSFiT} that included a shock heating prescription, finding $M_{\rm ej, B} = 0.01 \pm 0.001 M_{\odot}$, $M_{\rm ej, P} = 0.01 \pm 0.02 M_{\odot}$ and $M_{\rm ej, R} = 0.02^{+0.02}_{-0.01} M_{\odot}$ \citep{Rastinejad+22}. They find consistent results when performing joint afterglow and kilonova modeling \citep{Rastinejad+22}. Within errors, the blue and purple component masses are consistent with our results, but we find a larger median $M_{\rm ej, R}$ by a factor of two. Additional $M_{\rm ej, tot}$ estimates of this kilonova span $0.02 - 0.1 M_{\odot}$ \citep{Troja+22,Yang+22,Kunert+23}. For GRB\,230307A, $M_{\rm ej, tot}$ of $0.05^{+0.15}_{-0.05} M_{\odot}$ \citep{Levan+24} and $\approx 0.08 M_{\odot}$ \citep{Yang+24} are found in the literature. Taken together, our results are generally consistent with those found in the literature. Our median values are on the upper end of previous estimates, consistent with comparisons to AT\,2017gfo modeling.

\subsection{Caveats to Kilonova Model}
\label{subsec:opacity_err}
We find that in several cases the error bars on $M_{\rm ej}$ (particularly for the blue component) are unrealistically small. While a powerful tool to infer physical properties, \texttt{MOSFiT} makes a series of simplying assumptions that likely result in an underestimation of the true errors, similar to the conclusion made for \texttt{MOSFiT} modeling of tidal disruption events \citep{Mockler+19}.

In particular, the assumption of a constant grey opacity may significantly affect the model posteriors, as $M_{\rm ej}$, $V_{\rm ej}$ and $\kappa$ are degenerate parameters in predicting the shape of the light curve. Notably, $\kappa$ may vary up to an order of magnitude on the timescales of our observations ($0.1 \lesssim \delta t \lesssim 30$~days) as the ejecta temperature and density directly impact the elements' ionization states (e.g., \citealt{Tanaka+20,Banerjee+24}). We quantify the minimum systematic error introduced by our assumed opacities by running fits of AT\,2017gfo in which each component's opacity is a free parameter whilst holding the other two components' $\kappa$ values constant and keeping the same prior ranges listed in Table~\ref{tab:prior}. We then determine the difference in derived component $M_{\rm ej}$ relative to the masses inferred in Section~\ref{sec:KN_mod_description} with fiducial, fixed opacities. Specifically, we explore the effects of the range $\kappa_{\rm R} = 5-30$~cm$^2$~g$^{-1}$ (compared to $\kappa_{\rm R} = 10$~cm$^2$~g$^{-1}$ in our Section~\ref{sec:KN_mod_description} fits), $\kappa_{\rm P} = 1-5$~cm$^2$~g$^{-1}$ ($\kappa_{\rm P} = 3$~cm$^2$~g$^{-1}$), and $\kappa_{\rm B} = 0.2 - 2.5$~cm$^2$~g$^{-1}$ ($\kappa_{\rm B} = 0.5$~cm$^2$~g$^{-1}$). We use a minimum of $\kappa_{\rm B} = 0.2$~cm$^2$~g$^{-1}$ as it corresponds to the minimum value dictated by Thompson scattering of ionized elements \citep{Paczynski83}. We base the remaining ranges on calculations for kilonova opacities at $\delta t \approx 1$~day \citep{Banerjee+24,Tanaka+20}. 

Based on the three fits with free $\kappa$ values, we find the median errors on the component masses are $\Delta M_{\rm ej, R} = (-0.04, +0.0008)~M_{\odot}$, $\Delta M_{\rm ej, P} = (-0.004, +0.02)~M_{\odot}$ and $\Delta M_{\rm ej, B} = (-6.5 \times 10^{-5}, 0.0006)~M_{\odot}$. For the total mass, fitting with $\kappa$ as a free parameter results in only modestly lower total ejecta masses of which the median value is $\delta M_{\rm ej, tot} = -0.001 M_{\odot}$. We find that our fits varying $\kappa_{\rm P}$ result in the most significant source of uncertainty. This result can naturally be explained as values of $\kappa_{\rm P}$ more similar to $\kappa_{\rm B}$ or $\kappa_{\rm R}$ will divert a portion of the luminosity typically explained by $M_{\rm ej, B}$ or $M_{\rm ej, R}$ to $M_{\rm ej, P}$. This results in a different ratio between the two component masses but an overall similar $M_{\rm ej, tot}$. We conclude that our choice of opacity (within the range of values explored here) does not significantly impact our ejecta mass results. We do not include these uncertainties where we compare between our uniform fits.

We caution that, in addition to the effect explored above, the kilonova models used in this analysis do not account for the effect of jet-ejecta interaction, shock cooling, central engine activity, or magnetic fields, all of which may play a large role in determining ejecta masses (e.g., \citealt{Radice+18,Ciolfi+20,Nativi+21,Shrestha+23,Curtis+24,Hamidani+24_model}). In addition, several pieces of kilonova physics are still not well understood even in state-of-the-art simulations, such as the uncertainty in wavelength-dependent opacities, nuclear heating rate, and thermalization efficiencies \citep{Barnes+21,Bulla23,Brethauer+24,SarinRosswog24}. Further, our two-stage analysis of the afterglow and kilonova may result in additional uncertainty and bias in the derived kilonova parameters compared to a joint models (e.g., \citealt{WallaceSarin24}). To account for this point, we exclude data at $<0.5$~days in our kilonova fits, the timescale on which the afterglow is most likely to dominate, and propagate the uncertainties in our afterglow model to the data passed to the kilonova model (Sections~\ref{sec:ag_mod} and \ref{sec:KN_mod_description}). 

In light of these uncertainties, we emphasize that the aim of this work is to perform uniform modeling on a sample of kilonovae, allowing for an exploration of diversity and correlations with $\gamma$-ray and environment properties. For all objects in our sample besides AT\,2017gfo, the datasets are relatively sparse (e.g., Section~\ref{sec:obs} and Figure~\ref{fig:kn_models_subdata}), limiting our ability to investigate each uncertainty listed above. We provide all observations (Table~\ref{tab:obs}), including afterglow-subtracted photometry  (Table~\ref{tab:sub_obs}), for the community to model with other existing, or future, codes.

\section{Discussion}
\label{sec:discussion}

\begin{figure*}
\centering
\includegraphics[angle=0,width=\textwidth]{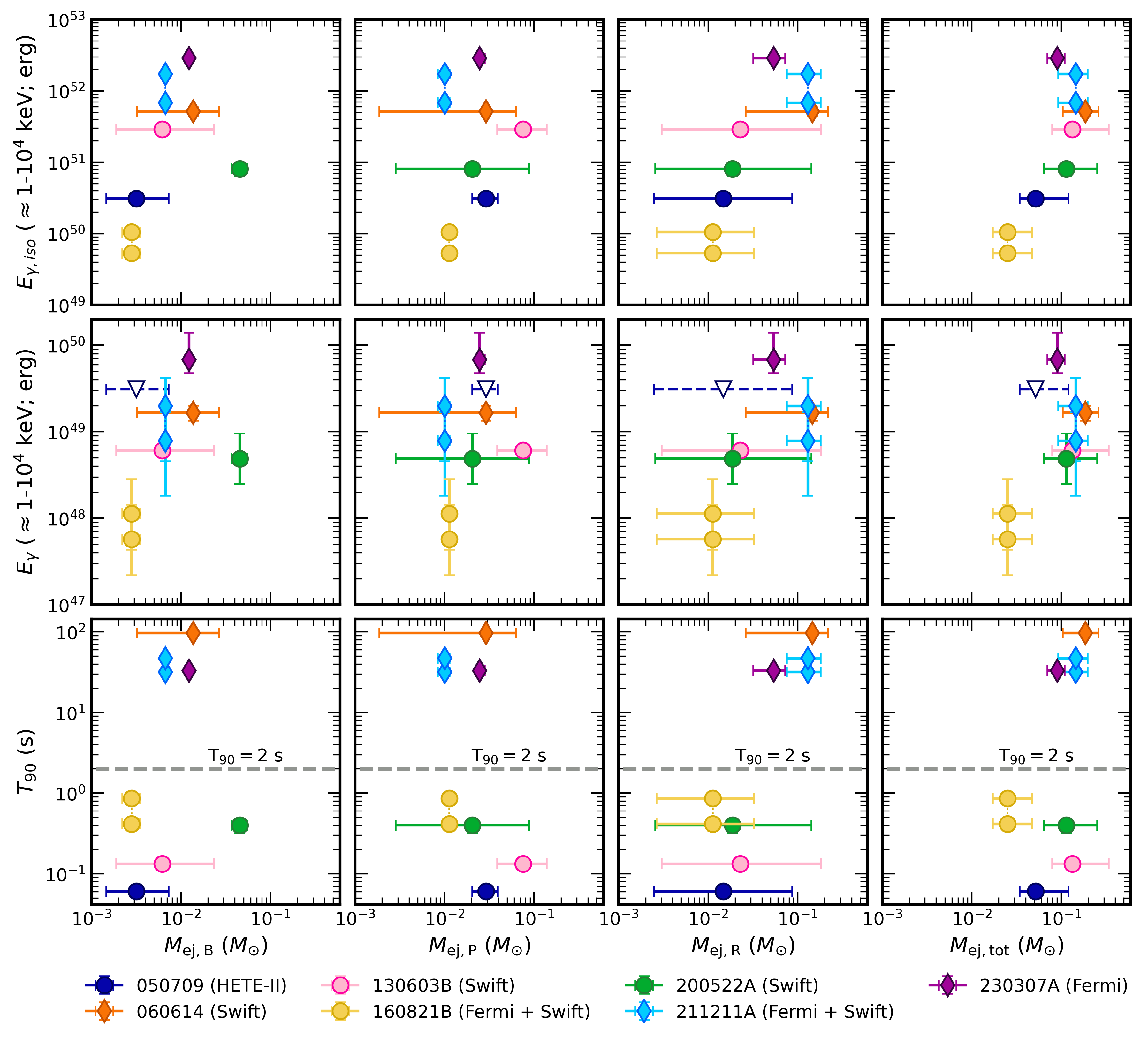}
\caption{The blue (first column), purple (second), red (third) and total (fourth) ejecta masses for each on-axis GRB event plotted against the respective $\gamma$-ray properties: $E_{\gamma, {\rm iso}}$ (top row), beaming-corrected $E_{\gamma}$ (middle row), and $T_{90}$ (bottom row). Long GRBs are marked with diamonds, while short GRBs are marked with circles. As there is only a lower limit on the jet half-opening angle of GRB\,050709 (Section~\ref{sec:long_short}) we mark the resulting upper limit on $E_{\gamma}$ with a triangle. We observe potential trends between $E_{\gamma, {\rm iso}}$, $E_{\gamma}$ and the red and total ejecta mass.}
\label{fig:mej_Eiso_t90}
\end{figure*}

\subsection{Comparison to $\gamma$-ray Properties and Implications for Long/Short Progenitors}
\label{sec:long_short}

Motivated by the unknown progenitor properties and/or mechanism driving merger-origin long GRBs, we next examine any trends between the kilonova ejecta and $\gamma$-ray properties. Historically, one factor in explaining the divide in $\gamma$-ray duration between BNS and stellar progenitors is their order-of-magnitude difference in the mass reservoir surrounding the central compact object. The significantly higher masses and larger physical size of long GRB progenitor stars lead to a longer timescale for accretion, translating to a longer-lived jet (e.g., \citealt{Tchekhovskoy+11}). To produce a long-lived GRB jet from a neutron star merger, previous works have posited the progenitors are a white dwarf-neutron star binary \citep{Yang+22,Wang+24} or an NSBH (e.g., \citealt{Rastinejad+22,Gompertz+22}), or that the merger remnant is a magnetar (e.g., \citealt{Rastinejad+22,Gompertz+22}). 

Here, we explore the theory that a long-lived, massive ($\approx 0.2 M_{\odot}$) accretion disk, a product of an asymmetric binary merger, is capable of powering longer-lived and more energetic GRBs \citep{Gottlieb+23}. The merger of an asymmetric binary, whether it is a BNS or an NSBH with a favorable mass ratio ($Q \approx 3-5$; e.g., \citealt{kawaguchi+16}), will produce a greater amount of lanthanide-rich tidal dynamical ejecta compared to a symmetric binary (e.g., \citealt{Hotokezaka+13,Kyutoku+18,Kawaguchi+20}). Within our modeling framework, this translates to an expected trend between $M_{\rm ej, R}$ and the duration and/or energy of the GRB.

To investigate these possible trends, we compare the kilonova ejecta masses (Table~\ref{tab:posterior}) with the values of $E_{\gamma,{\rm iso}}$ (described in Section~\ref{subsec:sample_sel}), beaming-corrected $E_{\gamma}$, and $T_{\rm 90, rest}$ (converted to the rest-frame using their respective redshifts; Table~\ref{tab:grbprops}). To calculate $E_{\gamma}$, we gather jet opening angle measurements ($\theta_{j}$) from the literature \citep{Xu+09,Rastinejad+22,Rouco+23}. All events in our sample have measured $\theta_j$ values from X-ray observations, with the exception of GRB\,050709 for which a lower limit is reported \citep{Rouco+23}. For GRB\,230307A we use 5000 random draws from our afterglow model 1 (Section~\ref{sec:ag_mod}), as it is fit to the combined X-ray light curves from the literature, therein providing a tighter constraint on the jet break than previous analyses \citep{Levan+24,Yang+24}. Our analysis finds $\theta_{j} = 3.95^{+1.72}_{-0.65}$ deg. We calculate the beaming-corrected energy, given by,
\begin{equation}
    E_{\gamma} = [1-{\rm cos}(\theta_{j})] \times E_{\gamma,{\rm iso}} \, .
\end{equation}
In Figure~\ref{fig:mej_Eiso_t90}, we show the results of this analysis. We observe that $M_{\rm ej, B}$ (with the exception of the outlier GRB\,200522A; Section~\ref{sec:results}), $M_{\rm ej, R}$ and $M_{\rm ej, tot}$ generally increase with higher values of $E_{\gamma,{\rm iso}}$ and $E_{\gamma}$. We further observe that $M_{\rm ej, R}$ increases with longer $T_{90}$. The trends with $E_{\gamma,{\rm iso}}$, $E_{\gamma}$, and $T_{\rm 90}$ are most apparent with $M_{\rm ej, R}$ (Figure~\ref{fig:mej_Eiso_t90}, third column), though the error bars for several short GRB masses and $\theta_j$ preclude a firm conclusion. We do not observe any apparent trends between $M_{\rm ej, P}$ (Figure~\ref{fig:mej_Eiso_t90}, second column) and $\gamma$-ray properties. 

We briefly test the statistical significance of any trends between the ejecta masses and the $\gamma$-ray properties. We apply the Pearson correlation coefficient (r-score) test to the datasets in each panel of Figure~\ref{fig:mej_Eiso_t90}. We caution that this test is agnostic to a model, and thus does not probe all underlying physical motivations. We randomly draw values from the ejecta mass posteriors and calculate r- and p-scores with the $E_{\gamma,{\rm iso}}$, $E_{\gamma}$, and $T_{\rm 90}$ values for 1000 iterations, producing distributions of 1000 r- and p-values. We then determine the fraction of random draws that imply a significant correlation between the ejecta mass and $\gamma$-ray properties using a threshold of p$< 0.05$. We do not find that a significant fraction of the p-scores favor a correlation between the ejecta masses and any $\gamma$-ray properties. The strongest correlation is between $M_{\rm ej, R}$ and $T_{\rm 90}$, where 32\% of p-scores indicate a significant correlation.

Though we do not find any statistically significant correlations, we observe that GRBs with $T_{90,{\rm rest}} \gtrsim 2$~s have higher median red ejecta masses ($M_{\rm ej, R} \gtrsim 0.05 M_{\odot}$) compared to typical short GRBs ($M_{\rm ej, R} \lesssim 0.02 M_{\odot}$) potentially hinting at a bimodality. We observe a similar pattern with $M_{\rm ej, B}$ and $E_{\gamma}$ (with the exception of GRB\,200522A), though this component may also originate in the post-merger disk wind (Section~\ref{sec:results}). The implication that $M_{\rm ej, B}$ seems to trend more strongly with $\gamma$-ray properties compared to $M_{\rm ej, P}$ may also suggest that long GRB mergers produce relatively blue disk winds, perhaps due to energy injection from the GRB. At present, our results could indicate asymmetric binaries as the progenitors of merger-driven long GRBs. However, to establish any statistically significant correlations, a larger population of joint GRB-kilonova detections with well-constrained ejecta masses is necessary. 

Finally, we note that while asymmetric binaries are uncommon in the observed population of BNS systems (e.g., \citealt{Tauris+17}),  LVK observations have revealed an asymmetric BNS (GW190425; \citealt{LVC_GW190425}) and an NSBH (GW230519; \citealt{LVK_230529}) merger. These detections may point to these asymmetric binaries being common in the Universe, with potential rates able to explain the increasing (but highly uncertain) rates of long GRBs from mergers.

\subsection{Kilonova Contribution to Universal $r$-Process}

At present, kilonovae are the only observationally-confirmed source of $r$-process production in the Universe. Indirect observational evidence may favor the existence of a second, ``faster'' heavy element nucleosynthesis channel. Specifically, to explain observations of $r$-process-enhanced metal poor stars in the Milky Way and dwarf galaxies (e.g., \citealt{Ji+16,Hansen+17,Frebel+18}) a significant fraction of NS mergers are required to have short delay times from enrichment to star formation (e.g., \citealt{Zevin+22}). Simulations have demonstrated that collapsar and magneto-hydrodynamical (jet-driven) supernovae could be this second channel (e.g., \citealt{SiegelBarnesMetzger2019,Mosta+18,HaleviMosta18}). At present, observations of these candidates are limited and those that exist do not support $r$-process enrichment \citep{Blanchard+23,Anand+24,Rastinejad+24}. Here, we explore if the average $r$-process yield from kilonovae, calculated using the median ejecta masses calculated in Section~\ref{sec:KN_mod} along with the current NS merger rates, is capable of producing the estimated $r$-process abundance in the Milky Way. In this analysis, we assume the kilonovae in our sample are created by BNS mergers only, though we note in Section~\ref{sec:long_short} the kilonovae following long GRBs are favored to be from asymmetric mergers that may be NSBH events. We make this assumption for simplicity, as only a small fraction of NSBH events are expected to produce kilonovae and the rate of NSBH events is sub-dominant compared to the rates of BNS mergers (e.g., \citealt{GWTC3_Rates,MandelBroekgaarden22}).

To calculate the $r$-process enrichment in the Milky Way from NS mergers, we employ the equation from \citet{Rosswog+18}:
\begin{equation} \small
    M_r \sim 17\,000 M_{\odot}\left[\frac{\mathcal{R}_{\rm BNS}}{500{\rm Gpc^{-3} yr ^{-1}}}\right]
    \left[\frac{\bar{m}_{\rm ej}}{0.03 M_{\odot}}\right]\left[\frac{\tau_{\rm gal}}{1.3 \times 10^{10} {\rm yr}}\right]
\end{equation}

\noindent where $\mathcal{R}_{\rm BNS}$ is the rate of NS mergers, $\bar{m}_{\rm ej}$ is the average kilonova ejecta mass, and $\tau_{\rm gal}$ is the age of the Milky Way, which we fix to $1.3 \times 10^{10}$~yr. We employ the range in BNS merger rate calculated from the Gravitational Wave Transient Catalog 3 (GWTC-3) of $\mathcal{R}_{\rm BNS} = 10 - 1700$ \citep{GWTC3_Rates}. For $\bar{m}_{\rm ej}$ we employ our median for the eight kilonovae in our sample, $M_{\rm ej, tot} = 0.085^{+0.108}_{-0.044} M_{\odot}$, including the uncertainties due to the opacity (Section~\ref{subsec:opacity_err}). Taking these values together, we find a wide range of $M_{r} \approx $ 500 - 400,000 $M_{\odot}$ for BNS mergers, mostly driven by the large uncertainty in BNS merger rate. 

This range for $M_{r}$ encompasses the estimate of total $r$-process mass in the Milky Way, $M_{r, {\rm MW}} \approx 23,000~M_{\odot}$ \citep{Venn+04,BattistiniBensby16,Hotokezaka+18} and, on the lower end, leaves room for the existence of a second $r$-process source. As discussed in Sections~\ref{subsec:uls_from_grbs}-\ref{subsec:opacity_err}, our $\bar{m}_{\rm ej}$ value is likely to be an overestimate of the true value for two reasons. First, less luminous, and thus less massive, kilonovae are likely to have been missed due to observational biases (Section~\ref{subsec:uls_from_grbs}). Second, in keeping with fits to AT\,2017gfo, our \texttt{MOSFiT} model derives ejecta mass values on the upper end of ranges from previous fits (Section~\ref{subsec:prev_fits}). However, we anticipate that the uncertainty in $\bar{m}_{\rm ej}$ in subdominant compared to the uncertainty in BNS merger rate, as our values are generally comparable with literature values, where they exist, (Section~\ref{subsec:prev_fits}) and likely do not vary beyond an order of magnitude.

The specific star formation rate of the host galaxy is a dominant factor in governing what fraction of ejected $r$-process elements enrich later generations of stars (Nugent et al., in prep.). Notably, in comparison to the quiescent host galaxy of GW\,170817/AT\,2017gfo \citep{Blanchard+17,Levan+17}, the hosts of the kilonovae in our on-axis GRB sample are all star-forming (\citealt{Nugent+22,Levan+24}; Table~\ref{tab:grbprops}). Though losses may still be significant for these latter events, it is more likely that they enriched their galaxies with heavy elements compared to AT\,2017gfo.

\begin{figure*}
\centering
\includegraphics[angle=0,width=\textwidth]{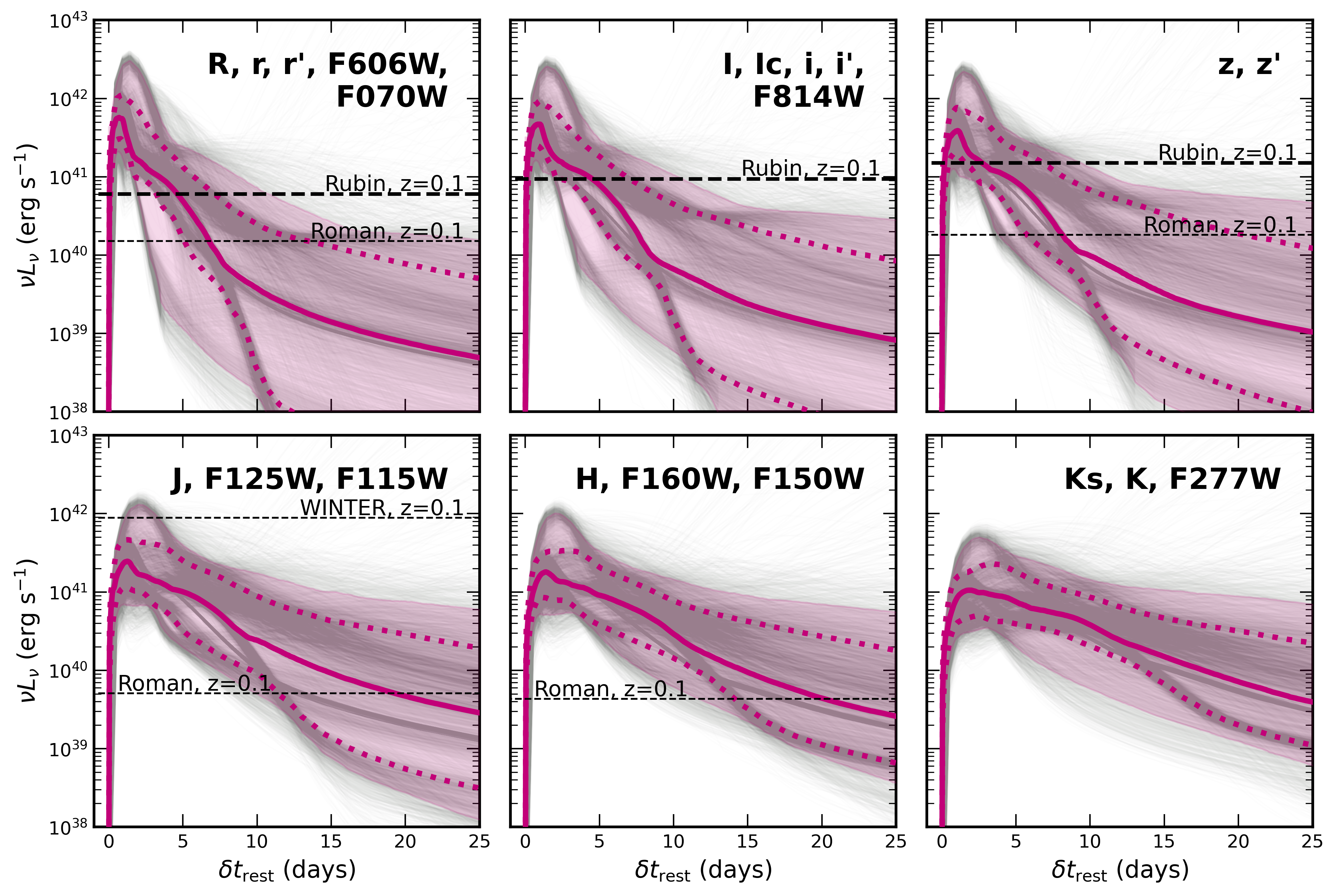}
\caption{The median (solid line), 68\% credible range (dotted lines) and 90\% credible range (shaded region) in luminosity space shown in the $rizJHK$-bands and calculated using random draws (light grey lines) from each event. We also show the expected depths of several current and upcoming wide-field observatories that plan to engage in kilonova searches \citep{Margutti+18_LSSTTOO,Rose+21,Bianco+22,Frostig+22,Andreoni_romankilonova}.}
\label{fig:lc_lum}
\end{figure*}

\subsection{Future Kilonova Observations}

We next consider the implications of our sample and model light curves for future wide-field kilonova searches, either triggered with a GRB or GW event or untriggered (e.g, \citealt{Smartt+17,Doctor+17,Kasliwal+17,Andreoni+20_ztfkn}). For these searches, an understanding of the span in kilonova light curve diversity is critical in the vetting process, which has yet to be enabled with observations beyond AT\,2017gfo.

For each of the seven on-axis GRB events in our sample, we create 900 light curves in the observed-frame $rizJHK$-bands. Each light curve is based on a random draw from the posterior (we use GRB\,230307A Model 1 in our fit given the reasoning in Section~\ref{sec:results}) and matched to the approximate rest-frame band in Figure~\ref{fig:lc_lum}. We create 50 light curves for GRB\,170817A/AT\,2017gfo due to computational constraints and iterate over these 18 times. Due to the tight posteriors on this event, we do not expect this process to affect our conclusions. Across the random draws, we calculate the median, 68\% and 90\% credible region in luminosity space for the sample of kilonovae. In Figure~\ref{fig:lc_lum} we show the median and kilonova luminosity range in each filter. Taken together, our results indicate that future kilonovae will span $\gtrsim$one order of magnitude in luminosity.

As we are motivated by both targeted and untargeted kilonova searches in large surveys, we also plot the single image 5$\sigma$ depths for the Rubin Observatory \citep{Bianco+22}, the WINTER $J$-band limiting magnitude \citep{Frostig+22} and the ``Wide Tier'' ($RZ$-band) and ``Deep Tier'' ($JH$-band) limiting magnitudes for Roman High Latitude Time Domain Survey (HLTDS; \citealt{Rose+21}), shifted to $z=0.1$. We caution that our results are based mostly on on-axis events, and the bolometric luminosity of kilonovae may vary up to factor of $\sim 10$ with viewing angle \citep{DarbhaKasen20}. In the $riz$-bands, Rubin observatory will be sensitive to the full and upper end of the 68\% credible range of kilonovae at $z=0.1$ out to $\delta t \approx 3$ and $\approx 7$~days, respectively. As shown with our analysis GRB\,200522A, in Section~\ref{sec:results}, a single epoch of simultaneous rest-frame optical observations on these timescales is sufficient for order-of-magnitude constraints on $M_{\rm ej, B}$. We therefore find that order-of-magnitude constraints  on $M_{\rm ej, B}$ are possible for kilonovae observed by Rubin at one epoch. We acknowledge that large outstanding challenges exist to distinguish these rare events from other transients, especially in just one or two epochs of observations, and encourage further development of automatic vetting tools. 

Roman's HLTDS will be a powerful tool in measuring the properties of low-redshift kilonovae if events can be identified. Indeed, Roman is capable of detecting $z=0.1$ kilonovae in the optical and near-IR out to $\delta t \approx 2$ and 3~weeks, respectively. If the Roman HLTDS observations occur at a five-day cadence, Roman is poised to observe $z=0.1$ kilonovae over $\approx$2-3 epochs in the optical and $\approx$3-5 epochs in the near-IR. Two to three epochs is sufficient for obtaining better than order-of-magnitude constraints on $M_{\rm ej}$-$V_{\rm ej}$, as we have demonstrated in Section~\ref{sec:KN_mod}. Finally, WINTER will be sensitive to the most luminous kilonovae, but not the 68\% credible range, at $z=0.1$. 

In addition to wide-field searches in the nearby Universe, well-localized \textit{Swift} GRBs continue to be a promising method to detect kilonovae. These events come with their own set of challenges, including higher redshifts, bright afterglows and low rates. Further, the more coarsely localized detections of GRBs by \textit{Fermi}, Space Variable Objects Monitor (SVOM; \citealt{SVOM}), the InterPlanetary Network (IPN; e.g., \citealt{IPNcat_1987,Svinkin+22}) and other $\gamma$-ray telescopes offer a second route to finding kilonovae using ``targeted'' wide-field surveys. However, for all GRB kilonovae, the higher average redshifts render follow-up with large-aperture ground-based telescopes and space-based observatories critical. Indeed, seven of the eight kilonovae in our sample had key observations with $HST$ that were critical to detections on $\gtrsim$~week timescales. Looking to the future, \textit{JWST} can obtain similar observations for events out to $z \approx 1$ (e.g., \citealt{Rastinejad+21}). For lower-redshift events \textit{JWST} has the power to capture the kilonova spectral energy distribution (SED) in the nebular phase, an important element in refining the mapping from observations to $M_{\rm ej}$.

\section{Conclusions}

We have compiled and collated the multi-wavelength light curves of eight kilonovae from the GCNs and literature. Five of these events follow short GRBs, while three events follow long GRBs, allowing us to explore trends between $\gamma$-ray and kilonova ejecta properties. We uniformly model the afterglows of seven events with on-axis GRBs, producing ``afterglow-subtracted'' light curves. We fit the afterglow-subtracted light curves with a three-component kilonova model in \texttt{MOSFiT} that accounts for geometric viewing effects. Our fits provide reasonable fits to the data, and we compare our posteriors to those in the literature. Our major conclusions are as follows: 

\begin{itemize}
    \item Our fits unveil a wide span in derived kilonova properties, namely $M_{\rm ej}$ and $V_{\rm ej}$, implying that the progenitors and/or remnants of these mergers are also diverse. We determine that the luminous kilonova of GRB\,200522A has a significantly more massive $M_{\rm ej, B}$ compared to the sample of events (or the luminosity may be boosted by non-radioactive heating like a magnetar; \citealt{Fong+21}), while the kilonova of GRB\,160821B is the least massive of the total sample.
    \item While well-sampled events provide the tightest constraints, we also find value in kilonovae with a single color measurement, particularly if their colors are unique (e.g., GRBs\,130603B and 200522A). 
    \item We discuss the main uncertainties in our modeling (Section~\ref{subsec:opacity_err}) and compare our results to previous fits with other modeling codes. We emphasize that all observations used in this work, including our ``afterglow-subtracted'' light curves, are provided for future modeling endeavors in Tables~\ref{tab:sub_obs} and \ref{tab:obs}.
    \item We demonstrate that GW\,170817/AT\,2017gfo is a ``representative'' kilonova in each components' $M_{\rm ej}$ and $V_{\rm ej}$ (Table~\ref{tab:posterior}), when compared to the seven kilonovae in our sample and deep upper limits from the literature. Given this, our estimate of the total Milky Way $r$-process mass produced by kilonovae does not change significantly when using the median ejecta mass of our sample compared to previous estimates made for AT\,2017gfo.
    \item We explore trends between our derived ejecta masses and $E_{\gamma, {\rm iso}}$, beaming-corrected $E_{\gamma}$ and $T_{90}$. Overall, we do not find any statistically significant correlations but observe that long GRB kilonovae have larger median $M_{\rm ej, R}$ compared to short GRB kilonovae. We hypothesize that this is indicative of an asymmetric binary merger origin for longer-lived GRBs. A larger sample of well-studied kilonovae following short and long GRBs will be critical to confirming this hypothesis.
    \item We produce median, 68\% and 90\% confidence range light curves in a variety of bands based on the posteriors of the eight events in our sample. Comparing these light curves to the expected depths of upcoming surveys, we anticipate that Rubin and Roman will be sensitive to the majority of the kilonova luminosity range for $z\lesssim0.1$ and are capable of order-of-magnitude mass constraints.
\end{itemize}

Here, we have shown that the existing sample of kilonovae, the majority of which are detected at a fixed viewing angle, demonstrates diversity and trends with $\gamma$-ray properties. Widening the sample of these events requires dedicated strategies for observational pointings (e.g., \citealt{Margutti+18_LSSTTOO,Coulter+24}) and kilonova candidate vetting (e.g., \citealt{Rastinejad+22_O3}) that take into account the full diversity of compact binary merger EM counterparts and environmental properties. The advent of next-generation GW detectors, deep wide-field surveys, and new $\gamma$-ray instruments, when combined with dedicated search strategies, opens the doors for unprecedented exploration into the physics of compact binary mergers, jets, and kilonovae. 

\section{Acknowledgements}

The authors thank Ore Gottlieb, Genevieve Schroeder, Anya Nugent, Tanmoy Laskar, Igor Andreoni, Sylvia Biscoveneau, Nick Kaaz, Ben Margalit and Michael Fausnaugh for helpful conversations regarding this manuscript.

J.C.R. acknowledges support from the Northwestern Presidential Fellowship. The Fong Group at Northwestern acknowledges support by the National Science Foundation under grant Nos. AST-2206494, AST-2308182, and CAREER grant No. AST-2047919. W.F. gratefully acknowledges support by the David and Lucile Packard Foundation, the Alfred P. Sloan Foundation, and the Research Corporation for Science Advancement through Cottrell Scholar Award 28284.

\renewcommand{\arraystretch}{1.}
\startlongtable
\begin{deluxetable*}{cCcccRC}
\savetablenum{4}
\tabletypesize{\footnotesize}
\centering
\tablecolumns{7}
\tabcolsep0.12in
\tablecaption{Afterglow-subtracted Observations
\label{tab:sub_obs}}
\tablehead {
\colhead {GRB}		&
\colhead {$\delta t$}	&
\colhead {Telescope}		&
\colhead {Instrument}		&
\colhead {Filter}		& 
\colhead {Subtracted Mag.}	&
\colhead {Uncertainty}  \\
\colhead {} &
\colhead {(days)} &
\colhead {} &
\colhead {}		& 
\colhead {}		& 
\colhead {(AB mag)} &
\colhead {(AB mag)}
}
\startdata
050709 & 1.24 & Swope-40 & Swope & $i$ & > 20.50 &  \\
 & 1.42 & Danish & DFOSC & $R$ & > 23.19 &  \\
 & 2.39 & Danish & DFOSC & $R$ & > 23.94 &  \\
 & 2.46 & VLT & FORS2 & $I$ & > 23.70 &  \\
 & 2.46 & VLT & FORS2 & $V$ & > 24.40 &  \\
 & 2.47 & VLT & FORS2 & $R$ & 25.07 & 0.58 \\
 & 4.36 & VLT & FORS1 & $V$ & > 25.02 &  \\
 & 4.37 & VLT & FORS1 & $I$ & > 24.55 &  \\
\enddata
\tablecomments{
Afterglow-subtracted observations are obtained using the methods described in Sections~\ref{sec:ag_mod} and \ref{sec:KN_mod}. \\
$^{\dagger}$Afterglow model 1 fit to X-ray and radio data only. \\
$^{\ddagger}$Afterglow model 2 fit to X-ray, radio and TESS data. \\ 
Observations are corrected for Galactic extinction. Observations of GRB\,130603B are also corrected for line-of-sight extinction measured from the early optical afterglow.}
\end{deluxetable*}

\bibliographystyle{aasjournal}
\bibliography{refs}

\begin{thebibliography}{}
\expandafter\ifx\csname natexlab\endcsname\relax\def\natexlab#1{#1}\fi
\providecommand{\url}[1]{\href{#1}{#1}}
\providecommand{\dodoi}[1]{doi:~\href{http://doi.org/#1}{\nolinkurl{#1}}}
\providecommand{\doeprint}[1]{\href{http://ascl.net/#1}{\nolinkurl{http://ascl.net/#1}}}
\providecommand{\doarXiv}[1]{\href{https://arxiv.org/abs/#1}{\nolinkurl{https://arxiv.org/abs/#1}}}

\bibitem[{{Abbott} {et~al.}(2016){Abbott}, {Abbott}, {Abbott}, {Abernathy}, {Acernese}, {Ackley}, {Adams}, {Adams}, {Addesso}, {Adhikari}, \& et~al.}]{lvc_2016_LRR}
{Abbott}, B.~P., {Abbott}, R., {Abbott}, T.~D., {et~al.} 2016, Living Reviews in Relativity, 19, \dodoi{10.1007/lrr-2016-1}

\bibitem[{{Abbott} {et~al.}(2017{\natexlab{a}}){Abbott}, {Abbott}, {Abbott}, {Acernese}, {Ackley}, {Adams}, {Adams}, {Addesso}, {Adhikari}, {Adya}, {Affeldt}, {Afrough}, {Agarwal}, {Agathos}, {Agatsuma}, {Aggarwal}, {Aguiar}, {Aiello}, {Ain}, {Ajith}, {Allen}, {Allen}, {Allocca}, {Altin}, {Amato}, {Ananyeva}, {Anderson}, \& {Anderson}}]{gw170817mma}
---. 2017{\natexlab{a}}, \apjl, 848, L12, \dodoi{10.3847/2041-8213/aa91c9}

\bibitem[{{Abbott} {et~al.}(2017{\natexlab{b}}){Abbott}, {Abbott}, {Abbott}, {Acernese}, {Ackley}, {Adams}, {Adams}, {Addesso}, {Adhikari}, {Adya}, {Affeldt}, {Afrough}, {Agarwal}, {Agathos}, {Agatsuma}, {Aggarwal}, {Aguiar}, {Aiello}, \& et~al.}]{gw170817}
---. 2017{\natexlab{b}}, \prl, 119, 161101, \dodoi{10.1103/PhysRevLett.119.161101}

\bibitem[{{Abbott} {et~al.}(2020{\natexlab{a}}){Abbott}, {Abbott}, {Abbott}, {Abraham}, {Acernese}, {Ackley}, {Adams}, {Adhikari}, {Adya}, {Affeldt}, {Agathos}, {Agatsuma}, {Aggarwal}, {Aguiar}, {Aiello}, {Ain}, {Ajith}, {Allen}, {Allocca}, {Aloy}, {Altin}, {Amato}, {Anand}, {Ananyeva}, {Anderson}, {Anderson}, {Angelova}, {Antier}, {Appert}, {Arai}, {Araya}, {Areeda}, {Ar{\`e}ne}, {Arnaud}, {Aronson}, {Arun}, {Ascenzi}, {Ashton}, {Aston}, {Astone}, {Aubin}, {Aufmuth}, {AultONeal}, {Austin}, {Avendano}, {Avila-Alvarez}, {Babak}, {Bacon}, {Badaracco}, {Bader}, {Bae}, {Baird}, {Baker}, {Baldaccini}, {Ballardin}, {Ballmer}, {Bals}, {Banagiri}, {Barayoga}, {Barbieri}, {Barclay}, {Barish}, {Barker}, {Barkett}, {Barnum}, {Barone}, {Barr}, {Barsotti}, {Barsuglia}, {Barta}, {Bartlett}, {Bartos}, {Bassiri}, {Basti}, {Bawaj}, {Bayley}, {Baylor}, {Bazzan}, {B{\'e}csy}, {Bejger}, {Belahcene}, {Bell}, {Beniwal}, {Benjamin}, {Berger}, {Bergmann}, {Bernuzzi}, \& {Berry}}]{LVC_GW190425}
---. 2020{\natexlab{a}}, \apjl, 892, L3, \dodoi{10.3847/2041-8213/ab75f5}

\bibitem[{{Abbott} {et~al.}(2020{\natexlab{b}}){Abbott}, {Abbott}, {Abraham}, {Acernese}, {Ackley}, {Adams}, {Adhikari}, {Adya}, {Affeldt}, {Agathos}, {Agatsuma}, {Aggarwal}, {Aguiar}, {Aich}, {Aiello}, {Ain}, {Ajith}, {Akcay}, {Allen}, {Allocca}, {Altin}, {Amato}, {Anand}, {Ananyeva}, {Anderson}, {Anderson}, {Angelova}, {Ansoldi}, {Antier}, {Appert}, {Arai}, {Araya}, {Areeda}, {Ar{\`e}ne}, {Arnaud}, {Aronson}, {Arun}, {Asali}, {Ascenzi}, {Ashton}, {Aston}, {Astone}, {Aubin}, {Aufmuth}, {AultONeal}, {Austin}, {Avendano}, {Babak}, {Bacon}, {Badaracco}, {Bader}, {Bae}, {Baer}, {Baird}, {Baldaccini}, {Ballardin}, {Ballmer}, {Bals}, {Balsamo}, {Baltus}, {Banagiri}, {Bankar}, {Bankar}, {Barayoga}, {Barbieri}, {Barish}, {Barker}, \& {Barkett}}]{LVC_GW190814}
{Abbott}, R., {Abbott}, T.~D., {Abraham}, S., {et~al.} 2020{\natexlab{b}}, \apjl, 896, L44, \dodoi{10.3847/2041-8213/ab960f}

\bibitem[{{Abbott} {et~al.}(2023){Abbott}, {Abbott}, {Acernese}, {Ackley}, {Adams}, {Adhikari}, {Adhikari}, {Adya}, {Affeldt}, {Agarwal}, {Agathos}, {Agatsuma}, {Aggarwal}, {Aguiar}, {Aiello}, {Ain}, {Ajith}, {Akutsu}, {de Alarc{\'o}n}, {Akcay}, {Albanesi}, {Allocca}, {Altin}, {Amato}, {Anand}, {Anand}, {Ananyeva}, {Anderson}, {Anderson}, {Ando}, {Andrade}, {Andres}, {Andri{\'c}}, {Angelova}, {Ansoldi}, {Antelis}, {Antier}, {Antonini}, {Appert}, {Arai}, {Arai}, {Arai}, {Araki}, {Araya}, {Araya}, {Areeda}, {Ar{\`e}ne}, {Aritomi}, {Arnaud}, {Arogeti}, {Aronson}, {Arun}, {Asada}, {Asali}, {Ashton}, {Aso}, {Assiduo}, {Aston}, {Astone}, {Aubin}, {Austin}, {Babak}, {Badaracco}, {Bader}, {Badger}, {Bae}, {Bae}, {Baer}, {Bagnasco}, {Bai}, {Baiotti}, {Baird}, {Bajpai}, {Ball}, {Ballardin}, {Ballmer}, {Balsamo}, {Baltus}, {Banagiri}, {Bankar}, {Barayoga}, {Barbieri}, {Barish}, {Barker}, {Barneo}, {Barone}, {Barr}, {Barsotti}, {Barsuglia}, {Barta}, {Bartlett}, {Barton}, {Bartos}, {Bassiri}, {Basti}, {Bawaj}, {Bayley},
  {Baylor}, {Bazzan}, {B{\'e}csy}, {Bedakihale}, {Bejger}, {Belahcene}, {Benedetto}, {Beniwal}, {Bennett}, {Bentley}, {Benyaala}, {Bergamin}, {Berger}, {Bernuzzi}, {Berry}, {Bersanetti}, {Bertolini}, {Betzwieser}, {Beveridge}, {Bhandare}, {Bhardwaj}, {Bhattacharjee}, {Bhaumik}, {Bilenko}, {Billingsley}, {Bini}, {Birney}, {Birnholtz}, {Biscans}, {Bischi}, {Biscoveanu}, {Bisht}, {Biswas}, {Bitossi}, {Bizouard}, {Blackburn}, {Blair}, {Blair}, {Blair}, {Bobba}, {Bode}, {Boer}, {Bogaert}, {Boldrini}, {Bonavena}, {Bondu}, {Bonilla}, {Bonnand}, {Booker}, {Boom}, {Bork}, {Boschi}, {Bose}, {Bose}, {Bossilkov}, {Boudart}, {Bouffanais}, {Bozzi}, {Bradaschia}, {Brady}, {Bramley}, {Branch}, {Branchesi}, {Brandt}, {Brau}, {Breschi}, {Briant}, {Briggs}, {Brillet}, {Brinkmann}, {Brockill}, {Brooks}, {Brooks}, {Brown}, {Brunett}, {Bruno}, {Bruntz}, {Bryant}, {Bulik}, {Bulten}, {Buonanno}, {Buscicchio}, {Buskulic}, {Buy}, {Byer}, {Cadonati}, {Cagnoli}, {Cahillane}, {Bustillo}, {Callaghan}, {Callister}, {Calloni}, {Cameron},
  {Camp}, {Canepa}, {Canevarolo}, {Cannavacciuolo}, {Cannon}, {Cao}, {Cao}, {Capocasa}, {Capote}, {Carapella}, {Carbognani}, {Carlin}, {Carney}, {Carpinelli}, {Carrillo}, {Carullo}, {Carver}, {Diaz}, {Casentini}, {Castaldi}, {Caudill}, {Cavagli{\`a}}, {Cavalier}, {Cavalieri}, {Ceasar}, {Cella}, {Cerd{\'a}-Dur{\'a}n}, {Cesarini}, {Chaibi}, {Chakravarti}, {Subrahmanya}, {Champion}, {Chan}, {Chan}, {Chan}, {Chan}, {Chan}, {Chandra}, {Chanial}, {Chao}, {Chapman-Bird}, {Charlton}, {Chase}, {Chassande-Mottin}, {Chatterjee}, {Chatterjee}, {Chatterjee}, {Chaturvedi}, {Chaty}, {Chatziioannou}, {Chen}, {Chen}, {Chen}, {Chen}, {Chen}, {Chen}, {Chen}, {Chen}, {Cheng}, {Cheong}, {Cheung}, {Chia}, {Chiadini}, {Chiang}, {Chiarini}, {Chierici}, {Chincarini}, {Chiofalo}, {Chiummo}, {Cho}, {Cho}, {Choudhary}, {Choudhary}, {Christensen}, {Chu}, {Chu}, {Chu}, {Chua}, {Chung}, {Ciani}, {Ciecielag}, {Cie{\'s}lar}, {Cifaldi}, {Ciobanu}, {Ciolfi}, {Cipriano}, {Cirone}, {Clara}, {Clark}, {Clark}, {Clarke}, {Clearwater}, {Clesse},
  {Cleva}, {Coccia}, {Codazzo}, {Cohadon}, {Cohen}, {Cohen}, {Colleoni}, {Collette}, {Colombo}, {Colpi}, {Compton}, {Constancio}, {Conti}, {Cooper}, {Corban}, {Corbitt}, {Cordero-Carri{\'o}n}, {Corezzi}, {Corley}, {Cornish}, {Corre}, {Corsi}, {Cortese}, {Costa}, {Cotesta}, {Coughlin}, {Coulon}, {Countryman}, {Cousins}, {Couvares}, {Coward}, {Cowart}, {Coyne}, {Coyne}, {Creighton}, {Creighton}, {Criswell}, {Croquette}, {Crowder}, {Cudell}, {Cullen}, {Cumming}, {Cummings}, {Cunningham}, {Cuoco}, {Cury{\l}o}, {Dabadie}, {Canton}, {Dall'Osso}, {D{\'a}lya}, {Dana}, {Daneshgaranbajastani}, {D'Angelo}, {Danila}, {Danilishin}, {D'Antonio}, {Danzmann}, {Darsow-Fromm}, {Dasgupta}, {Datrier}, {Datta}, {Dattilo}, {Dave}, {Davier}, {Davies}, {Davis}, {Davis}, {Daw}, {Dean}, {Debra}, {Deenadayalan}, {Degallaix}, {de Laurentis}, {Del{\'e}glise}, {Del Favero}, {de Lillo}, {de Lillo}, {Del Pozzo}, {Demarchi}, {de Matteis}, {D'Emilio}, {Demos}, {Dent}, {Depasse}, {de Pietri}, {De Rosa}, {de Rossi}, {Desalvo}, {de Simone},
  {Dhurandhar}, {D{\'\i}az}, {Diaz-Ortiz}, {Didio}, {Dietrich}, {di Fiore}, {di Fronzo}, {di Giorgio}, {di Giovanni}, {di Giovanni}, {di Girolamo}, {di Lieto}, {Ding}, {di Pace}, {di Palma}, {di Renzo}, {Divakarla}, {Dmitriev}, {Doctor}, {D'Onofrio}, {Donovan}, {Dooley}, {Doravari}, {Dorrington}, {Drago}, {Driggers}, {Drori}, {Ducoin}, {Dupej}, {Durante}, {D'Urso}, {Duverne}, {Dwyer}, {Eassa}, {Easter}, {Ebersold}, {Eckhardt}, {Eddolls}, {Edelman}, {Edo}, {Edy}, {Effler}, {Eguchi}, {Eichholz}, {Eikenberry}, {Eisenmann}, {Eisenstein}, {Ejlli}, {Engelby}, {Enomoto}, {Errico}, {Essick}, {Estell{\'e}s}, {Estevez}, {Etienne}, {Etzel}, {Evans}, {Evans}, {Ewing}, {Fafone}, {Fair}, {Fairhurst}, {Farah}, {Farinon}, {Farr}, {Farr}, {Farrow}, {Fauchon-Jones}, {Favaro}, {Favata}, {Fays}, {Fazio}, {Feicht}, {Fejer}, {Fenyvesi}, {Ferguson}, {Fernandez-Galiana}, {Ferrante}, {Ferreira}, {Fidecaro}, {Figura}, {Fiori}, {Fishbach}, {Fisher}, {Fittipaldi}, {Fiumara}, {Flaminio}, {Floden}, {Fong}, {Font}, {Fornal}, {Forsyth},
  {Franke}, {Frasca}, {Frasconi}, {Frederick}, {Freed}, {Frei}, {Freise}, {Frey}, {Fritschel}, {Frolov}, {Fronz{\'e}}, {Fujii}, {Fujikawa}, {Fukunaga}, {Fukushima}, {Fulda}, {Fyffe}, {Gabbard}, {Gadre}, {Gair}, {Gais}, {Galaudage}, {Gamba}, {Ganapathy}, {Ganguly}, {Gao}, {Gaonkar}, {Garaventa}, {Garc{\'\i}a}, {Garc{\'\i}a-N{\'u}{\~n}ez}, {Garc{\'\i}a-Quir{\'o}s}, {Garufi}, {Gateley}, {Gaudio}, {Gayathri}, {Ge}, {Gemme}, {Gennai}, {George}, {George}, {Gerberding}, {Gergely}, {Gewecke}, {Ghonge}, {Ghosh}, {Ghosh}, {Ghosh}, {Ghosh}, {Giacomazzo}, {Giacoppo}, {Giaime}, {Giardina}, {Gibson}, {Gier}, {Giesler}, {Giri}, {Gissi}, {Glanzer}, {Gleckl}, {Godwin}, {Golomb}, {Goetz}, {Goetz}, {Gohlke}, {Goncharov}, {Gonz{\'a}lez}, {Gopakumar}, {Gosselin}, {Gouaty}, {Gould}, {Grace}, {Grado}, {Granata}, {Granata}, {Grant}, {Gras}, {Grassia}, {Gray}, {Gray}, {Greco}, {Green}, {Green}, {Gretarsson}, {Gretarsson}, {Griffith}, {Griffiths}, {Griggs}, {Grignani}, {Grimaldi}, {Grimm}, {Grote}, {Grunewald}, {Gruning}, {Guerra},
  {Guidi}, {Guimaraes}, {Guix{\'e}}, {Gulati}, {Guo}, {Guo}, {Gupta}, {Gupta}, {Gupta}, {Gustafson}, {Gustafson}, {Guzman}, {Ha}, {Haegel}, {Hagiwara}, {Haino}, {Halim}, {Hall}, {Hamilton}, {Hammond}, {Han}, {Haney}, {Hanks}, {Hanna}, {Hannam}, {Hannuksela}, {Hansen}, {Hansen}, {Hanson}, {Harder}, {Hardwick}, {Haris}, {Harms}, {Harry}, {Harry}, {Hartwig}, {Hasegawa}, {Haskell}, {Hasskew}, {Haster}, {Hattori}, {Haughian}, {Hayakawa}, {Hayama}, {Hayes}, {Healy}, {Heidmann}, {Heidt}, {Heintze}, {Heinze}, {Heinzel}, {Heitmann}, {Hellman}, {Hello}, {Helmling-Cornell}, {Hemming}, {Hendry}, {Heng}, {Hennes}, {Hennig}, {Hennig}, {Hernandez}, {Vivanco}, {Heurs}, {Hild}, {Hill}, {Himemoto}, {Hines}, {Hiranuma}, {Hirata}, {Hirose}, {Hochheim}, {Hofman}, {Hohmann}, {Holcomb}, {Holland}, {Hollows}, {Holmes}, {Holt}, {Holz}, {Hong}, {Hopkins}, {Hough}, {Hourihane}, {Howell}, {Hoy}, {Hoyland}, {Hreibi}, {Hsieh}, {Hsu}, {Huang}, {Huang}, {Huang}, {Huang}, {Huang}, {Huang}, {H{\"u}bner}, {Huddart}, {Hughey}, {Hui}, {Hui},
  {Husa}, {Huttner}, {Huxford}, {Huynh-Dinh}, {Ide}, {Idzkowski}, {Iess}, {Ikenoue}, {Imam}, {Inayoshi}, {Ingram}, {Inoue}, {Ioka}, {Isi}, {Isleif}, {Ito}, {Itoh}, {Iyer}, {Izumi}, {Jaberianhamedan}, {Jacqmin}, {Jadhav}, {Jadhav}, {James}, {Jan}, {Jani}, {Janquart}, {Janssens}, {Janthalur}, {Jaranowski}, {Jariwala}, {Jaume}, {Jenkins}, {Jenner}, {Jeon}, {Jeunon}, {Jia}, {Jin}, {Johns}, {Jones}, {Jones}, {Jones}, {Jones}, {Jones}, {Jonker}, {Ju}, {Jung}, {Jung}, {Junker}, {Juste}, {Kaihotsu}, {Kajita}, {Kakizaki}, {Kalaghatgi}, {Kalogera}, {Kamai}, {Kamiizumi}, {Kanda}, {Kandhasamy}, {Kang}, {Kanner}, {Kao}, {Kapadia}, {Kapasi}, {Karat}, {Karathanasis}, {Karki}, {Kashyap}, {Kasprzack}, {Kastaun}, {Katsanevas}, {Katsavounidis}, {Katzman}, {Kaur}, {Kawabe}, {Kawaguchi}, {Kawai}, {Kawasaki}, {K{\'e}f{\'e}lian}, {Keitel}, {Key}, {Khadka}, {Khalili}, {Khan}, {Khazanov}, {Khetan}, {Khursheed}, {Kijbunchoo}, {Kim}, {Kim}, {Kim}, {Kim}, {Kim}, {Kim}, {Kimball}, {Kimura}, {Kinley-Hanlon}, {Kirchhoff}, {Kissel}, {Kita},
  {Kitazawa}, {Kleybolte}, {Klimenko}, {Knee}, {Knowles}, {Knyazev}, {Koch}, {Koekoek}, {Kojima}, {Kokeyama}, {Koley}, {Kolitsidou}, {Kolstein}, {Komori}, {Kondrashov}, {Kong}, {Kontos}, {Koper}, {Korobko}, {Kotake}, {Kovalam}, {Kozak}, {Kozakai}, {Kozu}, {Kringel}, {Krishnendu}, {Kr{\'o}lak}, {Kuehn}, {Kuei}, {Kuijer}, {Kulkarni}, {Kumar}, {Kumar}, {Kumar}, {Kumar}, {Kume}, {Kuns}, {Kuo}, {Kuo}, {Kuromiya}, {Kuroyanagi}, {Kusayanagi}, {Kuwahara}, {Kwak}, {Lagabbe}, {Laghi}, {Lalande}, {Lam}, {Lamberts}, {Landry}, {Landry}, {Lane}, {Lang}, {Lange}, {Lantz}, {La Rosa}, {Lartaux-Vollard}, {Lasky}, {Laxen}, {Lazzarini}, {Lazzaro}, {Leaci}, {Leavey}, {Lecoeuche}, {Lee}, {Lee}, {Lee}, {Lee}, {Lee}, {Lee}, {Lehmann}, {Lema{\^\i}tre}, {Leonardi}, {Leroy}, {Letendre}, {Levesque}, {Levin}, {Leviton}, {Leyde}, {Li}, {Li}, {Li}, {Li}, {Li}, {Li}, {Lin}, {Lin}, {Lin}, {Lin}, {Lin}, {Linde}, {Linker}, {Linley}, {Littenberg}, {Liu}, {Liu}, {Liu}, {Liu}, {Llamas}, {Llorens-Monteagudo}, {Lo}, {Lockwood}, {Loh}, {London},
  {Longo}, {Lopez}, {Portilla}, {Lorenzini}, {Loriette}, {Lormand}, {Losurdo}, {Lott}, {Lough}, {Lousto}, {Lovelace}, {Lucaccioni}, {L{\"u}ck}, {Lumaca}, {Lundgren}, {Luo}, {Lynam}, {Macas}, {Macinnis}, {MacLeod}, {MacMillan}, {Macquet}, {Hernandez}, {Magazz{\`u}}, {Magee}, {Maggiore}, {Magnozzi}, {Mahesh}, {Majorana}, {Makarem}, {Maksimovic}, {Maliakal}, {Malik}, {Man}, {Mandic}, {Mangano}, {Mango}, {Mansell}, {Manske}, {Mantovani}, {Mapelli}, {Marchesoni}, {Marchio}, {Marion}, {Mark}, {M{\'a}rka}, {M{\'a}rka}, {Markakis}, {Markosyan}, {Markowitz}, {Maros}, {Marquina}, {Marsat}, {Martelli}, {Martin}, {Martin}, {Martinez}, {Martinez}, {Martinez}, {Martinovic}, {Martynov}, {Marx}, {Masalehdan}, {Mason}, {Massera}, {Masserot}, {Massinger}, {Masso-Reid}, {Mastrogiovanni}, {Matas}, {Mateu-Lucena}, {Matichard}, {Matiushechkina}, {Mavalvala}, {McCann}, {McCarthy}, {McClelland}, {McClincy}, {McCormick}, {McCuller}, {McGhee}, {McGuire}, {McIsaac}, {McIver}, {McRae}, {McWilliams}, {Meacher}, {Mehmet}, {Mehta},
  {Meijer}, {Melatos}, {Melchor}, {Mendell}, {Menendez-Vazquez}, {Menoni}, {Mercer}, {Mereni}, {Merfeld}, {Merilh}, {Merritt}, {Merzougui}, {Meshkov}, {Messenger}, {Messick}, {Meyers}, {Meylahn}, {Mhaske}, {Miani}, {Miao}, {Michaloliakos}, {Michel}, {Michimura}, {Middleton}, {Milano}, {Miller}, {Miller}, {Miller}, {Miller}, {Millhouse}, {Mills}, {Milotti}, {Minazzoli}, {Minenkov}, {Mio}, {Mir}, {Miravet-Ten{\'e}s}, {Mishra}, {Mishra}, {Mistry}, {Mitra}, {Mitrofanov}, {Mitselmakher}, {Mittleman}, {Miyakawa}, {Miyamoto}, {Miyazaki}, {Miyo}, {Miyoki}, {Mo}, {Modafferi}, {Moguel}, {Mogushi}, {Mohapatra}, {Mohite}, {Molina}, {Molina-Ruiz}, {Mondin}, {Montani}, {Moore}, {Moraru}, {Morawski}, {More}, {Moreno}, {Moreno}, {Mori}, {Morisaki}, {Moriwaki}, {Morr{\'a}s}, {Mours}, {Mow-Lowry}, {Mozzon}, {Muciaccia}, {Mukherjee}, {Mukherjee}, {Mukherjee}, {Mukherjee}, {Mukherjee}, {Mukund}, {Mullavey}, {Munch}, {Mu{\~n}iz}, {Murray}, {Musenich}, {Muusse}, {Nadji}, {Nagano}, {Nagano}, {Nagar}, {Nakamura}, {Nakano}, {Nakano},
  {Nakashima}, {Nakayama}, {Napolano}, {Nardecchia}, {Narikawa}, {Naticchioni}, {Nayak}, {Nayak}, {Negishi}, {Neil}, {Neilson}, {Nelemans}, {Nelson}, {Nery}, {Neubauer}, {Neunzert}, {Ng}, {Ng}, {Nguyen}, {Nguyen}, {Nguyen}, {Quynh}, {Ni}, {Nichols}, {Nishizawa}, {Nissanke}, {Nitoglia}, {Nocera}, {Norman}, {North}, {Nozaki}, {Siles}, {Nuttall}, {Oberling}, {O'Brien}, {Obuchi}, {O'Dell}, {Oelker}, {Ogaki}, {Oganesyan}, {Oh}, {Oh}, {Oh}, {Ohashi}, {Ohishi}, {Ohkawa}, {Ohme}, {Ohta}, {Okada}, {Okutani}, {Okutomi}, {Olivetto}, {Oohara}, {Ooi}, {Oram}, {O'Reilly}, {Ormiston}, {Ormsby}, {Ortega}, {O'Shaughnessy}, {O'Shea}, {Oshino}, {Ossokine}, {Osthelder}, {Otabe}, {Ottaway}, {Overmier}, {Pace}, {Pagano}, {Page}, {Pagliaroli}, {Pai}, {Pai}, {Palamos}, {Palashov}, {Palomba}, {Pan}, {Pan}, {Panda}, {Pang}, {Pang}, {Pankow}, {Pannarale}, {Pant}, {Panther}, {Paoletti}, {Paoli}, {Paolone}, {Parisi}, {Park}, {Park}, {Parker}, {Pascucci}, {Pasqualetti}, {Passaquieti}, {Passuello}, {Patel}, {Pathak}, {Patricelli},
  {Patron}, {Paul}, {Payne}, {Pedraza}, {Pegoraro}, {Pele}, {Arellano}, {Penn}, {Perego}, {Pereira}, {Pereira}, {Perez}, {P{\'e}rigois}, {Perkins}, {Perreca}, {Perri{\`e}s}, {Petermann}, {Petterson}, {Pfeiffer}, {Pham}, {Phukon}, {Piccinni}, {Pichot}, {Piendibene}, {Piergiovanni}, {Pierini}, {Pierro}, {Pillant}, {Pillas}, {Pilo}, {Pinard}, {Pinto}, {Pinto}, {Piotrzkowski}, {Piotrzkowski}, {Pirello}, {Pitkin}, {Placidi}, {Planas}, {Plastino}, {Pluchar}, {Poggiani}, {Polini}, {Pong}, {Ponrathnam}, {Popolizio}, {Porter}, {Poulton}, {Powell}, {Pracchia}, {Pradier}, {Prajapati}, {Prasai}, {Prasanna}, {Pratten}, {Principe}, {Prodi}, {Prokhorov}, {Prosposito}, {Prudenzi}, {Puecher}, {Punturo}, {Puosi}, {Puppo}, {P{\"u}rrer}, {Qi}, {Quetschke}, {Quitzow-James}, {Raab}, {Raaijmakers}, {Radkins}, {Radulesco}, {Raffai}, {Rail}, {Raja}, {Rajan}, {Ramirez}, {Ramirez}, {Ramos-Buades}, {Rana}, {Rapagnani}, {Rapol}, {Ray}, {Raymond}, {Raza}, {Razzano}, {Read}, {Rees}, {Regimbau}, {Rei}, {Reid}, {Reid}, {Reitze}, {Relton},
  {Renzini}, {Rettegno}, {Reza}, {Rezac}, {Ricci}, {Richards}, {Richardson}, {Richardson}, {Riemenschneider}, {Riles}, {Rinaldi}, {Rink}, {Rizzo}, {Robertson}, {Robie}, {Robinet}, {Rocchi}, {Rodriguez}, {Rolland}, {Rollins}, {Romanelli}, {Romano}, {Romel}, {Romero-Rodr{\'\i}guez}, {Romero-Shaw}, {Romie}, {Ronchini}, {Rosa}, {Rose}, {Rosi{\'n}ska}, {Ross}, {Rowan}, {Rowlinson}, {Roy}, {Roy}, {Roy}, {Rozza}, {Ruggi}, {Ryan}, {Sachdev}, {Sadecki}, {Sadiq}, {Sago}, {Saito}, {Saito}, {Sakai}, {Sakai}, {Sakellariadou}, {Sakuno}, {Salafia}, {Salconi}, {Saleem}, {Salemi}, {Samajdar}, {Sanchez}, {Sanchez}, {Sanchez}, {Sanchis-Gual}, {Sanders}, {Sanuy}, {Saravanan}, {Sarin}, {Sassolas}, {Satari}, {Sathyaprakash}, {Sato}, {Sato}, {Sauter}, {Savage}, {Sawada}, {Sawant}, {Sawant}, {Sayah}, {Schaetzl}, {Scheel}, {Scheuer}, {Schiworski}, {Schmidt}, {Schmidt}, {Schnabel}, {Schneewind}, {Schofield}, {Sch{\"o}nbeck}, {Schulte}, {Schutz}, {Schwartz}, {Scott}, {Scott}, {Seglar-Arroyo}, {Sekiguchi}, {Sekiguchi}, {Sellers},
  {Sengupta}, {Sentenac}, {Seo}, {Sequino}, {Sergeev}, {Setyawati}, {Shaffer}, {Shahriar}, {Shams}, {Shao}, {Sharma}, {Sharma}, {Shawhan}, {Shcheblanov}, {Shibagaki}, {Shikauchi}, {Shimizu}, {Shimoda}, {Shimode}, {Shinkai}, {Shishido}, {Shoda}, {Shoemaker}, {Shoemaker}, {Shyamsundar}, {Sieniawska}, {Sigg}, {Singer}, {Singh}, {Singh}, {Singha}, {Sintes}, {Sipala}, {Skliris}, {Slagmolen}, {Slaven-Blair}, {Smetana}, {Smith}, {Smith}, {Soldateschi}, {Somala}, {Somiya}, {Son}, {Soni}, {Soni}, {Sordini}, {Sorrentino}, {Sorrentino}, {Sotani}, {Soulard}, {Souradeep}, {Sowell}, {Spagnuolo}, {Spencer}, {Spera}, {Srinivasan}, {Srivastava}, {Srivastava}, {Staats}, {Stachie}, {Steer}, {Steinhoff}, {Steinlechner}, {Steinlechner}, {Stevenson}, {Stops}, {Stover}, {Strain}, {Strang}, {Stratta}, {Strunk}, {Sturani}, {Stuver}, {Sudhagar}, {Sudhir}, {Sugimoto}, {Suh}, {Sullivan}, {Summerscales}, {Sun}, {Sun}, {Sunil}, {Sur}, {Suresh}, {Sutton}, {Suzuki}, {Suzuki}, {Swinkels}, {Szczepa{\'n}czyk}, {Szewczyk}, {Tacca}, {Tagoshi},
  {Tait}, {Takahashi}, {Takahashi}, {Takamori}, {Takano}, {Takeda}, {Takeda}, {Talbot}, {Talbot}, {Tanaka}, {Tanaka}, {Tanaka}, {Tanaka}, {Tanaka}, {Tanasijczuk}, {Tanioka}, {Tanner}, {Tao}, {Tao}, {Mart{\'\i}n}, {Taranto}, {Tasson}, {Telada}, {Tenorio}, {Terhune}, {Terkowski}, {Thirugnanasambandam}, {Thomas}, {Thomas}, {Thomas}, {Thompson}, {Thondapu}, {Thorne}, {Thrane}, {Tiwari}, {Tiwari}, {Tiwari}, {Toivonen}, {Toland}, {Tolley}, {Tomaru}, {Tomigami}, {Tomura}, {Tonelli}, {Torres-Forn{\'e}}, {Torrie}, {E Melo}, {T{\"o}yr{\"a}}, {Trapananti}, {Travasso}, {Traylor}, {Trevor}, {Tringali}, {Tripathee}, {Troiano}, {Trovato}, {Trozzo}, {Trudeau}, {Tsai}, {Tsai}, {Tsang}, {Tsang}, {Tsao}, {Tse}, {Tso}, {Tsubono}, {Tsuchida}, {Tsukada}, {Tsuna}, {Tsutsui}, {Tsuzuki}, {Turbang}, {Turconi}, {Tuyenbayev}, {Ubhi}, {Uchikata}, {Uchiyama}, {Udall}, {Ueda}, {Uehara}, {Ueno}, {Ueshima}, {Unnikrishnan}, {Uraguchi}, {Urban}, {Ushiba}, {Utina}, {Vahlbruch}, {Vajente}, {Vajpeyi}, {Valdes}, {Valentini}, {Valsan}, {van Bakel},
  {van Beuzekom}, {van den Brand}, {van den Broeck}, {Vander-Hyde}, {van der Schaaf}, {van Heijningen}, {Vanosky}, {van Putten}, {van Remortel}, {Vardaro}, {Vargas}, {Varma}, {Vas{\'u}th}, {Vecchio}, {Vedovato}, {Veitch}, {Veitch}, {Venneberg}, {Venugopalan}, {Verkindt}, {Verma}, {Verma}, {Veske}, {Vetrano}, {Vicer{\'e}}, {Vidyant}, {Viets}, {Vijaykumar}, {Villa-Ortega}, {Vinet}, {Virtuoso}, {Vitale}, {Vo}, {Vocca}, {von Reis}, {von Wrangel}, {Vorvick}, {Vyatchanin}, {Wade}, {Wade}, {Wagner}, {Walet}, {Walker}, {Wallace}, {Wallace}, {Walsh}, {Wang}, {Wang}, {Wang}, {Ward}, {Warner}, {Was}, {Washimi}, {Washington}, {Watchi}, {Weaver}, {Webster}, {Weinert}, {Weinstein}, {Weiss}, {Weller}, {Wellmann}, {Wen}, {We{\ss}els}, {Wette}, {Whelan}, {White}, {Whiting}, {Whittle}, {Wilken}, {Williams}, {Williams}, {Williamson}, {Willis}, {Willke}, {Wilson}, {Winkler}, {Wipf}, {Wlodarczyk}, {Woan}, {Woehler}, {Wofford}, {Wong}, {Wu}, {Wu}, {Wu}, {Wu}, {Wysocki}, {Xiao}, {Xu}, {Yamada}, {Yamamoto}, {Yamamoto}, {Yamamoto},
  {Yamamoto}, {Yamashita}, {Yamazaki}, {Yang}, {Yang}, {Yang}, {Yang}, {Yang}, {Yap}, {Yeeles}, {Yelikar}, {Ying}, {Yokogawa}, {Yokoyama}, {Yokozawa}, {Yoo}, {Yoshioka}, {Yu}, {Yu}, {Yuzurihara}, {Zadro{\.z}ny}, {Zanolin}, {Zeidler}, {Zelenova}, {Zendri}, {Zevin}, {Zhan}, {Zhang}, {Zhang}, {Zhang}, {Zhang}, {Zhang}, {Zhao}, {Zhao}, {Zhao}, {Zhao}, {Zheng}, {Zhou}, {Zhou}, {Zhu}, {Zhu}, {Zimmerman}, {Zlochower}, {Zucker}, {Zweizig}, {LIGO Scientific Collaboration}, {VIRGO Collaboration}, \& {KAGRA Collaboration}}]{GWTC3_Rates}
{Abbott}, R., {Abbott}, T.~D., {Acernese}, F., {et~al.} 2023, Physical Review X, 13, 011048, \dodoi{10.1103/PhysRevX.13.011048}

\bibitem[{{Alam} {et~al.}(2015){Alam}, {Albareti}, {Allende Prieto}, {Anders}, {Anderson}, {Anderton}, {Andrews}, {Armengaud}, {Aubourg}, {Bailey}, {Basu}, {Bautista}, {Beaton}, {Beers}, {Bender}, {Berlind}, {Beutler}, {Bhardwaj}, {Bird}, {Bizyaev}, {Blake}, {Blanton}, {Blomqvist}, {Bochanski}, {Bolton}, {Bovy}, {Shelden Bradley}, {Brandt}, {Brauer}, {Brinkmann}, {Brown}, {Brownstein}, {Burden}, {Burtin}, {Busca}, {Cai}, {Capozzi}, {Carnero Rosell}, {Carr}, {Carrera}, {Chambers}, {Chaplin}, {Chen}, {Chiappini}, {Chojnowski}, {Chuang}, {Clerc}, {Comparat}, {Covey}, {Croft}, {Cuesta}, {Cunha}, {da Costa}, {Da Rio}, {Davenport}, {Dawson}, {De Lee}, {Delubac}, {Deshpande}, {Dhital}, {Dutra-Ferreira}, {Dwelly}, {Ealet}, {Ebelke}, {Edmondson}, {Eisenstein}, {Ellsworth}, {Elsworth}, {Epstein}, {Eracleous}, {Escoffier}, {Esposito}, {Evans}, {Fan}, {Fern{\'a}ndez-Alvar}, {Feuillet}, {Filiz Ak}, {Finley}, {Finoguenov}, {Flaherty}, {Fleming}, {Font-Ribera}, {Foster}, {Frinchaboy}, {Galbraith-Frew}, {Garc{\'\i}a},
  {Garc{\'\i}a-Hern{\'a}ndez}, {Garc{\'\i}a P{\'e}rez}, {Gaulme}, {Ge}, {G{\'e}nova-Santos}, {Georgakakis}, {Ghezzi}, {Gillespie}, {Girardi}, {Goddard}, {Gontcho}, {Gonz{\'a}lez Hern{\'a}ndez}, {Grebel}, {Green}, {Grieb}, {Grieves}, {Gunn}, {Guo}, {Harding}, {Hasselquist}, {Hawley}, {Hayden}, {Hearty}, {Hekker}, {Ho}, {Hogg}, {Holley-Bockelmann}, {Holtzman}, {Honscheid}, {Huber}, {Huehnerhoff}, {Ivans}, {Jiang}, {Johnson}, {Kinemuchi}, {Kirkby}, {Kitaura}, {Klaene}, {Knapp}, {Kneib}, {Koenig}, {Lam}, {Lan}, {Lang}, {Laurent}, {Le Goff}, {Leauthaud}, {Lee}, {Lee}, {Licquia}, {Liu}, {Long}, {L{\'o}pez-Corredoira}, {Lorenzo-Oliveira}, {Lucatello}, {Lundgren}, {Lupton}, {Mack}, {Mahadevan}, {Maia}, {Majewski}, {Malanushenko}, {Malanushenko}, {Manchado}, {Manera}, {Mao}, {Maraston}, {Marchwinski}, {Margala}, {Martell}, {Martig}, {Masters}, {Mathur}, {McBride}, {McGehee}, {McGreer}, {McMahon}, {M{\'e}nard}, {Menzel}, {Merloni}, {M{\'e}sz{\'a}ros}, {Miller}, {Miralda-Escud{\'e}}, {Miyatake}, {Montero-Dorta}, {More},
  {Morganson}, {Morice-Atkinson}, {Morrison}, {Mosser}, {Muna}, {Myers}, {Nandra}, {Newman}, {Neyrinck}, {Nguyen}, {Nichol}, {Nidever}, {Noterdaeme}, {Nuza}, {O'Connell}, {O'Connell}, {O'Connell}, {Ogando}, {Olmstead}, {Oravetz}, {Oravetz}, {Osumi}, {Owen}, {Padgett}, {Padmanabhan}, {Paegert}, {Palanque-Delabrouille}, {Pan}, {Parejko}, {P{\^a}ris}, {Park}, {Pattarakijwanich}, {Pellejero-Ibanez}, {Pepper}, {Percival}, {P{\'e}rez-Fournon}, {P{\'e}rez-R{\`a}fols}, {Petitjean}, {Pieri}, {Pinsonneault}, {Porto de Mello}, {Prada}, {Prakash}, {Price-Whelan}, {Protopapas}, {Raddick}, {Rahman}, {Reid}, {Rich}, {Rix}, {Robin}, {Rockosi}, {Rodrigues}, {Rodr{\'\i}guez-Torres}, {Roe}, {Ross}, {Ross}, {Rossi}, {Ruan}, {Rubi{\~n}o-Mart{\'\i}n}, {Rykoff}, {Salazar-Albornoz}, {Salvato}, {Samushia}, {S{\'a}nchez}, {Santiago}, {Sayres}, {Schiavon}, {Schlegel}, {Schmidt}, {Schneider}, {Schultheis}, {Schwope}, {Sc{\'o}ccola}, {Scott}, {Sellgren}, {Seo}, {Serenelli}, {Shane}, {Shen}, {Shetrone}, {Shu}, {Silva Aguirre}, {Sivarani},
  {Skrutskie}, {Slosar}, {Smith}, {Sobreira}, {Souto}, {Stassun}, {Steinmetz}, {Stello}, {Strauss}, {Streblyanska}, {Suzuki}, {Swanson}, {Tan}, {Tayar}, {Terrien}, {Thakar}, {Thomas}, {Thomas}, {Thompson}, {Tinker}, {Tojeiro}, {Troup}, {Vargas-Maga{\~n}a}, {Vazquez}, {Verde}, {Viel}, {Vogt}, {Wake}, {Wang}, {Weaver}, {Weinberg}, {Weiner}, {White}, {Wilson}, {Wisniewski}, {Wood-Vasey}, {Ye`che}, {York}, {Zakamska}, {Zamora}, {Zasowski}, {Zehavi}, {Zhao}, {Zheng}, {Zhou}, {Zhou}, {Zou}, \& {Zhu}}]{Alam+15}
{Alam}, S., {Albareti}, F.~D., {Allende Prieto}, C., {et~al.} 2015, \apjs, 219, 12, \dodoi{10.1088/0067-0049/219/1/12}

\bibitem[{{Anand} {et~al.}(2023){Anand}, {Pang}, {Bulla}, {Coughlin}, {Dietrich}, {Healy}, {Hussenot-Desenonges}, {Jegou du Laz}, {Kasliwal}, {Kunert}, {Markin}, {Mooley}, {Nedora}, \& {Neuweiler}}]{Anand+23_170817}
{Anand}, S., {Pang}, P. T.~H., {Bulla}, M., {et~al.} 2023, arXiv e-prints, arXiv:2307.11080, \dodoi{10.48550/arXiv.2307.11080}

\bibitem[{{Anand} {et~al.}(2024){Anand}, {Barnes}, {Yang}, {Kasliwal}, {Coughlin}, {Sollerman}, {De}, {Fremling}, {Corsi}, {Ho}, {Balasubramanian}, {Omand}, {Srinivasaragavan}, {Cenko}, {Ahumada}, {Andreoni}, {Dahiwale}, {Das}, {Jencson}, {Karambelkar}, {Kumar}, {Metzger}, {Perley}, {Sarin}, {Schweyer}, {Schulze}, {Sharma}, {Sit}, {Stein}, {Tartaglia}, {Tinyanont}, {Tzanidakis}, {van Roestel}, {Yao}, {Bloom}, {Cook}, {Dekany}, {Graham}, {Groom}, {Kaplan}, {Masci}, {Medford}, {Riddle}, \& {Zhang}}]{Anand+24}
{Anand}, S., {Barnes}, J., {Yang}, S., {et~al.} 2024, \apj, 962, 68, \dodoi{10.3847/1538-4357/ad11df}

\bibitem[{{Andreoni} {et~al.}(2017){Andreoni}, {Ackley}, {Cooke}, {Acharyya}, {Allison}, {Anderson}, {Ashley}, {Baade}, {Bailes}, {Bannister}, {Beardsley}, {Bessell}, {Bian}, {Bland}, {Boer}, {Booler}, {Brandeker}, {Brown}, {Buckley}, {Chang}, {Coward}, {Crawford}, {Crisp}, {Crosse}, {Cucchiara}, {Cup{\'a}k}, {de Gois}, {Deller}, {Devillepoix}, {Dobie}, {Elmer}, {Emrich}, {Farah}, {Farrell}, {Franzen}, {Gaensler}, {Galloway}, {Gendre}, {Giblin}, {Goobar}, {Green}, {Hancock}, {Hartig}, {Howell}, {Horsley}, {Hotan}, {Howie}, {Hu}, {Hu}, {James}, {Johnston}, {Johnston-Hollitt}, {Kaplan}, {Kasliwal}, {Keane}, {Kenney}, {Klotz}, {Lau}, {Laugier}, {Lenc}, {Li}, {Liang}, {Lidman}, {Luvaul}, {Lynch}, {Ma}, {Macpherson}, {Mao}, {McClelland}, {McCully}, {M{\"o}ller}, {Morales}, {Morris}, {Murphy}, {Noysena}, {Onken}, {Orange}, {Os{\l}owski}, {Pallot}, {Paxman}, {Potter}, {Pritchard}, {Raja}, {Ridden-Harper}, {Romero-Colmenero}, {Sadler}, {Sansom}, {Scalzo}, {Schmidt}, {Scott}, {Seghouani}, {Shang}, {Shannon}, {Shao},
  {Shara}, {Sharp}, {Sokolowski}, {Sollerman}, {Staff}, {Steele}, {Sun}, {Suntzeff}, {Tao}, {Tingay}, {Towner}, {Thierry}, {Trott}, {Tucker}, {V{\"a}is{\"a}nen}, {Krishnan}, {Walker}, {Wang}, {Wang}, {Wayth}, {Whiting}, {Williams}, {Williams}, {Wolf}, {Wu}, {Wu}, {Yang}, {Yuan}, {Zhang}, {Zhou}, \& {Zovaro}}]{Andreoni+17}
{Andreoni}, I., {Ackley}, K., {Cooke}, J., {et~al.} 2017, \pasa, 34, e069, \dodoi{10.1017/pasa.2017.65}

\bibitem[{{Andreoni} {et~al.}(2020){Andreoni}, {Kool}, {Sagu{\'e}s Carracedo}, {Kasliwal}, {Bulla}, {Ahumada}, {Coughlin}, {Anand}, {Sollerman}, {Goobar}, {Kaplan}, {Loveridge}, {Karambelkar}, {Cooke}, {Bagdasaryan}, {Bellm}, {Cenko}, {Cook}, {De}, {Dekany}, {Delacroix}, {Drake}, {Duev}, {Fremling}, {Golkhou}, {Graham}, {Hale}, {Kulkarni}, {Kupfer}, {Laher}, {Mahabal}, {Masci}, {Rusholme}, {Smith}, {Tzanidakis}, {Van Sistine}, \& {Yao}}]{Andreoni+20_ztfkn}
{Andreoni}, I., {Kool}, E.~C., {Sagu{\'e}s Carracedo}, A., {et~al.} 2020, \apj, 904, 155, \dodoi{10.3847/1538-4357/abbf4c}

\bibitem[{{Andreoni} {et~al.}(2024){Andreoni}, {Coughlin}, {Criswell}, {Bulla}, {Toivonen}, {Singer}, {Palmese}, {Burns}, {Gezari}, {Kasliwal}, {Kiendrebeogo}, {Mahabal}, {Moriya}, {Rest}, {Scolnic}, {Simcoe}, {Soon}, {Stein}, \& {Travouillon}}]{Andreoni_romankilonova}
{Andreoni}, I., {Coughlin}, M.~W., {Criswell}, A.~W., {et~al.} 2024, Astroparticle Physics, 155, 102904, \dodoi{10.1016/j.astropartphys.2023.102904}

\bibitem[{{Arcavi}(2018)}]{Arcavi18}
{Arcavi}, I. 2018, \apjl, 855, L23, \dodoi{10.3847/2041-8213/aab267}

\bibitem[{{Arcavi} {et~al.}(2017){Arcavi}, {Hosseinzadeh}, {Howell}, {McCully}, {Poznanski}, {Kasen}, {Barnes}, {Zaltzman}, {Vasylyev}, {Maoz}, \& {Valenti}}]{Arcavi+17}
{Arcavi}, I., {Hosseinzadeh}, G., {Howell}, D.~A., {et~al.} 2017, \nat, 551, 64, \dodoi{10.1038/nature24291}

\bibitem[{{Arnett}(1982)}]{Arnett82}
{Arnett}, W.~D. 1982, \apj, 253, 785, \dodoi{10.1086/159681}

\bibitem[{{Ascenzi} {et~al.}(2019){Ascenzi}, {Coughlin}, {Dietrich}, {Foley}, {Ramirez-Ruiz}, {Piranomonte}, {Mockler}, {Murguia-Berthier}, {Fryer}, {Lloyd-Ronning}, \& {Rosswog}}]{Ascenzi+19}
{Ascenzi}, S., {Coughlin}, M.~W., {Dietrich}, T., {et~al.} 2019, \mnras, 486, 672, \dodoi{10.1093/mnras/stz891}

\bibitem[{{Atteia} {et~al.}(2022){Atteia}, {Cordier}, \& {Wei}}]{SVOM}
{Atteia}, J.~L., {Cordier}, B., \& {Wei}, J. 2022, International Journal of Modern Physics D, 31, 2230008, \dodoi{10.1142/S0218271822300087}

\bibitem[{{Atteia} {et~al.}(1987){Atteia}, {Barat}, {Hurley}, {Niel}, {Vedrenne}, {Evans}, {Fenimore}, {Klebesadel}, {Laros}, {Cline}, {Desai}, {Teegarden}, {Estulin}, {Zenchenko}, {Kusnetsov}, \& {Kurt}}]{IPNcat_1987}
{Atteia}, J.~L., {Barat}, C., {Hurley}, K., {et~al.} 1987, \apjs, 64, 305, \dodoi{10.1086/191198}

\bibitem[{{Banerjee} {et~al.}(2024){Banerjee}, {Tanaka}, {Kato}, \& {Gaigalas}}]{Banerjee+24}
{Banerjee}, S., {Tanaka}, M., {Kato}, D., \& {Gaigalas}, G. 2024, \apj, 968, 64, \dodoi{10.3847/1538-4357/ad4029}

\bibitem[{{Barnes} \& {Kasen}(2013)}]{barneskasen13}
{Barnes}, J., \& {Kasen}, D. 2013, \apj, 775, 18, \dodoi{10.1088/0004-637X/775/1/18}

\bibitem[{{Barnes} {et~al.}(2016){Barnes}, {Kasen}, {Wu}, \& {Mart{\'\i}nez-Pinedo}}]{barnes+16}
{Barnes}, J., {Kasen}, D., {Wu}, M.-R., \& {Mart{\'\i}nez-Pinedo}, G. 2016, \apj, 829, 110, \dodoi{10.3847/0004-637X/829/2/110}

\bibitem[{{Barnes} \& {Metzger}(2022)}]{BarnesMetzger22}
{Barnes}, J., \& {Metzger}, B.~D. 2022, \apjl, 939, L29, \dodoi{10.3847/2041-8213/ac9b41}

\bibitem[{{Barnes} \& {Metzger}(2023)}]{BarnesMetzger23}
---. 2023, \apj, 947, 55, \dodoi{10.3847/1538-4357/acc384}

\bibitem[{{Barnes} {et~al.}(2021){Barnes}, {Zhu}, {Lund}, {Sprouse}, {Vassh}, {McLaughlin}, {Mumpower}, \& {Surman}}]{Barnes+21}
{Barnes}, J., {Zhu}, Y.~L., {Lund}, K.~A., {et~al.} 2021, \apj, 918, 44, \dodoi{10.3847/1538-4357/ac0aec}

\bibitem[{{Barthelmy} {et~al.}(2006){Barthelmy}, {Barbier}, {Cummings}, {Fenimore}, {Gehrels}, {Hullinger}, {Krimm}, {Koss}, {Markwardt}, {Palmer}, {Parsons}, {Sakamoto}, {Sato}, {Stamatikos}, \& {Tueller}}]{Barthelmy_GCN5256}
{Barthelmy}, S., {Barbier}, L., {Cummings}, J., {et~al.} 2006, GRB Coordinates Network, 5256, 1

\bibitem[{{Barthelmy} {et~al.}(2005){Barthelmy}, {Barbier}, {Cummings}, {Fenimore}, {Gehrels}, {Hullinger}, {Krimm}, {Markwardt}, {Palmer}, {Parsons}, {Sato}, {Suzuki}, {Takahashi}, {Tashiro}, \& {Tueller}}]{Swift_BAT}
{Barthelmy}, S.~D., {Barbier}, L.~M., {Cummings}, J.~R., {et~al.} 2005, \ssr, 120, 143, \dodoi{10.1007/s11214-005-5096-3}

\bibitem[{{Battistini} \& {Bensby}(2016)}]{BattistiniBensby16}
{Battistini}, C., \& {Bensby}, T. 2016, \aap, 586, A49, \dodoi{10.1051/0004-6361/201527385}

\bibitem[{{Belles} {et~al.}(2021){Belles}, {D'Ai}, \& {Swift/UVOT Team}}]{grb211211a_uvot}
{Belles}, A., {D'Ai}, A., \& {Swift/UVOT Team}. 2021, GRB Coordinates Network, 31222, 1

\bibitem[{{Bennett} {et~al.}(2014){Bennett}, {Larson}, {Weiland}, \& {Hinshaw}}]{Bennett+14}
{Bennett}, C.~L., {Larson}, D., {Weiland}, J.~L., \& {Hinshaw}, G. 2014, \apj, 794, 135, \dodoi{10.1088/0004-637X/794/2/135}

\bibitem[{{Berger} {et~al.}(2013){Berger}, {Fong}, \& {Chornock}}]{berger+13}
{Berger}, E., {Fong}, W., \& {Chornock}, R. 2013, \apjl, 774, L23, \dodoi{10.1088/2041-8205/774/2/L23}

\bibitem[{{Bianco} {et~al.}(2022){Bianco}, {Ivezi{\'c}}, {Jones}, {Graham}, {Marshall}, {Saha}, {Strauss}, {Yoachim}, {Ribeiro}, {Anguita}, {Bauer}, {Bauer}, {Bellm}, {Blum}, {Brandt}, {Brough}, {Catelan}, {Clarkson}, {Connolly}, {Gawiser}, {Gizis}, {Hlo{\v{z}}ek}, {Kaviraj}, {Liu}, {Lochner}, {Mahabal}, {Mandelbaum}, {McGehee}, {Neilsen}, {Olsen}, {Peiris}, {Rhodes}, {Richards}, {Ridgway}, {Schwamb}, {Scolnic}, {Shemmer}, {Slater}, {Slosar}, {Smartt}, {Strader}, {Street}, {Trilling}, {Verma}, {Vivas}, {Wechsler}, \& {Willman}}]{Bianco+22}
{Bianco}, F.~B., {Ivezi{\'c}}, {\v{Z}}., {Jones}, R.~L., {et~al.} 2022, \apjs, 258, 1, \dodoi{10.3847/1538-4365/ac3e72}

\bibitem[{{Blanchard} {et~al.}(2017){Blanchard}, {Berger}, {Fong}, {Nicholl}, {Leja}, {Conroy}, {Alexander}, {Margutti}, {Williams}, {Doctor}, {Chornock}, {Villar}, {Cowperthwaite}, {Annis}, {Brout}, {Brown}, {Chen}, {Eftekhari}, {Frieman}, {Holz}, {Metzger}, {Rest}, {Sako}, \& {Soares-Santos}}]{Blanchard+17}
{Blanchard}, P.~K., {Berger}, E., {Fong}, W., {et~al.} 2017, \apjl, 848, L22, \dodoi{10.3847/2041-8213/aa9055}

\bibitem[{{Blanchard} {et~al.}(2023){Blanchard}, {Villar}, {Chornock}, {Laskar}, {Li}, {Leja}, {Pierel}, {Berger}, {Margutti}, {Alexander}, {Barnes}, {Cendes}, {Eftekhari}, {Kasen}, {LeBaron}, {Metzger}, {Muzerolle Page}, {Rest}, {Sears}, {Siegel}, \& {Karthik Yadavalli}}]{Blanchard+23}
{Blanchard}, P.~K., {Villar}, V.~A., {Chornock}, R., {et~al.} 2023, arXiv e-prints, arXiv:2308.14197, \dodoi{10.48550/arXiv.2308.14197}

\bibitem[{{Bloom} {et~al.}(2006){Bloom}, {Prochaska}, {Pooley}, {Blake}, {Foley}, {Jha}, {Ramirez-Ruiz}, {Granot}, {Filippenko}, {Sigurdsson}, {Barth}, {Chen}, {Cooper}, {Falco}, {Gal}, {Gerke}, {Gladders}, {Greene}, {Hennanwi}, {Ho}, {Hurley}, {Koester}, {Li}, {Lubin}, {Newman}, {Perley}, {Squires}, \& {Wood-Vasey}}]{Bloom+06}
{Bloom}, J.~S., {Prochaska}, J.~X., {Pooley}, D., {et~al.} 2006, \apj, 638, 354, \dodoi{10.1086/498107}

\bibitem[{{Breeveld} \& {Siegel}(2016)}]{Breeveld+16_GCN}
{Breeveld}, A.~A., \& {Siegel}, M.~H. 2016, GRB Coordinates Network, 19839, 1

\bibitem[{{Brethauer} {et~al.}(2024){Brethauer}, {Kasen}, {Margutti}, \& {Chornock}}]{Brethauer+24}
{Brethauer}, D., {Kasen}, D., {Margutti}, R., \& {Chornock}, R. 2024, arXiv e-prints, arXiv:2408.02731, \dodoi{10.48550/arXiv.2408.02731}

\bibitem[{{Bucciantini} {et~al.}(2012){Bucciantini}, {Metzger}, {Thompson}, \& {Quataert}}]{Bucciantini+12}
{Bucciantini}, N., {Metzger}, B.~D., {Thompson}, T.~A., \& {Quataert}, E. 2012, \mnras, 419, 1537, \dodoi{10.1111/j.1365-2966.2011.19810.x}

\bibitem[{{Bulla}(2019)}]{Bulla19}
{Bulla}, M. 2019, \mnras, 489, 5037, \dodoi{10.1093/mnras/stz2495}

\bibitem[{{Bulla}(2023)}]{Bulla23}
---. 2023, \mnras, 520, 2558, \dodoi{10.1093/mnras/stad232}

\bibitem[{{Burbidge} {et~al.}(1957){Burbidge}, {Burbidge}, {Fowler}, \& {Hoyle}}]{Burbidge+57}
{Burbidge}, E.~M., {Burbidge}, G.~R., {Fowler}, W.~A., \& {Hoyle}, F. 1957, Reviews of Modern Physics, 29, 547, \dodoi{10.1103/RevModPhys.29.547}

\bibitem[{{Burns} {et~al.}(2023){Burns}, {Goldstein}, {Lesage}, {Dalessi}, \& {Fermi-GBM Team.}}]{Burns_GCN_230307a}
{Burns}, E., {Goldstein}, A., {Lesage}, S., {Dalessi}, S., \& {Fermi-GBM Team.} 2023, GRB Coordinates Network, 33414, 1

\bibitem[{{Burrows} {et~al.}(2005){Burrows}, {Hill}, {Nousek}, {Kennea}, {Wells}, {Osborne}, {Abbey}, {Beardmore}, {Mukerjee}, {Short}, {Chincarini}, {Campana}, {Citterio}, {Moretti}, {Pagani}, {Tagliaferri}, {Giommi}, {Capalbi}, {Tamburelli}, {Angelini}, {Cusumano}, {Br{\"a}uninger}, {Burkert}, \& {Hartner}}]{Swift-XRT}
{Burrows}, D.~N., {Hill}, J.~E., {Nousek}, J.~A., {et~al.} 2005, \ssr, 120, 165, \dodoi{10.1007/s11214-005-5097-2}

\bibitem[{{Burrows} {et~al.}(2023){Burrows}, {Gropp}, {Osborne}, {Page}, {D'Elia}, {Sbarufatti}, {D'Ai}, {Dichiara}, {Evans}, \& {Swift-XRT Team}}]{XRT_GCN_230307a}
{Burrows}, D.~N., {Gropp}, J.~D., {Osborne}, J.~P., {et~al.} 2023, GRB Coordinates Network, 33429, 1

\bibitem[{{Cameron}(1957)}]{Cameron57}
{Cameron}, A.~G.~W. 1957, \aj, 62, 9, \dodoi{10.1086/107435}

\bibitem[{{Casentini} {et~al.}(2023){Casentini}, {Tavani}, {Pittori}, {Lucarelli}, {Verrecchia}, {Argan}, {Cardillo}, {Evangelista}, {Foffano}, {Piano}, {Addis}, {Baroncelli}, {Bulgarelli}, {di Piano}, {Fioretti}, {Panebianco}, {Parmiggiani}, {Longo}, {Romani}, {Marisaldi}, {Pilia}, {Trois}, {Donnarumma}, {Menegoni}, {Ursi}, {Giuliani}, {Tempesta}, \& {Agile Team}}]{AGILE_GCN_230307a}
{Casentini}, C., {Tavani}, M., {Pittori}, C., {et~al.} 2023, GRB Coordinates Network, 33412, 1

\bibitem[{{Castro-Tirado} {et~al.}(2005){Castro-Tirado}, {de Ugarte Postigo}, {Gorosabel}, {Fathkullin}, {Sokolov}, {Bremer}, {M{\'a}rquez}, {Mar{\'\i}n}, {Guziy}, {Jel{\'\i}nek}, {Kub{\'a}nek}, {Hudec}, {Vitek}, {Mateo Sanguino}, {Eigenbrod}, {P{\'e}rez-Ram{\'\i}rez}, {Sota}, {Masegosa}, {Prada}, \& {Moles}}]{Castro-Tirado+05}
{Castro-Tirado}, A.~J., {de Ugarte Postigo}, A., {Gorosabel}, J., {et~al.} 2005, \aap, 439, L15, \dodoi{10.1051/0004-6361:200500147}

\bibitem[{{Cenko} {et~al.}(2013){Cenko}, {Perley}, {Cucchiara}, {Fong}, \& {Levan}}]{GCN15121}
{Cenko}, S.~B., {Perley}, D.~A., {Cucchiara}, A., {Fong}, W., \& {Levan}, A.~J. 2013, GRB Coordinates Network, 15121, 1

\bibitem[{{Cenko} {et~al.}(2005){Cenko}, {Soifer}, {Bian}, {Desai}, {Kulkarni}, {Schmidt}, {Dey}, \& {Jannuzi}}]{GCN3409}
{Cenko}, S.~B., {Soifer}, B.~T., {Bian}, C., {et~al.} 2005, GRB Coordinates Network, 3409, 1

\bibitem[{{Chase} {et~al.}(2022){Chase}, {O'Connor}, {Fryer}, {Troja}, {Korobkin}, {Wollaeger}, {Ristic}, {Fontes}, {Hungerford}, \& {Herring}}]{Chase+22}
{Chase}, E.~A., {O'Connor}, B., {Fryer}, C.~L., {et~al.} 2022, \apj, 927, 163, \dodoi{10.3847/1538-4357/ac3d25}

\bibitem[{{Chatzopoulos} {et~al.}(2012){Chatzopoulos}, {Wheeler}, \& {Vinko}}]{Chatzopoulos+12}
{Chatzopoulos}, E., {Wheeler}, J.~C., \& {Vinko}, J. 2012, \apj, 746, 121, \dodoi{10.1088/0004-637X/746/2/121}

\bibitem[{{Chornock} {et~al.}(2017){Chornock}, {Berger}, {Kasen}, {Cowperthwaite}, {Nicholl}, {Villar}, {Alexander}, {Blanchard}, {Eftekhari}, {Fong}, {Margutti}, {Williams}, {Annis}, {Brout}, {Brown}, {Chen}, {Drout}, {Farr}, {Foley}, {Frieman}, {Fryer}, {Herner}, {Holz}, {Kessler}, {Matheson}, {Metzger}, {Quataert}, {Rest}, {Sako}, {Scolnic}, {Smith}, \& {Soares-Santos}}]{Chornock+17}
{Chornock}, R., {Berger}, E., {Kasen}, D., {et~al.} 2017, \apjl, 848, L19, \dodoi{10.3847/2041-8213/aa905c}

\bibitem[{{Ciolfi} \& {Kalinani}(2020)}]{Ciolfi+20}
{Ciolfi}, R., \& {Kalinani}, J.~V. 2020, \apjl, 900, L35, \dodoi{10.3847/2041-8213/abb240}

\bibitem[{{Cobb}(2006)}]{Cobb+06_GCN}
{Cobb}, B.~E. 2006, GRB Coordinates Network, 5259, 1

\bibitem[{{Combi} \& {Siegel}(2023)}]{Combi&Siegel23}
{Combi}, L., \& {Siegel}, D.~M. 2023, \prl, 131, 231402, \dodoi{10.1103/PhysRevLett.131.231402}

\bibitem[{{Coulter} {et~al.}(2017){Coulter}, {Foley}, {Kilpatrick}, {Drout}, {Piro}, {Shappee}, {Siebert}, {Simon}, {Ulloa}, {Kasen}, {Madore}, {Murguia-Berthier}, {Pan}, {Prochaska}, {Ramirez-Ruiz}, {Rest}, \& {Rojas-Bravo}}]{Coulter+17}
{Coulter}, D.~A., {Foley}, R.~J., {Kilpatrick}, C.~D., {et~al.} 2017, Science, 358, 1556, \dodoi{10.1126/science.aap9811}

\bibitem[{{Coulter} {et~al.}(2024){Coulter}, {Kilpatrick}, {Jones}, {Foley}, {Filippenko}, {Zheng}, {Swift}, {Rahman}, {Stacey}, {Piro}, {Rojas-Bravo}, {Anais Vilchez}, {Mu{\~n}oz-Elgueta}, {Arcavi}, {Dimitriadis}, {Siebert}, {Bloom}, {Bustamante-Rosell}, {Clever}, {Davis}, {Kutcka}, {Macias}, {McGill}, {Qui{\~n}onez}, {Ramirez-Ruiz}, {Siellez}, {Tinyanont}, {Cenko}, {Drout}, {Hausen}, {Jacobson-Gal{\'a}n}, {Howell}, {Kasen}, {McCully}, {Rest}, {Taggart}, \& {Valenti}}]{Coulter+24}
{Coulter}, D.~A., {Kilpatrick}, C.~D., {Jones}, D.~O., {et~al.} 2024, arXiv e-prints, arXiv:2404.15441, \dodoi{10.48550/arXiv.2404.15441}

\bibitem[{{Covino} {et~al.}(2006){Covino}, {Malesani}, {Israel}, {D'Avanzo}, {Antonelli}, {Chincarini}, {Fugazza}, {Conciatore}, {Della Valle}, {Fiore}, {Guetta}, {Hurley}, {Lazzati}, {Stella}, {Tagliaferri}, {Vietri}, {Campana}, {Burrows}, {D'Elia}, {Filliatre}, {Gehrels}, {Goldoni}, {Melandri}, {Mereghetti}, {Mirabel}, {Moretti}, {Nousek}, {O'Brien}, {Pellizza}, {Perna}, {Piranomonte}, {Romano}, \& {Zerbi}}]{Covino+06}
{Covino}, S., {Malesani}, D., {Israel}, G.~L., {et~al.} 2006, \aap, 447, L5, \dodoi{10.1051/0004-6361:200500228}

\bibitem[{{Cowperthwaite} {et~al.}(2017){Cowperthwaite}, {Berger}, {Villar}, {Metzger}, {Nicholl}, {Chornock}, {Blanchard}, {Fong}, {Margutti}, {Soares-Santos}, {Alexander}, {Allam}, {Annis}, {Brout}, {Brown}, {Butler}, {Chen}, {Diehl}, {Doctor}, {Drout}, {Eftekhari}, {Farr}, {Finley}, {Foley}, {Frieman}, {Fryer}, {Garc{\'{\i}}a-Bellido}, {Gill}, {Guillochon}, {Herner}, {Holz}, {Kasen}, {Kessler}, {Marriner}, {Matheson}, {Neilsen}, {Quataert}, {Palmese}, {Rest}, {Sako}, {Scolnic}, {Smith}, {Tucker}, {Williams}, {Balbinot}, {Carlin}, {Cook}, {Durret}, {Li}, {Lopes}, {Louren{\c c}o}, {Marshall}, {Medina}, {Muir}, {Mu{\~n}oz}, {Sauseda}, {Schlegel}, {Secco}, {Vivas}, {Wester}, {Zenteno}, {Zhang}, {Abbott}, {Banerji}, {Bechtol}, {Benoit-L{\'e}vy}, {Bertin}, {Buckley-Geer}, {Burke}, {Capozzi}, {Carnero Rosell}, {Carrasco Kind}, {Castander}, {Crocce}, {Cunha}, {D'Andrea}, {da Costa}, {Davis}, {DePoy}, {Desai}, {Dietrich}, {Drlica-Wagner}, {Eifler}, {Evrard}, {Fernandez}, {Flaugher}, {Fosalba}, {Gaztanaga},
  {Gerdes}, {Giannantonio}, {Goldstein}, {Gruen}, {Gruendl}, {Gutierrez}, {Honscheid}, {Jain}, {James}, {Jeltema}, {Johnson}, {Johnson}, {Kent}, {Krause}, {Kron}, {Kuehn}, {Nuropatkin}, {Lahav}, {Lima}, {Lin}, {Maia}, {March}, {Martini}, {McMahon}, {Menanteau}, {Miller}, {Miquel}, {Mohr}, {Neilsen}, {Nichol}, {Ogando}, {Plazas}, {Roe}, {Romer}, {Roodman}, {Rykoff}, {Sanchez}, {Scarpine}, {Schindler}, {Schubnell}, {Sevilla-Noarbe}, {Smith}, {Smith}, {Sobreira}, {Suchyta}, {Swanson}, {Tarle}, {Thomas}, {Thomas}, {Troxel}, {Vikram}, {Walker}, {Wechsler}, {Weller}, {Yanny}, \& {Zuntz}}]{Cowperthwaite+17}
{Cowperthwaite}, P.~S., {Berger}, E., {Villar}, V.~A., {et~al.} 2017, \apjl, 848, L17, \dodoi{10.3847/2041-8213/aa8fc7}

\bibitem[{{Cucchiara} {et~al.}(2013){Cucchiara}, {Prochaska}, {Perley}, {Cenko}, {Werk}, {Cardwell}, {Turner}, {Cao}, {Bloom}, \& {Cobb}}]{Cucchiara+13}
{Cucchiara}, A., {Prochaska}, J.~X., {Perley}, D., {et~al.} 2013, \apj, 777, 94, \dodoi{10.1088/0004-637X/777/2/94}

\bibitem[{{Curtis} {et~al.}(2024){Curtis}, {Bosch}, {M{\"o}sta}, {Radice}, {Bernuzzi}, {Perego}, {Haas}, \& {Schnetter}}]{Curtis+24}
{Curtis}, S., {Bosch}, P., {M{\"o}sta}, P., {et~al.} 2024, \apjl, 961, L26, \dodoi{10.3847/2041-8213/ad0fe1}

\bibitem[{{Curtis} {et~al.}(2023){Curtis}, {M{\"o}sta}, {Wu}, {Radice}, {Roberts}, {Ricigliano}, \& {Perego}}]{Curtis+23}
{Curtis}, S., {M{\"o}sta}, P., {Wu}, Z., {et~al.} 2023, \mnras, 518, 5313, \dodoi{10.1093/mnras/stac3128}

\bibitem[{{Dafcikova} {et~al.}(2023){Dafcikova}, {Ripa}, {Pal}, {Werner}, {Ohno}, {Takahashi}, {Meszaros}, {Csak}, {Husarikova}, {Munz}, {Topinka}, {Kolar}, {Breuer}, {Hroch}, {Urbanec}, {Kasal}, {Povalac}, {Hudec}, {Kapus}, {Frajt}, {Laszlo}, {Koleda}, {Smelko}, {Hanak}, {Lipovsky}, {Galgoczi}, {Uchida}, {Poon}, {Matake}, {Uchida}, {Bozoki}, {Dalya}, {Enoto}, {Frei}, {Friss}, {Fukazawa}, {Hirose}, {Hisadomi}, {Ichinohe}, {Kapas}, {Kiss}, {Mizuno}, {Nakazawa}, {Odaka}, {Takatsy}, {Torigoe}, {Kogiso}, {Yoneyama}, {Moritaki}, {Kano}, \& {GRBAlpha Collaboration.}}]{GRBAlpha_GCN_230307a}
{Dafcikova}, M., {Ripa}, J., {Pal}, A., {et~al.} 2023, GRB Coordinates Network, 33418, 1

\bibitem[{{Dalessi} {et~al.}(2023){Dalessi}, {Roberts}, {Meegan}, \& {Fermi GBM Team}}]{230307a_t90_GBM}
{Dalessi}, S., {Roberts}, O.~J., {Meegan}, C., \& {Fermi GBM Team}. 2023, GRB Coordinates Network, 33411, 1

\bibitem[{{Darbha} \& {Kasen}(2020)}]{DarbhaKasen20}
{Darbha}, S., \& {Kasen}, D. 2020, \apj, 897, 150, \dodoi{10.3847/1538-4357/ab9a34}

\bibitem[{{de Ugarte Postigo} {et~al.}(2014){de Ugarte Postigo}, {Th{\"o}ne}, {Rowlinson}, {Garc{\'\i}a-Benito}, {Levan}, {Gorosabel}, {Goldoni}, {Schulze}, {Zafar}, {Wiersema}, {S{\'a}nchez-Ram{\'\i}rez}, {Melandri}, {D'Avanzo}, {Oates}, {D'Elia}, {De Pasquale}, {Kr{\"u}hler}, {van der Horst}, {Xu}, {Watson}, {Piranomonte}, {Vergani}, {Milvang-Jensen}, {Kaper}, {Malesani}, {Fynbo}, {Cano}, {Covino}, {Flores}, {Greiss}, {Hammer}, {Hartoog}, {Hellmich}, {Heuser}, {Hjorth}, {Jakobsson}, {Mottola}, {Sparre}, {Sollerman}, {Tagliaferri}, {Tanvir}, {Vestergaard}, \& {Wijers}}]{deUP+13}
{de Ugarte Postigo}, A., {Th{\"o}ne}, C.~C., {Rowlinson}, A., {et~al.} 2014, \aap, 563, A62, \dodoi{10.1051/0004-6361/201322985}

\bibitem[{{Della Valle} {et~al.}(2006){Della Valle}, {Chincarini}, {Panagia}, {Tagliaferri}, {Malesani}, {Testa}, {Fugazza}, {Campana}, {Covino}, {Mangano}, {Antonelli}, {D'Avanzo}, {Hurley}, {Mirabel}, {Pellizza}, {Piranomonte}, \& {Stella}}]{DellaValle+06}
{Della Valle}, M., {Chincarini}, G., {Panagia}, N., {et~al.} 2006, \nat, 444, 1050, \dodoi{10.1038/nature05374}

\bibitem[{{D{\'\i}az} {et~al.}(2017){D{\'\i}az}, {Macri}, {Garcia Lambas}, {Mendes de Oliveira}, {Nilo Castell{\'o}n}, {Ribeiro}, {S{\'a}nchez}, {Schoenell}, {Abramo}, {Akras}, {Alcaniz}, {Artola}, {Beroiz}, {Bonoli}, {Cabral}, {Camuccio}, {Castillo}, {Chavushyan}, {Coelho}, {Colazo}, {Costa-Duarte}, {Cuevas Larenas}, {DePoy}, {Dom{\'\i}nguez Romero}, {Dultzin}, {Fern{\'a}ndez}, {Garc{\'\i}a}, {Girardini}, {Gon{\c{c}}alves}, {Gon{\c{c}}alves}, {Gurovich}, {Jim{\'e}nez-Teja}, {Kanaan}, {Lares}, {Lopes de Oliveira}, {L{\'o}pez-Cruz}, {Marshall}, {Melia}, {Molino}, {Padilla}, {Pe{\~n}uela}, {Placco}, {Qui{\~n}ones}, {Ram{\'\i}rez Rivera}, {Renzi}, {Riguccini}, {R{\'\i}os-L{\'o}pez}, {Rodriguez}, {Sampedro}, {Schneiter}, {Sodr{\'e}}, {Starck}, {Torres-Flores}, {Tornatore}, \& {Zadro{\.z}ny}}]{Diaz+17}
{D{\'\i}az}, M.~C., {Macri}, L.~M., {Garcia Lambas}, D., {et~al.} 2017, \apjl, 848, L29, \dodoi{10.3847/2041-8213/aa9060}

\bibitem[{{Doctor} {et~al.}(2017){Doctor}, {Kessler}, {Chen}, {Farr}, {Finley}, {Foley}, {Goldstein}, {Holz}, {Kim}, {Morganson}, {Sako}, {Scolnic}, {Smith}, {Soares-Santos}, {Spinka}, {Abbott}, {Abdalla}, {Allam}, {Annis}, {Bechtol}, {Benoit-L{\'e}vy}, {Bertin}, {Brooks}, {Buckley-Geer}, {Burke}, {Carnero Rosell}, {Carrasco Kind}, {Carretero}, {Cunha}, {D'Andrea}, {da Costa}, {DePoy}, {Desai}, {Diehl}, {Drlica-Wagner}, {Eifler}, {Frieman}, {Garc{\'\i}a-Bellido}, {Gaztanaga}, {Gerdes}, {Gruendl}, {Gschwend}, {Gutierrez}, {James}, {Krause}, {Kuehn}, {Kuropatkin}, {Lahav}, {Li}, {Lima}, {Maia}, {March}, {Marshall}, {Menanteau}, {Miquel}, {Neilsen}, {Nichol}, {Nord}, {Plazas}, {Romer}, {Sanchez}, {Scarpine}, {Schubnell}, {Sevilla-Noarbe}, {Smith}, {Sobreira}, {Suchyta}, {Swanson}, {Tarle}, {Walker}, {Wester}, \& {DES Collaboration}}]{Doctor+17}
{Doctor}, Z., {Kessler}, R., {Chen}, H.~Y., {et~al.} 2017, \apj, 837, 57, \dodoi{10.3847/1538-4357/aa5d09}

\bibitem[{{Drout} {et~al.}(2017){Drout}, {Piro}, {Shappee}, {Kilpatrick}, {Simon}, {Contreras}, {Coulter}, {Foley}, {Siebert}, {Morrell}, {Boutsia}, {Di Mille}, {Holoien}, {Kasen}, {Kollmeier}, {Madore}, {Monson}, {Murguia-Berthier}, {Pan}, {Prochaska}, {Ramirez-Ruiz}, {Rest}, {Adams}, {Alatalo}, {Ba{\~n}ados}, {Baughman}, {Beers}, {Bernstein}, {Bitsakis}, {Campillay}, {Hansen}, {Higgs}, {Ji}, {Maravelias}, {Marshall}, {Moni Bidin}, {Prieto}, {Rasmussen}, {Rojas-Bravo}, {Strom}, {Ulloa}, {Vargas-Gonz{\'a}lez}, {Wan}, \& {Whitten}}]{Drout+17}
{Drout}, M.~R., {Piro}, A.~L., {Shappee}, B.~J., {et~al.} 2017, Science, 358, 1570, \dodoi{10.1126/science.aaq0049}

\bibitem[{{Evans} {et~al.}(2016){Evans}, {Goad}, {Osborne}, \& {Beardmore}}]{Evans+16_GCN}
{Evans}, P.~A., {Goad}, M.~R., {Osborne}, J.~P., \& {Beardmore}, A.~P. 2016, GRB Coordinates Network, 19837, 1

\bibitem[{{Evans} {et~al.}(2020){Evans}, {Gropp}, {Laha}, {Lien}, {Page}, \& {Neil Gehrels Swift Observatory Team}}]{200522a_Swift}
{Evans}, P.~A., {Gropp}, J.~D., {Laha}, S., {et~al.} 2020, GRB Coordinates Network, 27778, 1

\bibitem[{{Evans} {et~al.}(2007){Evans}, {Beardmore}, {Page}, {Tyler}, {Osborne}, {Goad}, {O'Brien}, {Vetere}, {Racusin}, {Morris}, {Burrows}, {Capalbi}, {Perri}, {Gehrels}, \& {Romano}}]{Evans+07}
{Evans}, P.~A., {Beardmore}, A.~P., {Page}, K.~L., {et~al.} 2007, \aap, 469, 379, \dodoi{10.1051/0004-6361:20077530}

\bibitem[{{Evans} {et~al.}(2009){Evans}, {Beardmore}, {Page}, {Osborne}, {O'Brien}, {Willingale}, {Starling}, {Burrows}, {Godet}, {Vetere}, {Racusin}, {Goad}, {Wiersema}, {Angelini}, {Capalbi}, {Chincarini}, {Gehrels}, {Kennea}, {Margutti}, {Morris}, {Mountford}, {Pagani}, {Perri}, {Romano}, \& {Tanvir}}]{Evans+09}
---. 2009, \mnras, 397, 1177, \dodoi{10.1111/j.1365-2966.2009.14913.x}

\bibitem[{{Evans} {et~al.}(2017){Evans}, {Cenko}, {Kennea}, {Emery}, {Kuin}, {Korobkin}, {Wollaeger}, {Fryer}, {Madsen}, {Harrison}, {Xu}, {Nakar}, {Hotokezaka}, {Lien}, {Campana}, {Oates}, {Troja}, {Breeveld}, {Marshall}, {Barthelmy}, {Beardmore}, {Burrows}, {Cusumano}, {D'A{\`\i}}, {D'Avanzo}, {D'Elia}, {de Pasquale}, {Even}, {Fontes}, {Forster}, {Garcia}, {Giommi}, {Grefenstette}, {Gronwall}, {Hartmann}, {Heida}, {Hungerford}, {Kasliwal}, {Krimm}, {Levan}, {Malesani}, {Melandri}, {Miyasaka}, {Nousek}, {O'Brien}, {Osborne}, {Pagani}, {Page}, {Palmer}, {Perri}, {Pike}, {Racusin}, {Rosswog}, {Siegel}, {Sakamoto}, {Sbarufatti}, {Tagliaferri}, {Tanvir}, \& {Tohuvavohu}}]{Evans+17}
{Evans}, P.~A., {Cenko}, S.~B., {Kennea}, J.~A., {et~al.} 2017, Science, 358, 1565, \dodoi{10.1126/science.aap9580}

\bibitem[{{Fermi GBM Team}(2023)}]{230307a_GBM}
{Fermi GBM Team}. 2023, GRB Coordinates Network, 33405, 1

\bibitem[{{Fern{\'a}ndez} {et~al.}(2015){Fern{\'a}ndez}, {Kasen}, {Metzger}, \& {Quataert}}]{Fernandez+15}
{Fern{\'a}ndez}, R., {Kasen}, D., {Metzger}, B.~D., \& {Quataert}, E. 2015, \mnras, 446, 750, \dodoi{10.1093/mnras/stu2112}

\bibitem[{{Fern{\'a}ndez} \& {Metzger}(2016)}]{FernandezMetzger16}
{Fern{\'a}ndez}, R., \& {Metzger}, B.~D. 2016, Annual Review of Nuclear and Particle Science, 66, 23, \dodoi{10.1146/annurev-nucl-102115-044819}

\bibitem[{{Foley} {et~al.}(2013){Foley}, {Chornock}, {Fong}, {Berger}, \& {Jha}}]{foley+13_gcn}
{Foley}, R.~J., {Chornock}, R., {Fong}, W., {Berger}, E., \& {Jha}, S. 2013, GRB Coordinates Network, 14745, 1

\bibitem[{{Fong} {et~al.}(2016){Fong}, {Alexander}, \& {Laskar}}]{Fong+16_GCN}
{Fong}, W., {Alexander}, K.~D., \& {Laskar}, T. 2016, GRB Coordinates Network, 19854, 1

\bibitem[{{Fong} {et~al.}(2015){Fong}, {Berger}, {Margutti}, \& {Zauderer}}]{fong+15}
{Fong}, W., {Berger}, E., {Margutti}, R., \& {Zauderer}, B.~A. 2015, ApJ, 815, 102, \dodoi{10.1088/0004-637X/815/2/102}

\bibitem[{{Fong} {et~al.}(2020){Fong}, {Paterson}, \& {Berger}}]{Fong+20_GCN}
{Fong}, W., {Paterson}, K., \& {Berger}, E. 2020, GRB Coordinates Network, 27779, 1

\bibitem[{{Fong} {et~al.}(2014){Fong}, {Berger}, {Metzger}, {Margutti}, {Chornock}, {Migliori}, {Foley}, {Zauderer}, {Lunnan}, {Laskar}, {Desch}, {Meech}, {Sonnett}, {Dickey}, {Hedlund}, \& {Harding}}]{Fong+14}
{Fong}, W., {Berger}, E., {Metzger}, B.~D., {et~al.} 2014, \apj, 780, 118, \dodoi{10.1088/0004-637X/780/2/118}

\bibitem[{{Fong} {et~al.}(2019){Fong}, {Blanchard}, {Alexander}, {Strader}, {Margutti}, {Hajela}, {Villar}, {Wu}, {Ye}, {Berger}, {Chornock}, {Coppejans}, {Cowperthwaite}, {Eftekhari}, {Giannios}, {Guidorzi}, {Kathirgamaraju}, {Laskar}, {Macfadyen}, {Metzger}, {Nicholl}, {Paterson}, {Terreran}, {Sand}, {Sironi}, {Williams}, {Xie}, \& {Zrake}}]{Fong+19}
{Fong}, W., {Blanchard}, P.~K., {Alexander}, K.~D., {et~al.} 2019, \apjl, 883, L1, \dodoi{10.3847/2041-8213/ab3d9e}

\bibitem[{{Fong} {et~al.}(2021){Fong}, {Laskar}, {Rastinejad}, {Escorial}, {Schroeder}, {Barnes}, {Kilpatrick}, {Paterson}, {Berger}, {Metzger}, {Dong}, {Nugent}, {Strausbaugh}, {Blanchard}, {Goyal}, {Cucchiara}, {Terreran}, {Alexander}, {Eftekhari}, {Fryer}, {Margalit}, {Margutti}, \& {Nicholl}}]{Fong+21}
{Fong}, W., {Laskar}, T., {Rastinejad}, J., {et~al.} 2021, \apj, 906, 127, \dodoi{10.3847/1538-4357/abc74a}

\bibitem[{{Fong} {et~al.}(2022){Fong}, {Nugent}, {Dong}, {Berger}, {Paterson}, {Chornock}, {Levan}, {Blanchard}, {Alexander}, {Andrews}, {Cobb}, {Cucchiara}, {Fox}, {Fryer}, {Gordon}, {Kilpatrick}, {Lunnan}, {Margutti}, {Miller}, {Milne}, {Nicholl}, {Perley}, {Rastinejad}, {Escorial}, {Schroeder}, {Smith}, {Tanvir}, \& {Terreran}}]{Fong+22}
{Fong}, W.-f., {Nugent}, A.~E., {Dong}, Y., {et~al.} 2022, \apj, 940, 56, \dodoi{10.3847/1538-4357/ac91d0}

\bibitem[{{Foreman-Mackey} {et~al.}(2013){Foreman-Mackey}, {Hogg}, {Lang}, \& {Goodman}}]{emcee13}
{Foreman-Mackey}, D., {Hogg}, D.~W., {Lang}, D., \& {Goodman}, J. 2013, \pasp, 125, 306, \dodoi{10.1086/670067}

\bibitem[{{Foucart} {et~al.}(2014){Foucart}, {Deaton}, {Duez}, {O'Connor}, {Ott}, {Haas}, {Kidder}, {Pfeiffer}, {Scheel}, \& {Szilagyi}}]{Foucart+14}
{Foucart}, F., {Deaton}, M.~B., {Duez}, M.~D., {et~al.} 2014, \prd, 90, 024026, \dodoi{10.1103/PhysRevD.90.024026}

\bibitem[{{Fox} {et~al.}(2005){Fox}, {Frail}, {Price}, {Kulkarni}, {Berger}, {Piran}, {Soderberg}, {Cenko}, {Cameron}, {Gal-Yam}, {Kasliwal}, {Moon}, {Harrison}, {Nakar}, {Schmidt}, {Penprase}, {Chevalier}, {Kumar}, {Roth}, {Watson}, {Lee}, {Shectman}, {Phillips}, {Roth}, {McCarthy}, {Rauch}, {Cowie}, {Peterson}, {Rich}, {Kawai}, {Aoki}, {Kosugi}, {Totani}, {Park}, {MacFadyen}, \& {Hurley}}]{Fox+05}
{Fox}, D.~B., {Frail}, D.~A., {Price}, P.~A., {et~al.} 2005, \nat, 437, 845, \dodoi{10.1038/nature04189}

\bibitem[{{Frebel}(2018)}]{Frebel+18}
{Frebel}, A. 2018, Annual Review of Nuclear and Particle Science, 68, 237, \dodoi{10.1146/annurev-nucl-101917-021141}

\bibitem[{{Frostig} {et~al.}(2022){Frostig}, {Biscoveanu}, {Mo}, {Karambelkar}, {Dal Canton}, {Chen}, {Kasliwal}, {Katsavounidis}, {Lourie}, {Simcoe}, \& {Vitale}}]{Frostig+22}
{Frostig}, D., {Biscoveanu}, S., {Mo}, G., {et~al.} 2022, \apj, 926, 152, \dodoi{10.3847/1538-4357/ac4508}

\bibitem[{{Fugazza} {et~al.}(2006){Fugazza}, {Malesani}, {Romano}, {Tagliaferri}, {Covino}, {Chincarini}, {Della Valle}, {Fiore}, \& {Stella}}]{Fugazza+06_GCN}
{Fugazza}, D., {Malesani}, D., {Romano}, P., {et~al.} 2006, GRB Coordinates Network, 5276, 1

\bibitem[{{Fynbo} {et~al.}(2006){Fynbo}, {Watson}, {Th{\"o}ne}, {Sollerman}, {Bloom}, {Davis}, {Hjorth}, {Jakobsson}, {J{\o}rgensen}, {Graham}, {Fruchter}, {Bersier}, {Kewley}, {Cassan}, {Castro Cer{\'o}n}, {Foley}, {Gorosabel}, {Hinse}, {Horne}, {Jensen}, {Klose}, {Kocevski}, {Marquette}, {Perley}, {Ramirez-Ruiz}, {Stritzinger}, {Vreeswijk}, {Wijers}, {Woller}, {Xu}, \& {Zub}}]{Fynbo+06}
{Fynbo}, J. P.~U., {Watson}, D., {Th{\"o}ne}, C.~C., {et~al.} 2006, \nat, 444, 1047, \dodoi{10.1038/nature05375}

\bibitem[{{Gal-Yam} {et~al.}(2006){Gal-Yam}, {Fox}, {Price}, {Ofek}, {Davis}, {Leonard}, {Soderberg}, {Schmidt}, {Lewis}, {Peterson}, {Kulkarni}, {Berger}, {Cenko}, {Sari}, {Sharon}, {Frail}, {Moon}, {Brown}, {Cucchiara}, {Harrison}, {Piran}, {Persson}, {McCarthy}, {Penprase}, {Chevalier}, \& {MacFadyen}}]{GalYam+06}
{Gal-Yam}, A., {Fox}, D.~B., {Price}, P.~A., {et~al.} 2006, \nat, 444, 1053, \dodoi{10.1038/nature05373}

\bibitem[{{Gao} {et~al.}(2017){Gao}, {Zhang}, {L{\"u}}, \& {Li}}]{Gao+17}
{Gao}, H., {Zhang}, B., {L{\"u}}, H.-J., \& {Li}, Y. 2017, \apj, 837, 50, \dodoi{10.3847/1538-4357/aa5be3}

\bibitem[{{Gehrels} {et~al.}(2006){Gehrels}, {Norris}, {Barthelmy}, {Granot}, {Kaneko}, {Kouveliotou}, {Markwardt}, {M{\'e}sz{\'a}ros}, {Nakar}, {Nousek}, {O'Brien}, {Page}, {Palmer}, {Parsons}, {Roming}, {Sakamoto}, {Sarazin}, {Schady}, {Stamatikos}, \& {Woosley}}]{Gehrels+06}
{Gehrels}, N., {Norris}, J.~P., {Barthelmy}, S.~D., {et~al.} 2006, \nat, 444, 1044, \dodoi{10.1038/nature05376}

\bibitem[{{Gillanders} \& {Smartt}(2024)}]{Gillanders+24}
{Gillanders}, J.~H., \& {Smartt}, S.~J. 2024, arXiv e-prints, arXiv:2408.11093.
\newblock \doarXiv{2408.11093}

\bibitem[{{Gillanders} {et~al.}(2022){Gillanders}, {Smartt}, {Sim}, {Bauswein}, \& {Goriely}}]{Gillanders+22}
{Gillanders}, J.~H., {Smartt}, S.~J., {Sim}, S.~A., {Bauswein}, A., \& {Goriely}, S. 2022, \mnras, 515, 631, \dodoi{10.1093/mnras/stac1258}

\bibitem[{{Gillanders} {et~al.}(2023){Gillanders}, {Troja}, {Fryer}, {Ristic}, {O'Connor}, {Fontes}, {Yang}, {Domoto}, {Rahmouni}, {Tanaka}, {Fox}, \& {Dichiara}}]{Gillanders+23}
{Gillanders}, J.~H., {Troja}, E., {Fryer}, C.~L., {et~al.} 2023, arXiv e-prints, arXiv:2308.00633, \dodoi{10.48550/arXiv.2308.00633}

\bibitem[{Goldstein {et~al.}(2017)Goldstein, Veres, Burns, Briggs, Hamburg, Kocevski, Wilson-Hodge, Preece, Poolakkil, Roberts, \& et~al.}]{Goldstein+2017}
Goldstein, A., Veres, P., Burns, E., {et~al.} 2017, The Astrophysical Journal, 848, L14, \dodoi{10.3847/2041-8213/aa8f41}

\bibitem[{{Golenetskii} {et~al.}(2013){Golenetskii}, {Aptekar}, {Frederiks}, {Mazets}, {Pal'Shin}, {Oleynik}, {Ulanov}, {Svinkin}, \& {Cline}}]{konus_130603b}
{Golenetskii}, S., {Aptekar}, R., {Frederiks}, D., {et~al.} 2013, GRB Coordinates Network, 14771, 1

\bibitem[{{Gompertz} {et~al.}(2023){Gompertz}, {Nicholl}, {Smith}, {Harisankar}, {Pratten}, {Schmidt}, \& {Smith}}]{Gompertz+23_NSBH}
{Gompertz}, B.~P., {Nicholl}, M., {Smith}, J.~C., {et~al.} 2023, \mnras, 526, 4585, \dodoi{10.1093/mnras/stad2990}

\bibitem[{{Gompertz} {et~al.}(2018){Gompertz}, {Levan}, {Tanvir}, {Hjorth}, {Covino}, {Evans}, {Fruchter}, {Gonz{\'a}lez-Fern{\'a}ndez}, {Jin}, {Lyman}, {Oates}, {O'Brien}, \& {Wiersema}}]{Gompertz+18}
{Gompertz}, B.~P., {Levan}, A.~J., {Tanvir}, N.~R., {et~al.} 2018, \apj, 860, 62, \dodoi{10.3847/1538-4357/aac206}

\bibitem[{{Gompertz} {et~al.}(2022){Gompertz}, {Ravasio}, {Nicholl}, {Levan}, {Metzger}, {Oates}, {Lamb}, {Fong}, {Malesani}, {Rastinejad}, {Tanvir}, {Evans}, {Jonker}, {Page}, \& {Pe'er}}]{Gompertz+22}
{Gompertz}, B.~P., {Ravasio}, M.~E., {Nicholl}, M., {et~al.} 2022, arXiv e-prints, arXiv:2205.05008.
\newblock \doarXiv{2205.05008}

\bibitem[{{Gottlieb} {et~al.}(2018){Gottlieb}, {Nakar}, {Piran}, \& {Hotokezaka}}]{Gottlieb+18}
{Gottlieb}, O., {Nakar}, E., {Piran}, T., \& {Hotokezaka}, K. 2018, \mnras, 479, 588, \dodoi{10.1093/mnras/sty1462}

\bibitem[{{Gottlieb} {et~al.}(2023){Gottlieb}, {Metzger}, {Quataert}, {Issa}, {Martineau}, {Foucart}, {Duez}, {Kidder}, {Pfeiffer}, \& {Scheel}}]{Gottlieb+23}
{Gottlieb}, O., {Metzger}, B.~D., {Quataert}, E., {et~al.} 2023, \apjl, 958, L33, \dodoi{10.3847/2041-8213/ad096e}

\bibitem[{{Granot} \& {Sari}(2002)}]{GranotSari02}
{Granot}, J., \& {Sari}, R. 2002, \apj, 568, 820, \dodoi{10.1086/338966}

\bibitem[{{Guillochon} {et~al.}(2017){Guillochon}, {Parrent}, {Kelley}, \& {Margutti}}]{Guillochon+17}
{Guillochon}, J., {Parrent}, J., {Kelley}, L.~Z., \& {Margutti}, R. 2017, \apj, 835, 64, \dodoi{10.3847/1538-4357/835/1/64}

\bibitem[{{Gupta} {et~al.}(2021){Gupta}, {Pandey}, {Ror}, {Kumar}, {Aryan}, {Dimple}, {Ghosh}, {Kumar}, \& {Misra}}]{GCN31299}
{Gupta}, R., {Pandey}, S.~B., {Ror}, A., {et~al.} 2021, GRB Coordinates Network, 31299, 1

\bibitem[{{Halevi} \& {M{\"o}sta}(2018)}]{HaleviMosta18}
{Halevi}, G., \& {M{\"o}sta}, P. 2018, \mnras, 477, 2366, \dodoi{10.1093/mnras/sty797}

\bibitem[{{Hamidani} {et~al.}(2024{\natexlab{a}}){Hamidani}, {Kimura}, {Tanaka}, \& {Ioka}}]{Hamidani+24}
{Hamidani}, H., {Kimura}, S.~S., {Tanaka}, M., \& {Ioka}, K. 2024{\natexlab{a}}, \apj, 963, 137, \dodoi{10.3847/1538-4357/ad20d0}

\bibitem[{{Hamidani} {et~al.}(2024{\natexlab{b}}){Hamidani}, {Kimura}, {Tanaka}, \& {Ioka}}]{Hamidani+24_model}
---. 2024{\natexlab{b}}, \apj, 963, 137, \dodoi{10.3847/1538-4357/ad20d0}

\bibitem[{{Hansen} {et~al.}(2017){Hansen}, {Simon}, {Marshall}, {Li}, {Carollo}, {DePoy}, {Nagasawa}, {Bernstein}, {Drlica-Wagner}, {Abdalla}, {Allam}, {Annis}, {Bechtol}, {Benoit-L{\'e}vy}, {Brooks}, {Buckley-Geer}, {Carnero Rosell}, {Carrasco Kind}, {Carretero}, {Cunha}, {da Costa}, {Desai}, {Eifler}, {Fausti Neto}, {Flaugher}, {Frieman}, {Garc{\'\i}a-Bellido}, {Gaztanaga}, {Gerdes}, {Gruen}, {Gruendl}, {Gschwend}, {Gutierrez}, {James}, {Krause}, {Kuehn}, {Kuropatkin}, {Lahav}, {Miquel}, {Plazas}, {Romer}, {Sanchez}, {Santiago}, {Scarpine}, {Smith}, {Soares-Santos}, {Sobreira}, {Suchyta}, {Swanson}, {Tarle}, {Walker}, \& {DES Collaboration}}]{Hansen+17}
{Hansen}, T.~T., {Simon}, J.~D., {Marshall}, J.~L., {et~al.} 2017, \apj, 838, 44, \dodoi{10.3847/1538-4357/aa634a}

\bibitem[{{Hjorth} {et~al.}(2005){Hjorth}, {Sollerman}, {Gorosabel}, {Granot}, {Klose}, {Kouveliotou}, {Melinder}, {Ramirez-Ruiz}, {Starling}, {Thomsen}, {Andersen}, {Fynbo}, {Jensen}, {Vreeswijk}, {Castro Cer{\'o}n}, {Jakobsson}, {Levan}, {Pedersen}, {Rhoads}, {Tanvir}, {Watson}, \& {Wijers}}]{Hjorth+05}
{Hjorth}, J., {Sollerman}, J., {Gorosabel}, J., {et~al.} 2005, \apjl, 630, L117, \dodoi{10.1086/491733}

\bibitem[{{Hotokezaka} {et~al.}(2018){Hotokezaka}, {Beniamini}, \& {Piran}}]{Hotokezaka+18}
{Hotokezaka}, K., {Beniamini}, P., \& {Piran}, T. 2018, International Journal of Modern Physics D, 27, 1842005, \dodoi{10.1142/S0218271818420051}

\bibitem[{{Hotokezaka} {et~al.}(2013){Hotokezaka}, {Kyutoku}, {Tanaka}, {Kiuchi}, {Sekiguchi}, {Shibata}, \& {Wanajo}}]{Hotokezaka+13}
{Hotokezaka}, K., {Kyutoku}, K., {Tanaka}, M., {et~al.} 2013, \apjl, 778, L16, \dodoi{10.1088/2041-8205/778/1/L16}

\bibitem[{{Hotokezaka} {et~al.}(2021){Hotokezaka}, {Tanaka}, {Kato}, \& {Gaigalas}}]{Hotokezaka+21}
{Hotokezaka}, K., {Tanaka}, M., {Kato}, D., \& {Gaigalas}, G. 2021, \mnras, 506, 5863, \dodoi{10.1093/mnras/stab1975}

\bibitem[{{Hotokezaka} {et~al.}(2023){Hotokezaka}, {Tanaka}, {Kato}, \& {Gaigalas}}]{Hotokezaka+23}
---. 2023, arXiv e-prints, arXiv:2307.00988, \dodoi{10.48550/arXiv.2307.00988}

\bibitem[{{Hu} {et~al.}(2017){Hu}, {Wu}, {Andreoni}, {Ashley}, {Cooke}, {Cui}, {Du}, {Dai}, {Gu}, {Hu}, {Lu}, {Li}, {Li}, {Liang}, {Liu}, {Ma}, {Shang}, {Sun}, {Suntzeff}, {Tao}, {Udden}, {Wang}, {Wang}, {Wen}, {Xiao}, {Su}, {Yang}, {Yang}, {Yuan}, {Zhou}, {Zhang}, {Zhou}, \& {Zhu}}]{Hu+17}
{Hu}, L., {Wu}, X., {Andreoni}, I., {et~al.} 2017, Science Bulletin, 62, 1433, \dodoi{10.1016/j.scib.2017.10.006}

\bibitem[{{Ioka} \& {Nakamura}(2019)}]{Ioka+19}
{Ioka}, K., \& {Nakamura}, T. 2019, \mnras, 487, 4884, \dodoi{10.1093/mnras/stz1650}

\bibitem[{{Ito} {et~al.}(2021){Ito}, {Hosokawa}, {Murata}, {Imai}, {Takamatsu}, {Takaku}, {Niwano}, {Noto}, {Sato}, {Yamaguchi}, {Yatsu}, {Kawai}, \& {MITSuME Collaboration}}]{GCN31217}
{Ito}, N., {Hosokawa}, R., {Murata}, K.~L., {et~al.} 2021, GRB Coordinates Network, 31217, 1

\bibitem[{{Ji} {et~al.}(2016){Ji}, {Frebel}, {Chiti}, \& {Simon}}]{Ji+16}
{Ji}, A.~P., {Frebel}, A., {Chiti}, A., \& {Simon}, J.~D. 2016, \nat, 531, 610, \dodoi{10.1038/nature17425}

\bibitem[{{Jin} {et~al.}(2020){Jin}, {Covino}, {Liao}, {Li}, {D'Avanzo}, {Fan}, \& {Wei}}]{Jin+20}
{Jin}, Z.-P., {Covino}, S., {Liao}, N.-H., {et~al.} 2020, Nature Astronomy, 4, 77, \dodoi{10.1038/s41550-019-0892-y}

\bibitem[{{Jin} {et~al.}(2016){Jin}, {Hotokezaka}, {Li}, {Tanaka}, {D'Avanzo}, {Fan}, {Covino}, {Wei}, \& {Piran}}]{jin+16}
{Jin}, Z.-P., {Hotokezaka}, K., {Li}, X., {et~al.} 2016, Nature Communications, 7, 12898, \dodoi{10.1038/ncomms12898}

\bibitem[{{Just} {et~al.}(2015){Just}, {Bauswein}, {Ardevol Pulpillo}, {Goriely}, \& {Janka}}]{just+15}
{Just}, O., {Bauswein}, A., {Ardevol Pulpillo}, R., {Goriely}, S., \& {Janka}, H.~T. 2015, \mnras, 448, 541, \dodoi{10.1093/mnras/stv009}

\bibitem[{{Kasen} {et~al.}(2017){Kasen}, {Metzger}, {Barnes}, {Quataert}, \& {Ramirez-Ruiz}}]{kasen+17}
{Kasen}, D., {Metzger}, B., {Barnes}, J., {Quataert}, E., \& {Ramirez-Ruiz}, E. 2017, Nature, 551, 80, \dodoi{10.1038/nature24453}

\bibitem[{{Kasliwal} {et~al.}(2017){Kasliwal}, {Nakar}, {Singer}, {Kaplan}, {Cook}, {Van Sistine}, {Lau}, {Fremling}, {Gottlieb}, {Jencson}, {Adams}, {Feindt}, {Hotokezaka}, {Ghosh}, {Perley}, {Yu}, {Piran}, {Allison}, {Anupama}, {Balasubramanian}, {Bannister}, {Bally}, {Barnes}, {Barway}, {Bellm}, {Bhalerao}, {Bhattacharya}, {Blagorodnova}, {Bloom}, {Brady}, {Cannella}, {Chatterjee}, {Cenko}, {Cobb}, {Copperwheat}, {Corsi}, {De}, {Dobie}, {Emery}, {Evans}, {Fox}, {Frail}, {Frohmaier}, {Goobar}, {Hallinan}, {Harrison}, {Helou}, {Hinderer}, {Ho}, {Horesh}, {Ip}, {Itoh}, {Kasen}, {Kim}, {Kuin}, {Kupfer}, {Lynch}, {Madsen}, {Mazzali}, {Miller}, {Mooley}, {Murphy}, {Ngeow}, {Nichols}, {Nissanke}, {Nugent}, {Ofek}, {Qi}, {Quimby}, {Rosswog}, {Rusu}, {Sadler}, {Schmidt}, {Sollerman}, {Steele}, {Williamson}, {Xu}, {Yan}, {Yatsu}, {Zhang}, \& {Zhao}}]{Kasliwal+17}
{Kasliwal}, M.~M., {Nakar}, E., {Singer}, L.~P., {et~al.} 2017, Science, 358, 1559, \dodoi{10.1126/science.aap9455}

\bibitem[{{Kasliwal} {et~al.}(2022){Kasliwal}, {Kasen}, {Lau}, {Perley}, {Rosswog}, {Ofek}, {Hotokezaka}, {Chary}, {Sollerman}, {Goobar}, \& {Kaplan}}]{Kasliwal+22}
{Kasliwal}, M.~M., {Kasen}, D., {Lau}, R.~M., {et~al.} 2022, \mnras, 510, L7, \dodoi{10.1093/mnrasl/slz007}

\bibitem[{{Kawaguchi} {et~al.}(2016){Kawaguchi}, {Kyutoku}, {Shibata}, \& {Tanaka}}]{kawaguchi+16}
{Kawaguchi}, K., {Kyutoku}, K., {Shibata}, M., \& {Tanaka}, M. 2016, ApJ, 825, 52, \dodoi{10.3847/0004-637X/825/1/52}

\bibitem[{{Kawaguchi} {et~al.}(2020{\natexlab{a}}){Kawaguchi}, {Shibata}, \& {Tanaka}}]{Kawaguchi+20b}
{Kawaguchi}, K., {Shibata}, M., \& {Tanaka}, M. 2020{\natexlab{a}}, \apj, 889, 171, \dodoi{10.3847/1538-4357/ab61f6}

\bibitem[{{Kawaguchi} {et~al.}(2020{\natexlab{b}}){Kawaguchi}, {Shibata}, \& {Tanaka}}]{Kawaguchi+20}
---. 2020{\natexlab{b}}, \apj, 893, 153, \dodoi{10.3847/1538-4357/ab8309}

\bibitem[{{Kilpatrick} {et~al.}(2017){Kilpatrick}, {Foley}, {Kasen}, {Murguia-Berthier}, {Ramirez-Ruiz}, {Coulter}, {Drout}, {Piro}, {Shappee}, {Boutsia}, {Contreras}, {Di Mille}, {Madore}, {Morrell}, {Pan}, {Prochaska}, {Rest}, {Rojas-Bravo}, {Siebert}, {Simon}, \& {Ulloa}}]{kilpatrick+17}
{Kilpatrick}, C.~D., {Foley}, R.~J., {Kasen}, D., {et~al.} 2017, Science, 358, 1583, \dodoi{10.1126/science.aaq0073}

\bibitem[{{Kilpatrick} {et~al.}(2021){Kilpatrick}, {Coulter}, {Arcavi}, {Brink}, {Dimitriadis}, {Filippenko}, {Foley}, {Howell}, {Jones}, {Kasen}, {Makler}, {Piro}, {Rojas-Bravo}, {Sand}, {Swift}, {Tucker}, {Zheng}, {Allam}, {Annis}, {Antilen}, {Bachmann}, {Bloom}, {Bom}, {Bostroem}, {Brout}, {Burke}, {Butler}, {Butner}, {Campillay}, {Clever}, {Conselice}, {Cooke}, {Dage}, {de Carvalho}, {de Jaeger}, {Desai}, {Garcia}, {Garcia-Bellido}, {Gill}, {Girish}, {Hallakoun}, {Herner}, {Hiramatsu}, {Holz}, {Huber}, {Kawash}, {McCully}, {Medallon}, {Metzger}, {Modak}, {Morgan}, {Mu{\~n}oz}, {Mu{\~n}oz-Elgueta}, {Murakami}, {Felipe Olivares}, {Palmese}, {Patra}, {Pereira}, {Pessi}, {Pineda-Garcia}, {Quirola-V{\'a}squez}, {Ramirez-Ruiz}, {Rembold}, {Rest}, {Rodr{\'\i}guez}, {Santana-Silva}, {Sherman}, {Siebert}, {Smith}, {Smith}, {Soares-Santos}, {Stacey}, {Stahl}, {Strader}, {Strasburger}, {Sunseri}, {Tinyanont}, {Tucker}, {Ulloa}, {Valenti}, {Vasylyev}, {Wiesner}, \& {Zhang}}]{Kilpatrick+21}
{Kilpatrick}, C.~D., {Coulter}, D.~A., {Arcavi}, I., {et~al.} 2021, \apj, 923, 258, \dodoi{10.3847/1538-4357/ac23c6}

\bibitem[{{Korobkin} {et~al.}(2012){Korobkin}, {Rosswog}, {Arcones}, \& {Winteler}}]{Korobkin+12}
{Korobkin}, O., {Rosswog}, S., {Arcones}, A., \& {Winteler}, C. 2012, \mnras, 426, 1940, \dodoi{10.1111/j.1365-2966.2012.21859.x}

\bibitem[{{Kozyrev} {et~al.}(2023){Kozyrev}, {Golovin}, {Litvak}, {Mitrofanov}, {Sanin}, {Hend/Mars Odyssey Team}, {Svinkin}, {Lysenko}, {Ridnaia}, {Ipn}, {Goldstein}, {Briggs}, {Wilson-Hodge}, {Burns}, {Fermi Gbm Team}, {Bozzo}, {Ferrigno}, {INTEGRAL SPI-ACS Grb Team}, {Barthelmy}, {Cummings}, {Krimm}, {Palmer}, {Tohuvavohu}, {Swift-Bat Team}, {Boynton}, {Fellows}, {Harshman}, {Enos}, {Starr}, {Gardner}, \& {Grs-Odyssey Grb Team}}]{IPN_GCN_230307a}
{Kozyrev}, A.~S., {Golovin}, D.~V., {Litvak}, M.~L., {et~al.} 2023, GRB Coordinates Network, 33413, 1

\bibitem[{{Kumar} {et~al.}(2021){Kumar}, {Bhalerao}, {Gupta}, {Norbu}, {Anupama}, {Sahu}, {Barway}, {Dutta}, {Kumar}, {Dimple}, {Pandey}, {Mishra}, {HCT}, \& {GIT Team}}]{GCN31227}
{Kumar}, H., {Bhalerao}, V., {Gupta}, R., {et~al.} 2021, GRB Coordinates Network, 31227, 1

\bibitem[{{Kunert} {et~al.}(2024){Kunert}, {Antier}, {Nedora}, {Bulla}, {Pang}, {Anand}, {Coughlin}, {Tews}, {Barnes}, {Hussenot-Desenonges}, {Healy}, {Jegou du Laz}, {Pilloix}, {Kiendrebeogo}, \& {Dietrich}}]{Kunert+23}
{Kunert}, N., {Antier}, S., {Nedora}, V., {et~al.} 2024, \mnras, 527, 3900, \dodoi{10.1093/mnras/stad3463}

\bibitem[{{Kyutoku} {et~al.}(2015){Kyutoku}, {Ioka}, {Okawa}, {Shibata}, \& {Taniguchi}}]{Kyutoku+15}
{Kyutoku}, K., {Ioka}, K., {Okawa}, H., {Shibata}, M., \& {Taniguchi}, K. 2015, \prd, 92, 044028, \dodoi{10.1103/PhysRevD.92.044028}

\bibitem[{{Kyutoku} {et~al.}(2018){Kyutoku}, {Kiuchi}, {Sekiguchi}, {Shibata}, \& {Taniguchi}}]{Kyutoku+18}
{Kyutoku}, K., {Kiuchi}, K., {Sekiguchi}, Y., {Shibata}, M., \& {Taniguchi}, K. 2018, \prd, 97, 023009, \dodoi{10.1103/PhysRevD.97.023009}

\bibitem[{{Lamb} {et~al.}(2019){Lamb}, {Tanvir}, {Levan}, {de Ugarte Postigo}, {Kawaguchi}, {Corsi}, {Evans}, {Gompertz}, {Malesani}, {Page}, {Wiersema}, {Rosswog}, {Shibata}, {Tanaka}, {van der Horst}, {Cano}, {Fynbo}, {Fruchter}, {Greiner}, {Heintz}, {Higgins}, {Hjorth}, {Izzo}, {Jakobsson}, {Kann}, {O'Brien}, {Perley}, {Pian}, {Pugliese}, {Starling}, {Th{\"o}ne}, {Watson}, {Wijers}, \& {Xu}}]{lamb+19}
{Lamb}, G.~P., {Tanvir}, N.~R., {Levan}, A.~J., {et~al.} 2019, ApJ, 883, 48, \dodoi{10.3847/1538-4357/ab38bb}

\bibitem[{{Laskar} {et~al.}(2015){Laskar}, {Berger}, {Margutti}, {Perley}, {Zauderer}, {Sari}, \& {Fong}}]{Laskar+15}
{Laskar}, T., {Berger}, E., {Margutti}, R., {et~al.} 2015, \apj, 814, 1, \dodoi{10.1088/0004-637X/814/1/1}

\bibitem[{{Laskar} {et~al.}(2014){Laskar}, {Berger}, {Tanvir}, {Zauderer}, {Margutti}, {Levan}, {Perley}, {Fong}, {Wiersema}, {Menten}, \& {Hrudkova}}]{Laskar+14}
{Laskar}, T., {Berger}, E., {Tanvir}, N., {et~al.} 2014, \apj, 781, 1, \dodoi{10.1088/0004-637X/781/1/1}

\bibitem[{{Lattimer} \& {Schramm}(1974)}]{LattimerSchramm74}
{Lattimer}, J.~M., \& {Schramm}, D.~N. 1974, \apjl, 192, L145, \dodoi{10.1086/181612}

\bibitem[{{Lazzati} {et~al.}(2017){Lazzati}, {L{\'o}pez-C{\'a}mara}, {Cantiello}, {Morsony}, {Perna}, \& {Workman}}]{Lazzati+17}
{Lazzati}, D., {L{\'o}pez-C{\'a}mara}, D., {Cantiello}, M., {et~al.} 2017, \apjl, 848, L6, \dodoi{10.3847/2041-8213/aa8f3d}

\bibitem[{{Levan} {et~al.}(2013){Levan}, {Tanvir}, {Wiersema}, {Hartoog}, {Kolle}, {Mendez}, \& {Kupfer}}]{levan+13_gcn}
{Levan}, A.~J., {Tanvir}, N.~R., {Wiersema}, K., {et~al.} 2013, GRB Coordinates Network, 14742, 1

\bibitem[{{Levan} {et~al.}(2016){Levan}, {Wiersema}, {Tanvir}, {Malesani}, {Xu}, \& {de Ugarte Postigo}}]{Levan+16_GCN}
{Levan}, A.~J., {Wiersema}, K., {Tanvir}, N.~R., {et~al.} 2016, GRB Coordinates Network, 19846, 1

\bibitem[{{Levan} {et~al.}(2017){Levan}, {Lyman}, {Tanvir}, {Hjorth}, {Mandel}, {Stanway}, {Steeghs}, {Fruchter}, {Troja}, {Schr{\o}der}, {Wiersema}, {Bruun}, {Cano}, {Cenko}, {de Ugarte Postigo}, {Evans}, {Fairhurst}, {Fox}, {Fynbo}, {Gompertz}, {Greiner}, {Im}, {Izzo}, {Jakobsson}, {Kangas}, {Khandrika}, {Lien}, {Malesani}, {O'Brien}, {Osborne}, {Palazzi}, {Pian}, {Perley}, {Rosswog}, {Ryan}, {Schulze}, {Sutton}, {Th{\"o}ne}, {Watson}, \& {Wijers}}]{Levan+17}
{Levan}, A.~J., {Lyman}, J.~D., {Tanvir}, N.~R., {et~al.} 2017, \apjl, 848, L28, \dodoi{10.3847/2041-8213/aa905f}

\bibitem[{{Levan} {et~al.}(2024){Levan}, {Gompertz}, {Salafia}, {Bulla}, {Burns}, {Hotokezaka}, {Izzo}, {Lamb}, {Malesani}, {Oates}, {Ravasio}, {Rouco Escorial}, {Schneider}, {Sarin}, {Schulze}, {Tanvir}, {Ackley}, {Anderson}, {Brammer}, {Christensen}, {Dhillon}, {Evans}, {Fausnaugh}, {Fong}, {Fruchter}, {Fryer}, {Fynbo}, {Gaspari}, {Heintz}, {Hjorth}, {Kennea}, {Kennedy}, {Laskar}, {Leloudas}, {Mandel}, {Martin-Carrillo}, {Metzger}, {Nicholl}, {Nugent}, {Palmerio}, {Pugliese}, {Rastinejad}, {Rhodes}, {Rossi}, {Saccardi}, {Smartt}, {Stevance}, {Tohuvavohu}, {van der Horst}, {Vergani}, {Watson}, {Barclay}, {Bhirombhakdi}, {Breedt}, {Breeveld}, {Brown}, {Campana}, {Chrimes}, {D'Avanzo}, {D'Elia}, {De Pasquale}, {Dyer}, {Galloway}, {Garbutt}, {Green}, {Hartmann}, {Jakobsson}, {Kerry}, {Kouveliotou}, {Langeroodi}, {Le Floc'h}, {Leung}, {Littlefair}, {Munday}, {O'Brien}, {Parsons}, {Pelisoli}, {Sahman}, {Salvaterra}, {Sbarufatti}, {Steeghs}, {Tagliaferri}, {Th{\"o}ne}, {de Ugarte Postigo}, \& {Kann}}]{Levan+24}
{Levan}, A.~J., {Gompertz}, B.~P., {Salafia}, O.~S., {et~al.} 2024, \nat, 626, 737, \dodoi{10.1038/s41586-023-06759-1}

\bibitem[{{Li} \& {Paczy{\'n}ski}(1998)}]{Li&Paczynski98}
{Li}, L.-X., \& {Paczy{\'n}ski}, B. 1998, \apjl, 507, L59, \dodoi{10.1086/311680}

\bibitem[{{Lien} {et~al.}(2016){Lien}, {Sakamoto}, {Barthelmy}, {Baumgartner}, {Cannizzo}, {Chen}, {Collins}, {Cummings}, {Gehrels}, {Krimm}, {Markwardt}, {Palmer}, {Stamatikos}, {Troja}, \& {Ukwatta}}]{Lien+16}
{Lien}, A., {Sakamoto}, T., {Barthelmy}, S.~D., {et~al.} 2016, \apj, 829, 7, \dodoi{10.3847/0004-637X/829/1/7}

\bibitem[{{Lippuner} {et~al.}(2017){Lippuner}, {Fern{\'a}ndez}, {Roberts}, {Foucart}, {Kasen}, {Metzger}, \& {Ott}}]{Lippuner+17}
{Lippuner}, J., {Fern{\'a}ndez}, R., {Roberts}, L.~F., {et~al.} 2017, \mnras, 472, 904, \dodoi{10.1093/mnras/stx1987}

\bibitem[{{Lipunov} {et~al.}(2017){Lipunov}, {Gorbovskoy}, {Kornilov}, {.~Tyurina}, {Balanutsa}, {Kuznetsov}, {Vlasenko}, {Kuvshinov}, {Gorbunov}, {Buckley}, {Krylov}, {Podesta}, {Lopez}, {Podesta}, {Levato}, {Saffe}, {Mallamachi}, {Potter}, {Budnev}, {Gress}, {Ishmuhametova}, {Vladimirov}, {Zimnukhov}, {Yurkov}, {Sergienko}, {Gabovich}, {Rebolo}, {Serra-Ricart}, {Israelyan}, {Chazov}, {Wang}, {Tlatov}, \& {Panchenko}}]{Lipunov+17}
{Lipunov}, V.~M., {Gorbovskoy}, E., {Kornilov}, V.~G., {et~al.} 2017, \apjl, 850, L1, \dodoi{10.3847/2041-8213/aa92c0}

\bibitem[{{Londish} {et~al.}(2006){Londish}, {Wieringa}, \& {Frail}}]{Londish+06}
{Londish}, D., {Wieringa}, M.~H., \& {Frail}, D.~A. 2006, GRB Coordinates Network, 5359, 1

\bibitem[{{Lyman} {et~al.}(2018){Lyman}, {Lamb}, {Levan}, {Mandel}, {Tanvir}, {Kobayashi}, {Gompertz}, {Hjorth}, {Fruchter}, {Kangas}, {Steeghs}, {Steele}, {Cano}, {Copperwheat}, {Evans}, {Fynbo}, {Gall}, {Im}, {Izzo}, {Jakobsson}, {Milvang-Jensen}, {O'Brien}, {Osborne}, {Palazzi}, {Perley}, {Pian}, {Rosswog}, {Rowlinson}, {Schulze}, {Stanway}, {Sutton}, {Th{\"o}ne}, {de Ugarte Postigo}, {Watson}, {Wiersema}, \& {Wijers}}]{Lyman+18}
{Lyman}, J.~D., {Lamb}, G.~P., {Levan}, A.~J., {et~al.} 2018, Nature Astronomy, 2, 751, \dodoi{10.1038/s41550-018-0511-3}

\bibitem[{{Mandel} \& {Broekgaarden}(2022)}]{MandelBroekgaarden22}
{Mandel}, I., \& {Broekgaarden}, F.~S. 2022, Living Reviews in Relativity, 25, 1, \dodoi{10.1007/s41114-021-00034-3}

\bibitem[{{Mangan} {et~al.}(2021){Mangan}, {Dunwoody}, {Meegan}, \& {Fermi GBM Team}}]{grb211211a_gbm}
{Mangan}, J., {Dunwoody}, R., {Meegan}, C., \& {Fermi GBM Team}. 2021, GRB Coordinates Network, 31210, 1

\bibitem[{{Mangano} {et~al.}(2007){Mangano}, {Holland}, {Malesani}, {Troja}, {Chincarini}, {Zhang}, {La Parola}, {Brown}, {Burrows}, {Campana}, {Capalbi}, {Cusumano}, {Della Valle}, {Gehrels}, {Giommi}, {Grupe}, {Guidorzi}, {Mineo}, {Moretti}, {Osborne}, {Pandey}, {Perri}, {Romano}, {Roming}, \& {Tagliaferri}}]{Mangano+07}
{Mangano}, V., {Holland}, S.~T., {Malesani}, D., {et~al.} 2007, \aap, 470, 105, \dodoi{10.1051/0004-6361:20077232}

\bibitem[{{Mao} {et~al.}(2021){Mao}, {Xin}, \& {Bai}}]{GCN31232}
{Mao}, J., {Xin}, Y.~X., \& {Bai}, J.~M. 2021, GRB Coordinates Network, 31232, 1

\bibitem[{{Margutti} \& {Chornock}(2020)}]{MarguttiChornock20}
{Margutti}, R., \& {Chornock}, R. 2020, arXiv e-prints, arXiv:2012.04810.
\newblock \doarXiv{2012.04810}

\bibitem[{{Margutti} {et~al.}(2018){Margutti}, {Cowperthwaite}, {Doctor}, {Mortensen}, {Pankow}, {Salafia}, {Villar}, {Alexander}, {Annis}, {Andreoni}, {Baldeschi}, {Balmaverde}, {Berger}, {Bernardini}, {Berry}, {Bianco}, {Blanchard}, {Brocato}, {Carnerero}, {Cartier}, {Cenko}, {Chornock}, {Chomiuk}, {Copperwheat}, {Coughlin}, {Coppejans}, {Corsi}, {D'Ammando}, {Datrier}, {D'Avanzo}, {Dimitriadis}, {Drout}, {Foley}, {Fong}, {Fox}, {Ghirlanda}, {Goldstein}, {Grindlay}, {Guidorzi}, {Haiman}, {Hendry}, {Holz}, {Hung}, {Inserra}, {Jones}, {Kalogera}, {Kilpatrick}, {Lamb}, {Laskar}, {Levan}, {Mason}, {Maguire}, {Melandri}, {Milisavljevic}, {Miller}, {Narayan}, {Nielsen}, {Nicholl}, {Nissanke}, {Nugent}, {Pan}, {Pasham}, {Paterson}, {Piranomonte}, {Racusin}, {Rest}, {Righi}, {Sand}, {Seaman}, {Scolnic}, {Siellez}, {Singer}, {Szkody}, {Smith}, {Steeghs}, {Sullivan}, {Tanvir}, {Terreran}, {Trimble}, {Valenti}, {LSST Transient}, \& {Variable Stars Collaboration}}]{Margutti+18_LSSTTOO}
{Margutti}, R., {Cowperthwaite}, P., {Doctor}, Z., {et~al.} 2018, arXiv e-prints, arXiv:1812.04051, \dodoi{10.48550/arXiv.1812.04051}

\bibitem[{{McCully} {et~al.}(2017){McCully}, {Hiramatsu}, {Howell}, {Hosseinzadeh}, {Arcavi}, {Kasen}, {Barnes}, {Shara}, {Williams}, {V{\"a}is{\"a}nen}, {Potter}, {Romero-Colmenero}, {Crawford}, {Buckley}, {Cooke}, {Andreoni}, {Pritchard}, {Mao}, {Gromadzki}, \& {Burke}}]{McCully+17}
{McCully}, C., {Hiramatsu}, D., {Howell}, D.~A., {et~al.} 2017, \apjl, 848, L32, \dodoi{10.3847/2041-8213/aa9111}

\bibitem[{{Meegan} {et~al.}(2009){Meegan}, {Lichti}, {Bhat}, {Bissaldi}, {Briggs}, {Connaughton}, {Diehl}, {Fishman}, {Greiner}, {Hoover}, {van der Horst}, {von Kienlin}, {Kippen}, {Kouveliotou}, {McBreen}, {Paciesas}, {Preece}, {Steinle}, {Wallace}, {Wilson}, \& {Wilson-Hodge}}]{Meegan+09}
{Meegan}, C., {Lichti}, G., {Bhat}, P.~N., {et~al.} 2009, \apj, 702, 791, \dodoi{10.1088/0004-637X/702/1/791}

\bibitem[{{Mei} {et~al.}(2022){Mei}, {Banerjee}, {Oganesyan}, {Salafia}, {Giarratana}, {Branchesi}, {D'Avanzo}, {Campana}, {Ghirlanda}, {Ronchini}, {Shukla}, \& {Tiwari}}]{Mei+22}
{Mei}, A., {Banerjee}, B., {Oganesyan}, G., {et~al.} 2022, \nat, 612, 236, \dodoi{10.1038/s41586-022-05404-7}

\bibitem[{{Melandri} {et~al.}(2013){Melandri}, {Baumgartner}, {Burrows}, {Cummings}, {Gehrels}, {Gronwall}, {Page}, {Palmer}, {Starling}, \& {Ukwatta}}]{bat130603b_Gcn}
{Melandri}, A., {Baumgartner}, W.~H., {Burrows}, D.~N., {et~al.} 2013, GRB Coordinates Network, 14735, 1

\bibitem[{{Metzger}(2019)}]{metzger19}
{Metzger}, B.~D. 2019, Living Reviews in Relativity, 23, 1, \dodoi{10.1007/s41114-019-0024-0}

\bibitem[{{Metzger} {et~al.}(2015){Metzger}, {Bauswein}, {Goriely}, \& {Kasen}}]{Metzger+15}
{Metzger}, B.~D., {Bauswein}, A., {Goriely}, S., \& {Kasen}, D. 2015, \mnras, 446, 1115, \dodoi{10.1093/mnras/stu2225}

\bibitem[{{Metzger} \& {Fern{\'a}ndez}(2014)}]{metzgerfernandez14}
{Metzger}, B.~D., \& {Fern{\'a}ndez}, R. 2014, \mnras, 441, 3444, \dodoi{10.1093/mnras/stu802}

\bibitem[{{Metzger} {et~al.}(2018){Metzger}, {Thompson}, \& {Quataert}}]{Metzger+18}
{Metzger}, B.~D., {Thompson}, T.~A., \& {Quataert}, E. 2018, \apj, 856, 101, \dodoi{10.3847/1538-4357/aab095}

\bibitem[{{Metzger} {et~al.}(2010){Metzger}, {Mart{\'\i}nez-Pinedo}, {Darbha}, {Quataert}, {Arcones}, {Kasen}, {Thomas}, {Nugent}, {Panov}, \& {Zinner}}]{Metzger+10}
{Metzger}, B.~D., {Mart{\'\i}nez-Pinedo}, G., {Darbha}, S., {et~al.} 2010, \mnras, 406, 2650, \dodoi{10.1111/j.1365-2966.2010.16864.x}

\bibitem[{{Miller} {et~al.}(2019){Miller}, {Ryan}, {Dolence}, {Burrows}, {Fontes}, {Fryer}, {Korobkin}, {Lippuner}, {Mumpower}, \& {Wollaeger}}]{Miller+18}
{Miller}, J.~M., {Ryan}, B.~R., {Dolence}, J.~C., {et~al.} 2019, \prd, 100, 023008, \dodoi{10.1103/PhysRevD.100.023008}

\bibitem[{{Minaev} {et~al.}(2021){Minaev}, {Pozanenko}, \& {GRB IKI FuN}}]{grb211211a_integral}
{Minaev}, P., {Pozanenko}, A., \& {GRB IKI FuN}. 2021, GRB Coordinates Network, 31230, 1

\bibitem[{{Mockler} {et~al.}(2019){Mockler}, {Guillochon}, \& {Ramirez-Ruiz}}]{Mockler+19}
{Mockler}, B., {Guillochon}, J., \& {Ramirez-Ruiz}, E. 2019, \apj, 872, 151, \dodoi{10.3847/1538-4357/ab010f}

\bibitem[{{Mooley} {et~al.}(2022){Mooley}, {Anderson}, \& {Lu}}]{Mooley+22}
{Mooley}, K.~P., {Anderson}, J., \& {Lu}, W. 2022, \nat, 610, 273, \dodoi{10.1038/s41586-022-05145-7}

\bibitem[{{Moskvitin} {et~al.}(2021){Moskvitin}, {Spiridonova}, {Belkin}, {Pozanenko}, {Pankov}, \& {GRB IKI FuN}}]{GCN31234}
{Moskvitin}, A., {Spiridonova}, O., {Belkin}, S., {et~al.} 2021, GRB Coordinates Network, 31234, 1

\bibitem[{{M{\"o}sta} {et~al.}(2018){M{\"o}sta}, {Roberts}, {Halevi}, {Ott}, {Lippuner}, {Haas}, \& {Schnetter}}]{Mosta+18}
{M{\"o}sta}, P., {Roberts}, L.~F., {Halevi}, G., {et~al.} 2018, \apj, 864, 171, \dodoi{10.3847/1538-4357/aad6ec}

\bibitem[{{Nativi} {et~al.}(2021){Nativi}, {Bulla}, {Rosswog}, {Lundman}, {Kowal}, {Gizzi}, {Lamb}, \& {Perego}}]{Nativi+21}
{Nativi}, L., {Bulla}, M., {Rosswog}, S., {et~al.} 2021, \mnras, 500, 1772, \dodoi{10.1093/mnras/staa3337}

\bibitem[{{Navaneeth} {et~al.}(2023){Navaneeth}, {Waratkar}, {Vibhute}, {Bhalerao}, {Bhattacharya}, {Rao}, {Vadawale}, \& {AstroSat CZTI Collaboration}}]{AstroSAT_GCN_230307a}
{Navaneeth}, P.~K., {Waratkar}, G., {Vibhute}, A., {et~al.} 2023, GRB Coordinates Network, 33415, 1

\bibitem[{{Nedora} {et~al.}(2021){Nedora}, {Bernuzzi}, {Radice}, {Daszuta}, {Endrizzi}, {Perego}, {Prakash}, {Safarzadeh}, {Schianchi}, \& {Logoteta}}]{Nedora+21}
{Nedora}, V., {Bernuzzi}, S., {Radice}, D., {et~al.} 2021, \apj, 906, 98, \dodoi{10.3847/1538-4357/abc9be}

\bibitem[{{Nicholl} {et~al.}(2021){Nicholl}, {Margalit}, {Schmidt}, {Smith}, {Ridley}, \& {Nuttall}}]{Nicholl+21}
{Nicholl}, M., {Margalit}, B., {Schmidt}, P., {et~al.} 2021, \mnras, 505, 3016, \dodoi{10.1093/mnras/stab1523}

\bibitem[{{Nicholl} {et~al.}(2017){Nicholl}, {Berger}, {Kasen}, {Metzger}, {Elias}, {Brice{\~n}o}, {Alexander}, {Blanchard}, {Chornock}, {Cowperthwaite}, {Eftekhari}, {Fong}, {Margutti}, {Villar}, {Williams}, {Brown}, {Annis}, {Bahramian}, {Brout}, {Brown}, {Chen}, {Clemens}, {Dennihy}, {Dunlap}, {Holz}, {Marchesini}, {Massaro}, {Moskowitz}, {Pelisoli}, {Rest}, {Ricci}, {Sako}, {Soares-Santos}, \& {Strader}}]{Nicholl+17}
{Nicholl}, M., {Berger}, E., {Kasen}, D., {et~al.} 2017, \apjl, 848, L18, \dodoi{10.3847/2041-8213/aa9029}

\bibitem[{{Nicuesa Guelbenzu} {et~al.}(2012){Nicuesa Guelbenzu}, {Klose}, {Greiner}, {Kann}, {Kr{\"u}hler}, {Rossi}, {Schulze}, {Afonso}, {Elliott}, {Filgas}, {Hartmann}, {K{\"u}pc{\"u} Yolda{\textcommabelow s}}, {McBreen}, {Nardini}, {Olivares E.}, {Rau}, {Schmidl}, {Schady}, {Sudilovsky}, {Updike}, \& {Yolda{\textcommabelow s}}}]{Guelbenzu+12}
{Nicuesa Guelbenzu}, A., {Klose}, S., {Greiner}, J., {et~al.} 2012, \aap, 548, A101, \dodoi{10.1051/0004-6361/201219551}

\bibitem[{{Nugent} {et~al.}(2022){Nugent}, {Fong}, {Dong}, {Leja}, {Berger}, {Zevin}, {Chornock}, {Cobb}, {Kelley}, {Kilpatrick}, {Levan}, {Margutti}, {Paterson}, {Perley}, {Escorial}, {Smith}, \& {Tanvir}}]{Nugent+22}
{Nugent}, A.~E., {Fong}, W.-F., {Dong}, Y., {et~al.} 2022, \apj, 940, 57, \dodoi{10.3847/1538-4357/ac91d1}

\bibitem[{{O'Connor} {et~al.}(2021){O'Connor}, {Troja}, {Dichiara}, {Chase}, {Ryan}, {Cenko}, {Fryer}, {Ricci}, {Marshall}, {Kouveliotou}, {Wollaeger}, {Fontes}, {Korobkin}, {Gatkine}, {Kutyrev}, {Veilleux}, {Kawai}, \& {Sakamoto}}]{O'Connor+21}
{O'Connor}, B., {Troja}, E., {Dichiara}, S., {et~al.} 2021, \mnras, 502, 1279, \dodoi{10.1093/mnras/stab132}

\bibitem[{{Oechslin} {et~al.}(2007){Oechslin}, {Janka}, \& {Marek}}]{Oechslin+07}
{Oechslin}, R., {Janka}, H.~T., \& {Marek}, A. 2007, \aap, 467, 395, \dodoi{10.1051/0004-6361:20066682}

\bibitem[{{Osborne} {et~al.}(2021){Osborne}, {Page}, {Ambrosi}, {Capalbi}, {Perri}, {Burrows}, {Gropp}, {Kennea}, {Beardmore}, {D'Ai}, \& {Swift-XRT Team}}]{grb211211a_xrt}
{Osborne}, J.~P., {Page}, K.~L., {Ambrosi}, E., {et~al.} 2021, GRB Coordinates Network, 31212, 1

\bibitem[{{Paczynski}(1983)}]{Paczynski83}
{Paczynski}, B. 1983, \apj, 267, 315, \dodoi{10.1086/160870}

\bibitem[{{Perley} {et~al.}(2012){Perley}, {Modjaz}, {Morgan}, {Cenko}, {Bloom}, {Butler}, {Filippenko}, \& {Miller}}]{Perley+12}
{Perley}, D.~A., {Modjaz}, M., {Morgan}, A.~N., {et~al.} 2012, \apj, 758, 122, \dodoi{10.1088/0004-637X/758/2/122}

\bibitem[{{Perley} {et~al.}(2009){Perley}, {Metzger}, {Granot}, {Butler}, {Sakamoto}, {Ramirez-Ruiz}, {Levan}, {Bloom}, {Miller}, {Bunker}, {Chen}, {Filippenko}, {Gehrels}, {Glazebrook}, {Hall}, {Hurley}, {Kocevski}, {Li}, {Lopez}, {Norris}, {Piro}, {Poznanski}, {Prochaska}, {Quataert}, \& {Tanvir}}]{Perley09}
{Perley}, D.~A., {Metzger}, B.~D., {Granot}, J., {et~al.} 2009, \apj, 696, 1871, \dodoi{10.1088/0004-637X/696/2/1871}

\bibitem[{{Pian} {et~al.}(2017){Pian}, {D'Avanzo}, {Benetti}, {Branchesi}, {Brocato}, {Campana}, {Cappellaro}, {Covino}, {D'Elia}, {Fynbo}, {Getman}, {Ghirlanda}, {Ghisellini}, {Grado}, {Greco}, {Hjorth}, {Kouveliotou}, {Levan}, {Limatola}, {Malesani}, {Mazzali}, {Melandri}, {M{\o}ller}, {Nicastro}, {Palazzi}, {Piranomonte}, {Rossi}, {Salafia}, {Selsing}, {Stratta}, {Tanaka}, {Tanvir}, {Tomasella}, {Watson}, {Yang}, {Amati}, {Antonelli}, {Ascenzi}, {Bernardini}, {Bo{\"e}r}, {Bufano}, {Bulgarelli}, {Capaccioli}, {Casella}, {Castro-Tirado}, {Chassande-Mottin}, {Ciolfi}, {Copperwheat}, {Dadina}, {De Cesare}, {di Paola}, {Fan}, {Gendre}, {Giuffrida}, {Giunta}, {Hunt}, {Israel}, {Jin}, {Kasliwal}, {Klose}, {Lisi}, {Longo}, {Maiorano}, {Mapelli}, {Masetti}, {Nava}, {Patricelli}, {Perley}, {Pescalli}, {Piran}, {Possenti}, {Pulone}, {Razzano}, {Salvaterra}, {Schipani}, {Spera}, {Stamerra}, {Stella}, {Tagliaferri}, {Testa}, {Troja}, {Turatto}, {Vergani}, \& {Vergani}}]{Pian+17}
{Pian}, E., {D'Avanzo}, P., {Benetti}, S., {et~al.} 2017, \nat, 551, 67, \dodoi{10.1038/nature24298}

\bibitem[{{Pozanenko} {et~al.}(2018){Pozanenko}, {Barkov}, {Minaev}, {Volnova}, {Mazaeva}, {Moskvitin}, {Krugov}, {Samodurov}, {Loznikov}, \& {Lyutikov}}]{Pozanenko+17}
{Pozanenko}, A.~S., {Barkov}, M.~V., {Minaev}, P.~Y., {et~al.} 2018, \apjl, 852, L30, \dodoi{10.3847/2041-8213/aaa2f6}

\bibitem[{{Price} {et~al.}(2006){Price}, {Berger}, \& {Fox}}]{Price+06}
{Price}, P.~A., {Berger}, E., \& {Fox}, D.~B. 2006, GRB Coordinates Network, 5275, 1

\bibitem[{{Radice} {et~al.}(2018){Radice}, {Perego}, {Bernuzzi}, \& {Zhang}}]{Radice+18}
{Radice}, D., {Perego}, A., {Bernuzzi}, S., \& {Zhang}, B. 2018, \mnras, 481, 3670, \dodoi{10.1093/mnras/sty2531}

\bibitem[{{Rastinejad} {et~al.}(2021){Rastinejad}, {Fong}, {Kilpatrick}, {Paterson}, {Tanvir}, {Levan}, {Metzger}, {Berger}, {Chornock}, {Cobb}, {Laskar}, {Milne}, {Nugent}, \& {Smith}}]{Rastinejad+21}
{Rastinejad}, J.~C., {Fong}, W., {Kilpatrick}, C.~D., {et~al.} 2021, \apj, 916, 89, \dodoi{10.3847/1538-4357/ac04b4}

\bibitem[{{Rastinejad} {et~al.}(2022{\natexlab{a}}){Rastinejad}, {Gompertz}, {Levan}, {Fong}, {Nicholl}, {Lamb}, {Malesani}, {Nugent}, {Oates}, {Tanvir}, {de Ugarte Postigo}, {Kilpatrick}, {Moore}, {Metzger}, {Ravasio}, {Rossi}, {Schroeder}, {Jencson}, {Sand}, {Smith}, {Ag{\"u}{\'\i} Fern{\'a}ndez}, {Berger}, {Blanchard}, {Chornock}, {Cobb}, {De Pasquale}, {Fynbo}, {Izzo}, {Kann}, {Laskar}, {Marini}, {Paterson}, {Escorial}, {Sears}, \& {Th{\"o}ne}}]{Rastinejad+22}
{Rastinejad}, J.~C., {Gompertz}, B.~P., {Levan}, A.~J., {et~al.} 2022{\natexlab{a}}, \nat, 612, 223, \dodoi{10.1038/s41586-022-05390-w}

\bibitem[{{Rastinejad} {et~al.}(2022{\natexlab{b}}){Rastinejad}, {Paterson}, {Fong}, {Sand}, {Lundquist}, {Hosseinzadeh}, {Christensen}, {Daly}, {Gibbs}, {Hall}, {Shelly}, \& {Yang}}]{Rastinejad+22_O3}
{Rastinejad}, J.~C., {Paterson}, K., {Fong}, W., {et~al.} 2022{\natexlab{b}}, \apj, 927, 50, \dodoi{10.3847/1538-4357/ac4d34}

\bibitem[{{Rastinejad} {et~al.}(2024){Rastinejad}, {Fong}, {Levan}, {Tanvir}, {Kilpatrick}, {Fruchter}, {Anand}, {Bhirombhakdi}, {Covino}, {Fynbo}, {Halevi}, {Hartmann}, {Heintz}, {Izzo}, {Jakobsson}, {Kangas}, {Lamb}, {Malesani}, {Melandri}, {Metzger}, {Milvang-Jensen}, {Pian}, {Pugliese}, {Rossi}, {Siegel}, {Singh}, \& {Stratta}}]{Rastinejad+24}
{Rastinejad}, J.~C., {Fong}, W., {Levan}, A.~J., {et~al.} 2024, \apj, 968, 14, \dodoi{10.3847/1538-4357/ad409c}

\bibitem[{{Rhoads}(1999)}]{Rhoads99}
{Rhoads}, J.~E. 1999, \apj, 525, 737, \dodoi{10.1086/307907}

\bibitem[{{Ricker} {et~al.}(2003){Ricker}, {Atteia}, {Crew}, {Doty}, {Fenimore}, {Galassi}, {Graziani}, {Hurley}, {Jernigan}, {Kawai}, {Lamb}, {Matsuoka}, {Pizzichini}, {Shirasaki}, {Tamagawa}, {Vanderspek}, {Vedrenne}, {Villasenor}, {Woosley}, \& {Yoshida}}]{HETE}
{Ricker}, G.~R., {Atteia}, J.~L., {Crew}, G.~B., {et~al.} 2003, in American Institute of Physics Conference Series, Vol. 662, Gamma-Ray Burst and Afterglow Astronomy 2001: A Workshop Celebrating the First Year of the HETE Mission, ed. G.~R. {Ricker} \& R.~K. {Vanderspek} (AIP), 3--16, \dodoi{10.1063/1.1579291}

\bibitem[{{Roming} {et~al.}(2005){Roming}, {Kennedy}, {Mason}, {Nousek}, {Ahr}, {Bingham}, {Broos}, {Carter}, {Hancock}, {Huckle}, {Hunsberger}, {Kawakami}, {Killough}, {Koch}, {McLelland}, {Smith}, {Smith}, {Soto}, {Boyd}, {Breeveld}, {Holland}, {Ivanushkina}, {Pryzby}, {Still}, \& {Stock}}]{Swift-UVOT}
{Roming}, P. W.~A., {Kennedy}, T.~E., {Mason}, K.~O., {et~al.} 2005, \ssr, 120, 95, \dodoi{10.1007/s11214-005-5095-4}

\bibitem[{{Rose} {et~al.}(2021){Rose}, {Baltay}, {Hounsell}, {Macias}, {Rubin}, {Scolnic}, {Aldering}, {Bohlin}, {Dai}, {Deustua}, {Foley}, {Fruchter}, {Galbany}, {Jha}, {Jones}, {Joshi}, {Kelly}, {Kessler}, {Kirshner}, {Mandel}, {Perlmutter}, {Pierel}, {Qu}, {Rabinowitz}, {Rest}, {Riess}, {Rodney}, {Sako}, {Siebert}, {Strolger}, {Suzuki}, {Thorp}, {Van Dyk}, {Wang}, {Ward}, \& {Wood-Vasey}}]{Rose+21}
{Rose}, B.~M., {Baltay}, C., {Hounsell}, R., {et~al.} 2021, arXiv e-prints, arXiv:2111.03081, \dodoi{10.48550/arXiv.2111.03081}

\bibitem[{{Rossi} {et~al.}(2020){Rossi}, {Stratta}, {Maiorano}, {Spighi}, {Masetti}, {Palazzi}, {Gardini}, {Melandri}, {Nicastro}, {Pian}, {Branchesi}, {Dadina}, {Testa}, {Brocato}, {Benetti}, {Ciolfi}, {Covino}, {D'Elia}, {Grado}, {Izzo}, {Perego}, {Piranomonte}, {Salvaterra}, {Selsing}, {Tomasella}, {Yang}, {Vergani}, {Amati}, \& {Stephen}}]{Rossi+20}
{Rossi}, A., {Stratta}, G., {Maiorano}, E., {et~al.} 2020, \mnras, 493, 3379, \dodoi{10.1093/mnras/staa479}

\bibitem[{{Rossi} {et~al.}(2022){Rossi}, {Rothberg}, {Palazzi}, {Kann}, {D'Avanzo}, {Amati}, {Klose}, {Perego}, {Pian}, {Guidorzi}, {Pozanenko}, {Savaglio}, {Stratta}, {Agapito}, {Covino}, {Cusano}, {D'Elia}, {De Pasquale}, {Della Valle}, {Kuhn}, {Izzo}, {Loffredo}, {Masetti}, {Melandri}, {Minaev}, {Guelbenzu}, {Paris}, {Paiano}, {Plantet}, {Rossi}, {Salvaterra}, {Schulze}, {Veillet}, \& {Volnova}}]{Rossi+22}
{Rossi}, A., {Rothberg}, B., {Palazzi}, E., {et~al.} 2022, \apj, 932, 1, \dodoi{10.3847/1538-4357/ac60a2}

\bibitem[{{Rosswog} {et~al.}(1999){Rosswog}, {Liebend{\"o}rfer}, {Thielemann}, {Davies}, {Benz}, \& {Piran}}]{Rosswog+99}
{Rosswog}, S., {Liebend{\"o}rfer}, M., {Thielemann}, F.~K., {et~al.} 1999, \aap, 341, 499, \dodoi{10.48550/arXiv.astro-ph/9811367}

\bibitem[{{Rosswog} {et~al.}(2018){Rosswog}, {Sollerman}, {Feindt}, {Goobar}, {Korobkin}, {Wollaeger}, {Fremling}, \& {Kasliwal}}]{Rosswog+18}
{Rosswog}, S., {Sollerman}, J., {Feindt}, U., {et~al.} 2018, \aap, 615, A132, \dodoi{10.1051/0004-6361/201732117}

\bibitem[{{Rouco Escorial} {et~al.}(2023){Rouco Escorial}, {Fong}, {Berger}, {Laskar}, {Margutti}, {Schroeder}, {Rastinejad}, {Cornish}, {Popp}, {Lally}, {Nugent}, {Paterson}, {Metzger}, {Chornock}, {Alexander}, {Cendes}, \& {Eftekhari}}]{Rouco+23}
{Rouco Escorial}, A., {Fong}, W., {Berger}, E., {et~al.} 2023, \apj, 959, 13, \dodoi{10.3847/1538-4357/acf830}

\bibitem[{{Rowlinson} {et~al.}(2010{\natexlab{a}}){Rowlinson}, {Wiersema}, {Levan}, {Tanvir}, {O'Brien}, {Rol}, {Hjorth}, {Th{\"o}ne}, {de Ugarte Postigo}, {Fynbo}, {Jakobsson}, {Pagani}, \& {Stamatikos}}]{Rowlinson+10b}
{Rowlinson}, A., {Wiersema}, K., {Levan}, A.~J., {et~al.} 2010{\natexlab{a}}, \mnras, 408, 383, \dodoi{10.1111/j.1365-2966.2010.17115.x}

\bibitem[{{Rowlinson} {et~al.}(2010{\natexlab{b}}){Rowlinson}, {O'Brien}, {Tanvir}, {Zhang}, {Evans}, {Lyons}, {Levan}, {Willingale}, {Page}, {Onal}, {Burrows}, {Beardmore}, {Ukwatta}, {Berger}, {Hjorth}, {Fruchter}, {Tunnicliffe}, {Fox}, \& {Cucchiara}}]{Rowlinson+10}
{Rowlinson}, A., {O'Brien}, P.~T., {Tanvir}, N.~R., {et~al.} 2010{\natexlab{b}}, \mnras, 409, 531, \dodoi{10.1111/j.1365-2966.2010.17354.x}

\bibitem[{{Sari} \& {Esin}(2001)}]{SariEsin01}
{Sari}, R., \& {Esin}, A.~A. 2001, \apj, 548, 787, \dodoi{10.1086/319003}

\bibitem[{{Sari} {et~al.}(1999){Sari}, {Piran}, \& {Halpern}}]{Sari+99}
{Sari}, R., {Piran}, T., \& {Halpern}, J.~P. 1999, \apjl, 519, L17, \dodoi{10.1086/312109}

\bibitem[{{Sari} {et~al.}(1998){Sari}, {Piran}, \& {Narayan}}]{Sari+98}
{Sari}, R., {Piran}, T., \& {Narayan}, R. 1998, \apjl, 497, L17, \dodoi{10.1086/311269}

\bibitem[{{Sarin} \& {Rosswog}(2024)}]{SarinRosswog24}
{Sarin}, N., \& {Rosswog}, S. 2024, arXiv e-prints, arXiv:2404.07271, \dodoi{10.48550/arXiv.2404.07271}

\bibitem[{{Savchenko} {et~al.}(2017){Savchenko}, {Ferrigno}, {Kuulkers}, {Bazzano}, {Bozzo}, {Brandt}, {Chenevez}, {Courvoisier}, {Diehl}, {Domingo}, {Hanlon}, {Jourdain}, {von Kienlin}, {Laurent}, {Lebrun}, {Lutovinov}, {Martin-Carrillo}, {Mereghetti}, {Natalucci}, {Rodi}, {Roques}, {Sunyaev}, \& {Ubertini}}]{Savchenko+17}
{Savchenko}, V., {Ferrigno}, C., {Kuulkers}, E., {et~al.} 2017, \apjl, 848, L15, \dodoi{10.3847/2041-8213/aa8f94}

\bibitem[{{Schlafly} \& {Finkbeiner}(2011)}]{SchlaflyFinkbeiner11}
{Schlafly}, E.~F., \& {Finkbeiner}, D.~P. 2011, \apj, 737, 103, \dodoi{10.1088/0004-637X/737/2/103}

\bibitem[{{Schmidt} {et~al.}(2006){Schmidt}, {Peterson}, \& {Lewis}}]{Schmidt+06_GCN}
{Schmidt}, B., {Peterson}, B., \& {Lewis}, K. 2006, GRB Coordinates Network, 5258, 1

\bibitem[{{Schroeder} {et~al.}(2020){Schroeder}, {Fong}, \& {Alexander}}]{200522a_VLA}
{Schroeder}, G., {Fong}, W., \& {Alexander}, K.~D. 2020, GRB Coordinates Network, 27786, 1

\bibitem[{{Schroeder} {et~al.}(2024){Schroeder}, {Fong}, {Kilpatrick}, {Rouco Escorial}, {Laskar}, {Nugent}, {Rastinejad}, {Alexander}, {Berger}, {Brink}, {Chornock}, {de Bom}, {Dong}, {Eftekhari}, {Filippenko}, {Fuentes-Carvajal}, {Jacobson-Galan}, {Malkan}, {Margutti}, {Pearson}, {Rhodes}, {Salinas}, {Sand}, {Santana-Silva}, {Santos}, {Sears}, {Shrestha}, {Smith}, {Webb}, {de Wet}, \& {Yang}}]{Schroeder+24}
{Schroeder}, G., {Fong}, W.-f., {Kilpatrick}, C.~D., {et~al.} 2024, arXiv e-prints, arXiv:2407.13822, \dodoi{10.48550/arXiv.2407.13822}

\bibitem[{{Sekiguchi} {et~al.}(2015){Sekiguchi}, {Kiuchi}, {Kyutoku}, \& {Shibata}}]{sekiguchi+16}
{Sekiguchi}, Y., {Kiuchi}, K., {Kyutoku}, K., \& {Shibata}, M. 2015, \prd, 91, 064059, \dodoi{10.1103/PhysRevD.91.064059}

\bibitem[{{Shappee} {et~al.}(2017){Shappee}, {Simon}, {Drout}, {Piro}, {Morrell}, {Prieto}, {Kasen}, {Holoien}, {Kollmeier}, {Kelson}, {Coulter}, {Foley}, {Kilpatrick}, {Siebert}, {Madore}, {Murguia-Berthier}, {Pan}, {Prochaska}, {Ramirez-Ruiz}, {Rest}, {Adams}, {Alatalo}, {Ba{\~n}ados}, {Baughman}, {Bernstein}, {Bitsakis}, {Boutsia}, {Bravo}, {Di Mille}, {Higgs}, {Ji}, {Maravelias}, {Marshall}, {Placco}, {Prieto}, \& {Wan}}]{Shappee+17}
{Shappee}, B.~J., {Simon}, J.~D., {Drout}, M.~R., {et~al.} 2017, Science, 358, 1574, \dodoi{10.1126/science.aaq0186}

\bibitem[{{Shibata} \& {Hotokezaka}(2019)}]{Shibata+19}
{Shibata}, M., \& {Hotokezaka}, K. 2019, Annual Review of Nuclear and Particle Science, 69, 41, \dodoi{10.1146/annurev-nucl-101918-023625}

\bibitem[{{Shrestha} {et~al.}(2023){Shrestha}, {Bulla}, {Nativi}, {Markin}, {Rosswog}, \& {Dietrich}}]{Shrestha+23}
{Shrestha}, M., {Bulla}, M., {Nativi}, L., {et~al.} 2023, \mnras, 523, 2990, \dodoi{10.1093/mnras/stad1583}

\bibitem[{{Siegel} {et~al.}(2019){Siegel}, {Barnes}, \& {Metzger}}]{SiegelBarnesMetzger2019}
{Siegel}, D.~M., {Barnes}, J., \& {Metzger}, B.~D. 2019, \nat, 569, 241, \dodoi{10.1038/s41586-019-1136-0}

\bibitem[{{Smartt} {et~al.}(2017){Smartt}, {Chen}, {Jerkstrand}, {Coughlin}, {Kankare}, {Sim}, {Fraser}, {Inserra}, {Maguire}, {Chambers}, {Huber}, {Kr{\"u}hler}, {Leloudas}, {Magee}, {Shingles}, {Smith}, {Young}, {Tonry}, {Kotak}, {Gal-Yam}, {Lyman}, {Homan}, {Agliozzo}, {Anderson}, {Angus}, {Ashall}, {Barbarino}, {Bauer}, {Berton}, {Botticella}, {Bulla}, {Bulger}, {Cannizzaro}, {Cano}, {Cartier}, {Cikota}, {Clark}, {De Cia}, {Della Valle}, {Denneau}, {Dennefeld}, {Dessart}, {Dimitriadis}, {Elias-Rosa}, {Firth}, {Flewelling}, {Fl{\"o}rs}, {Franckowiak}, {Frohmaier}, {Galbany}, {Gonz{\'a}lez-Gait{\'a}n}, {Greiner}, {Gromadzki}, {Guelbenzu}, {Guti{\'e}rrez}, {Hamanowicz}, {Hanlon}, {Harmanen}, {Heintz}, {Heinze}, {Hernandez}, {Hodgkin}, {Hook}, {Izzo}, {James}, {Jonker}, {Kerzendorf}, {Klose}, {Kostrzewa-Rutkowska}, {Kowalski}, {Kromer}, {Kuncarayakti}, {Lawrence}, {Lowe}, {Magnier}, {Manulis}, {Martin-Carrillo}, {Mattila}, {McBrien}, {M{\"u}ller}, {Nordin}, {O'Neill}, {Onori}, {Palmerio}, {Pastorello},
  {Patat}, {Pignata}, {Podsiadlowski}, {Pumo}, {Prentice}, {Rau}, {Razza}, {Rest}, {Reynolds}, {Roy}, {Ruiter}, {Rybicki}, {Salmon}, {Schady}, {Schultz}, {Schweyer}, {Seitenzahl}, {Smith}, {Sollerman}, {Stalder}, {Stubbs}, {Sullivan}, {Szegedi}, {Taddia}, {Taubenberger}, {Terreran}, {van Soelen}, {Vos}, {Wainscoat}, {Walton}, {Waters}, {Weiland}, {Willman}, {Wiseman}, {Wright}, {Wyrzykowski}, \& {Yaron}}]{Smartt+17}
{Smartt}, S.~J., {Chen}, T.~W., {Jerkstrand}, A., {et~al.} 2017, \nat, 551, 75, \dodoi{10.1038/nature24303}

\bibitem[{{Soares-Santos} {et~al.}(2017){Soares-Santos}, {Holz}, {Annis}, {Chornock}, {Herner}, {Berger}, {Brout}, {Chen}, {Kessler}, {Sako}, {Allam}, {Tucker}, {Butler}, {Palmese}, {Doctor}, {Diehl}, {Frieman}, {Yanny}, {Lin}, {Scolnic}, {Cowperthwaite}, {Neilsen}, {Marriner}, {Kuropatkin}, {Hartley}, {Paz-Chinch{\'o}n}, {Alexander}, {Balbinot}, {Blanchard}, {Brown}, {Carlin}, {Conselice}, {Cook}, {Drlica-Wagner}, {Drout}, {Durret}, {Eftekhari}, {Farr}, {Finley}, {Foley}, {Fong}, {Fryer}, {Garc{\'\i}a-Bellido}, {Gill}, {Gruendl}, {Hanna}, {Kasen}, {Li}, {Lopes}, {Louren{\c{c}}o}, {Margutti}, {Marshall}, {Matheson}, {Medina}, {Metzger}, {Mu{\~n}oz}, {Muir}, {Nicholl}, {Quataert}, {Rest}, {Sauseda}, {Schlegel}, {Secco}, {Sobreira}, {Stebbins}, {Villar}, {Vivas}, {Walker}, {Wester}, {Williams}, {Zenteno}, {Zhang}, {Abbott}, {Abdalla}, {Banerji}, {Bechtol}, {Benoit-L{\'e}vy}, {Bertin}, {Brooks}, {Buckley-Geer}, {Burke}, {Carnero Rosell}, {Carrasco Kind}, {Carretero}, {Castander}, {Crocce}, {Cunha}, {D'Andrea},
  {da Costa}, {Davis}, {Desai}, {Dietrich}, {Doel}, {Eifler}, {Fernand ez}, {Flaugher}, {Fosalba}, {Gaztanaga}, {Gerdes}, {Giannantonio}, {Goldstein}, {Gruen}, {Gschwend}, {Gutierrez}, {Honscheid}, {Jain}, {James}, {Jeltema}, {Johnson}, {Johnson}, {Kent}, {Krause}, {Kron}, {Kuehn}, {Kuhlmann}, {Lahav}, {Lima}, {Maia}, {March}, {McMahon}, {Menanteau}, {Miquel}, {Mohr}, {Nichol}, {Nord}, {Ogand o}, {Petravick}, {Plazas}, {Romer}, {Roodman}, {Rykoff}, {Sanchez}, {Scarpine}, {Schubnell}, {Sevilla-Noarbe}, {Smith}, {Smith}, {Suchyta}, {Swanson}, {Tarle}, {Thomas}, {Thomas}, {Troxel}, {Vikram}, {Wechsler}, {Weller}, {Dark Energy Survey}, \& {Dark Energy Camera GW-EM Collaboration}}]{Soares-Santos+17}
{Soares-Santos}, M., {Holz}, D.~E., {Annis}, J., {et~al.} 2017, \apjl, 848, L16, \dodoi{10.3847/2041-8213/aa9059}

\bibitem[{{Speagle}(2020)}]{Speagle+20}
{Speagle}, J.~S. 2020, \mnras, 493, 3132, \dodoi{10.1093/mnras/staa278}

\bibitem[{{Stamatikos} {et~al.}(2021){Stamatikos}, {Barthelmy}, {D'Ai}, {Krimm}, {Laha}, {Lien}, {Markwardt}, {Palmer}, {Parsotan}, \& {Sakamoto}}]{GRB211211a_bat}
{Stamatikos}, M., {Barthelmy}, S.~D., {D'Ai}, A., {et~al.} 2021, GRB Coordinates Network, 31209, 1

\bibitem[{{Stanbro} \& {Meegan}(2016)}]{Fermi_160821b}
{Stanbro}, M., \& {Meegan}, C. 2016, GRB Coordinates Network, 19843, 1

\bibitem[{{Strausbaugh} \& {Cucchiara}(2020)}]{LCO_GCN_200522a}
{Strausbaugh}, R., \& {Cucchiara}, A. 2020, GRB Coordinates Network, 27792, 1

\bibitem[{{Strausbaugh} \& {Cucchiara}(2021)}]{GCN31214}
---. 2021, GRB Coordinates Network, 31214, 1

\bibitem[{{Svinkin} {et~al.}(2022){Svinkin}, {Hurley}, {Ridnaia}, {Lysenko}, {Frederiks}, {Golenetskii}, {Tsvetkova}, {Ulanov}, {Kokomov}, {Cline}, {Mitrofanov}, {Golovin}, {Kozyrev}, {Litvak}, {Sanin}, {Goldstein}, {Briggs}, {Wilson-Hodge}, {Burns}, {von Kienlin}, {Zhang}, {Rau}, {Savchenko}, {Bozzo}, {Ferrigno}, {Barthelmy}, {Cummings}, {Krimm}, {Palmer}, {Tohuvavohu}, {Yamaoka}, {Ohno}, {Fukazawa}, {Hanabata}, {Takahashi}, {Tashiro}, {Terada}, {Murakami}, {Makishima}, {Boynton}, {Fellows}, {Harshman}, {Enos}, {Starr}, {Goldsten}, {Gold}, {Ursi}, {Tavani}, {Bulgarelli}, {Casentini}, {Del Monte}, {Evangelista}, {Galli}, {Longo}, {Marisaldi}, {Parmiggiani}, {Pittori}, {Romani}, {Verrecchia}, {Smith}, {Hajdas}, {Xiao}, {Cai}, {Yi}, {Zhang}, {Xiong}, {Li}, {Huang}, {Li}, {Zhang}, {Song}, {Liu}, {Li}, {Peng}, \& {Martinez-Castellanos}}]{Svinkin+22}
{Svinkin}, D.~S., {Hurley}, K., {Ridnaia}, A.~V., {et~al.} 2022, \apjs, 259, 34, \dodoi{10.3847/1538-4365/ac4607}

\bibitem[{{Tamura} {et~al.}(2021){Tamura}, {Yoshida}, {Sakamoto}, {Pal'Shin}, {Sugita}, {Kawakubo}, {Yamaoka}, {Nakahira}, {Asaoka}, {Torii}, {Akaike}, {Kobayashi}, {Shimizu}, {Cannady}, {Cherry}, {Ricciarini}, {Marrocchesi}, \& {Calet Collaboration}}]{grb211211a_CALET}
{Tamura}, T., {Yoshida}, A., {Sakamoto}, T., {et~al.} 2021, GRB Coordinates Network, 31226, 1

\bibitem[{{Tanaka} \& {Hotokezaka}(2013)}]{tanaka+13}
{Tanaka}, M., \& {Hotokezaka}, K. 2013, ApJ, 775, 113, \dodoi{10.1088/0004-637X/775/2/113}

\bibitem[{{Tanaka} {et~al.}(2020){Tanaka}, {Kato}, {Gaigalas}, \& {Kawaguchi}}]{Tanaka+20}
{Tanaka}, M., {Kato}, D., {Gaigalas}, G., \& {Kawaguchi}, K. 2020, \mnras, 496, 1369, \dodoi{10.1093/mnras/staa1576}

\bibitem[{{Tanaka} {et~al.}(2017){Tanaka}, {Utsumi}, {Mazzali}, {Tominaga}, {Yoshida}, {Sekiguchi}, {Morokuma}, {Motohara}, {Ohta}, {Kawabata}, {Abe}, {Aoki}, {Asakura}, {Baar}, {Barway}, {Bond}, {Doi}, {Fujiyoshi}, {Furusawa}, {Honda}, {Itoh}, {Kawabata}, {Kawai}, {Kim}, {Lee}, {Miyazaki}, {Morihana}, {Nagashima}, {Nagayama}, {Nakaoka}, {Nakata}, {Ohsawa}, {Ohshima}, {Okita}, {Saito}, {Sumi}, {Tajitsu}, {Takahashi}, {Takayama}, {Tamura}, {Tanaka}, {Terai}, {Tristram}, {Yasuda}, \& {Zenko}}]{Tanaka+17}
{Tanaka}, M., {Utsumi}, Y., {Mazzali}, P.~A., {et~al.} 2017, \pasj, 69, 102, \dodoi{10.1093/pasj/psx121}

\bibitem[{{Tanvir} {et~al.}(2013){Tanvir}, {Levan}, {Fruchter}, {Hjorth}, {Hounsell}, {Wiersema}, \& {Tunnicliffe}}]{Tanvir+13}
{Tanvir}, N.~R., {Levan}, A.~J., {Fruchter}, A.~S., {et~al.} 2013, \nat, 500, 547, \dodoi{10.1038/nature12505}

\bibitem[{{Tanvir} {et~al.}(2017){Tanvir}, {Levan}, {Gonz{\'a}lez-Fern{\'a}ndez}, {Korobkin}, {Mandel}, {Rosswog}, {Hjorth}, {D'Avanzo}, {Fruchter}, {Fryer}, {Kangas}, {Milvang-Jensen}, {Rosetti}, {Steeghs}, {Wollaeger}, {Cano}, {Copperwheat}, {Covino}, {D'Elia}, {de Ugarte Postigo}, {Evans}, {Even}, {Fairhurst}, {Figuera Jaimes}, {Fontes}, {Fujii}, {Fynbo}, {Gompertz}, {Greiner}, {Hodosan}, {Irwin}, {Jakobsson}, {J{\o}rgensen}, {Kann}, {Lyman}, {Malesani}, {McMahon}, {Melandri}, {O'Brien}, {Osborne}, {Palazzi}, {Perley}, {Pian}, {Piranomonte}, {Rabus}, {Rol}, {Rowlinson}, {Schulze}, {Sutton}, {Th{\"o}ne}, {Ulaczyk}, {Watson}, {Wiersema}, \& {Wijers}}]{Tanvir+17}
{Tanvir}, N.~R., {Levan}, A.~J., {Gonz{\'a}lez-Fern{\'a}ndez}, C., {et~al.} 2017, \apjl, 848, L27, \dodoi{10.3847/2041-8213/aa90b6}

\bibitem[{{Tauris} {et~al.}(2017){Tauris}, {Kramer}, {Freire}, {Wex}, {Janka}, {Langer}, {Podsiadlowski}, {Bozzo}, {Chaty}, {Kruckow}, {van den Heuvel}, {Antoniadis}, {Breton}, \& {Champion}}]{Tauris+17}
{Tauris}, T.~M., {Kramer}, M., {Freire}, P.~C.~C., {et~al.} 2017, \apj, 846, 170, \dodoi{10.3847/1538-4357/aa7e89}

\bibitem[{{Tchekhovskoy} {et~al.}(2011){Tchekhovskoy}, {Narayan}, \& {McKinney}}]{Tchekhovskoy+11}
{Tchekhovskoy}, A., {Narayan}, R., \& {McKinney}, J.~C. 2011, \mnras, 418, L79, \dodoi{10.1111/j.1745-3933.2011.01147.x}

\bibitem[{{The LIGO Scientific Collaboration} {et~al.}(2024){The LIGO Scientific Collaboration}, {the Virgo Collaboration}, \& {the KAGRA Collaboration}}]{LVK_230529}
{The LIGO Scientific Collaboration}, {the Virgo Collaboration}, \& {the KAGRA Collaboration}. 2024, arXiv e-prints, arXiv:2404.04248, \dodoi{10.48550/arXiv.2404.04248}

\bibitem[{{The LIGO Scientific Collaboration} {et~al.}(2021){The LIGO Scientific Collaboration}, {the Virgo Collaboration}, {the KAGRA Collaboration}, {Abbott}, {Abbott}, {Acernese}, {Ackley}, {Adams}, {Adhikari}, {Adhikari}, {Adya}, {Affeldt}, {Agarwal}, {Agathos}, {Agatsuma}, {Aggarwal}, {Aguiar}, {Aiello}, {Ain}, {Ajith}, {Akcay}, {Akutsu}, {Albanesi}, {Allocca}, {Altin}, {Amato}, {Anand}, {Anand}, {Ananyeva}, {Anderson}, {Anderson}, {Ando}, {Andrade}, {Andres}, {Andri{\'c}}, \& {Angelova}}]{GWTC-3}
{The LIGO Scientific Collaboration}, {the Virgo Collaboration}, {the KAGRA Collaboration}, {et~al.} 2021, arXiv e-prints, arXiv:2111.03606.
\newblock \doarXiv{2111.03606}

\bibitem[{{Thone} {et~al.}(2013){Thone}, {de Ugarte Postigo}, {Gorosabel}, {Tanvir}, \& {Fynbo}}]{thone+13_gcn}
{Thone}, C.~C., {de Ugarte Postigo}, A., {Gorosabel}, J., {Tanvir}, N., \& {Fynbo}, J.~P.~U. 2013, GRB Coordinates Network, 14744, 1

\bibitem[{{Troja} {et~al.}(2017){Troja}, {Piro}, {van Eerten}, {Wollaeger}, {Im}, {Fox}, {Butler}, {Cenko}, {Sakamoto}, {Fryer}, {Ricci}, {Lien}, {Ryan}, {Korobkin}, {Lee}, {Burgess}, {Lee}, {Watson}, {Choi}, {Covino}, {D'Avanzo}, {Fontes}, {Gonz{\'a}lez}, {Khandrika}, {Kim}, {Kim}, {Lee}, {Lee}, {Kutyrev}, {Lim}, {S{\'a}nchez-Ram{\'\i}rez}, {Veilleux}, {Wieringa}, \& {Yoon}}]{Troja+17}
{Troja}, E., {Piro}, L., {van Eerten}, H., {et~al.} 2017, \nat, 551, 71, \dodoi{10.1038/nature24290}

\bibitem[{{Troja} {et~al.}(2018){Troja}, {Ryan}, {Piro}, {van Eerten}, {Cenko}, {Yoon}, {Lee}, {Im}, {Sakamoto}, {Gatkine}, {Kutyrev}, \& {Veilleux}}]{troja+18}
{Troja}, E., {Ryan}, G., {Piro}, L., {et~al.} 2018, Nature Communications, 9, 4089, \dodoi{10.1038/s41467-018-06558-7}

\bibitem[{{Troja} {et~al.}(2019){Troja}, {Castro-Tirado}, {Becerra Gonz{\'a}lez}, {Hu}, {Ryan}, {Cenko}, {Ricci}, {Novara}, {S{\'a}nchez-R{\'a}mirez}, {Acosta-Pulido}, {Ackley}, {Caballero Garc{\'\i}a}, {Eikenberry}, {Guziy}, {Jeong}, {Lien}, {M{\'a}rquez}, {Pand ey}, {Park}, {Sakamoto}, {Tello}, {Sokolov}, {Sokolov}, {Tiengo}, {Valeev}, {Zhang}, \& {Veilleux}}]{troja+19}
{Troja}, E., {Castro-Tirado}, A.~J., {Becerra Gonz{\'a}lez}, J., {et~al.} 2019, MNRAS, 489, 2104, \dodoi{10.1093/mnras/stz2255}

\bibitem[{{Troja} {et~al.}(2022){Troja}, {Fryer}, {O'Connor}, {Ryan}, {Dichiara}, {Kumar}, {Ito}, {Gupta}, {Wollaeger}, {Norris}, {Kawai}, {Butler}, {Aryan}, {Misra}, {Hosokawa}, {Murata}, {Niwano}, {Pandey}, {Kutyrev}, {van Eerten}, {Chase}, {Hu}, {Caballero-Garcia}, \& {Castro-Tirado}}]{Troja+22}
{Troja}, E., {Fryer}, C.~L., {O'Connor}, B., {et~al.} 2022, \nat, 612, 228, \dodoi{10.1038/s41586-022-05327-3}

\bibitem[{{Utsumi} {et~al.}(2017){Utsumi}, {Tanaka}, {Tominaga}, {Yoshida}, {Barway}, {Nagayama}, {Zenko}, {Aoki}, {Fujiyoshi}, {Furusawa}, {Kawabata}, {Koshida}, {Lee}, {Morokuma}, {Motohara}, {Nakata}, {Ohsawa}, {Ohta}, {Okita}, {Tajitsu}, {Tanaka}, {Terai}, {Yasuda}, {Abe}, {Asakura}, {Bond}, {Miyazaki}, {Sumi}, {Tristram}, {Honda}, {Itoh}, {Itoh}, {Kawabata}, {Morihana}, {Nagashima}, {Nakaoka}, {Ohshima}, {Takahashi}, {Takayama}, {Aoki}, {Baar}, {Doi}, {Finet}, {Kanda}, {Kawai}, {Kim}, {Kuroda}, {Liu}, {Matsubayashi}, {Murata}, {Nagai}, {Saito}, {Saito}, {Sako}, {Sekiguchi}, {Tamura}, {Tanaka}, {Uemura}, \& {Yamaguchi}}]{Utsumi+17}
{Utsumi}, Y., {Tanaka}, M., {Tominaga}, N., {et~al.} 2017, \pasj, 69, 101, \dodoi{10.1093/pasj/psx118}

\bibitem[{{Valenti} {et~al.}(2017){Valenti}, {Sand}, {Yang}, {Cappellaro}, {Tartaglia}, {Corsi}, {Jha}, {Reichart}, {Haislip}, \& {Kouprianov}}]{Valenti+17}
{Valenti}, S., {Sand}, D.~J., {Yang}, S., {et~al.} 2017, \apjl, 848, L24, \dodoi{10.3847/2041-8213/aa8edf}

\bibitem[{{Venn} {et~al.}(2004){Venn}, {Irwin}, {Shetrone}, {Tout}, {Hill}, \& {Tolstoy}}]{Venn+04}
{Venn}, K.~A., {Irwin}, M., {Shetrone}, M.~D., {et~al.} 2004, \aj, 128, 1177, \dodoi{10.1086/422734}

\bibitem[{{Veres} {et~al.}(2023){Veres}, {Bhat}, {Burns}, {Hamburg}, {Fraija}, {Kocevski}, {Preece}, {Poolakkil}, {Christensen}, {Bizouard}, {Dal Canton}, {Bala}, {Bissaldi}, {Briggs}, {Cleveland}, {Goldstein}, {Hristov}, {Hui}, {Lesage}, {Mailyan}, {Roberts}, \& {Wilson-Hodge}}]{Veres+23}
{Veres}, P., {Bhat}, P.~N., {Burns}, E., {et~al.} 2023, \apjl, 954, L5, \dodoi{10.3847/2041-8213/ace82d}

\bibitem[{{Villar} {et~al.}(2017{\natexlab{a}}){Villar}, {Berger}, {Metzger}, \& {Guillochon}}]{Villar+17_model}
{Villar}, V.~A., {Berger}, E., {Metzger}, B.~D., \& {Guillochon}, J. 2017{\natexlab{a}}, \apj, 849, 70, \dodoi{10.3847/1538-4357/aa8fcb}

\bibitem[{{Villar} {et~al.}(2017{\natexlab{b}}){Villar}, {Guillochon}, {Berger}, {Metzger}, {Cowperthwaite}, {Nicholl}, {Alexand er}, {Blanchard}, {Chornock}, {Eftekhari}, {Fong}, {Margutti}, \& {Williams}}]{villar+17}
{Villar}, V.~A., {Guillochon}, J., {Berger}, E., {et~al.} 2017{\natexlab{b}}, ApJL, 851, L21, \dodoi{10.3847/2041-8213/aa9c84}

\bibitem[{{Wallace} \& {Sarin}(2024)}]{WallaceSarin24}
{Wallace}, W.~F., \& {Sarin}, N. 2024, arXiv e-prints, arXiv:2409.07539.
\newblock \doarXiv{2409.07539}

\bibitem[{{Wang} {et~al.}(2024){Wang}, {Yu}, {Ren}, {Yang}, {Zou}, \& {Zhu}}]{Wang+24}
{Wang}, X.~I., {Yu}, Y.-W., {Ren}, J., {et~al.} 2024, \apjl, 964, L9, \dodoi{10.3847/2041-8213/ad2df6}

\bibitem[{{Watson} {et~al.}(2019){Watson}, {Hansen}, {Selsing}, {Koch}, {Malesani}, {Andersen}, {Fynbo}, {Arcones}, {Bauswein}, {Covino}, {Grado}, {Heintz}, {Hunt}, {Kouveliotou}, {Leloudas}, {Levan}, {Mazzali}, \& {Pian}}]{Watson+19}
{Watson}, D., {Hansen}, C.~J., {Selsing}, J., {et~al.} 2019, \nat, 574, 497, \dodoi{10.1038/s41586-019-1676-3}

\bibitem[{{Waxman} {et~al.}(2022){Waxman}, {Ofek}, \& {Kushnir}}]{Waxman+22}
{Waxman}, E., {Ofek}, E.~O., \& {Kushnir}, D. 2022, arXiv e-prints, arXiv:2206.10710.
\newblock \doarXiv{2206.10710}

\bibitem[{{Winkler} {et~al.}(2003){Winkler}, {Courvoisier}, {Di Cocco}, {Gehrels}, {Gim{\'e}nez}, {Grebenev}, {Hermsen}, {Mas-Hesse}, {Lebrun}, {Lund}, {Palumbo}, {Paul}, {Roques}, {Schnopper}, {Sch{\"o}nfelder}, {Sunyaev}, {Teegarden}, {Ubertini}, {Vedrenne}, \& {Dean}}]{INTEGRAL+03}
{Winkler}, C., {Courvoisier}, T.~J.~L., {Di Cocco}, G., {et~al.} 2003, \aap, 411, L1, \dodoi{10.1051/0004-6361:20031288}

\bibitem[{{Wollaeger} {et~al.}(2021){Wollaeger}, {Fryer}, {Chase}, {Fontes}, {Ristic}, {Hungerford}, {Korobkin}, {O'Shaughnessy}, \& {Herring}}]{Wollaeger+21}
{Wollaeger}, R.~T., {Fryer}, C.~L., {Chase}, E.~A., {et~al.} 2021, \apj, 918, 10, \dodoi{10.3847/1538-4357/ac0d03}

\bibitem[{{Xiao} {et~al.}(2024){Xiao}, {Zhang}, {Zhu}, {Xiong}, {Gao}, {Xu}, {Zhang}, {Peng}, {Li}, {Zhang}, {Lu}, {Lin}, {Liu}, {Zhang}, {Ge}, {Tuo}, {Xue}, {Fu}, {Liu}, {Liu}, {Li}, {Wang}, {Zheng}, {Wang}, {Jiang}, {Li}, {Liu}, {Cao}, {Luo}, {Yang}, {Yi}, {Wang}, {Cai}, {Yi}, {Zhao}, {Xie}, {Li}, {Luo}, {Song}, {Zhang}, {Qu}, {Liu}, {Li}, {Xu}, \& {Li}}]{Xiao+24}
{Xiao}, S., {Zhang}, Y.-Q., {Zhu}, Z.-P., {et~al.} 2024, \apj, 970, 6, \dodoi{10.3847/1538-4357/ad4ee1}

\bibitem[{{Xiong} {et~al.}(2023){Xiong}, {Wang}, {Huang}, \& {Gecam Team}}]{GECAM_GCN_230307a}
{Xiong}, S., {Wang}, C., {Huang}, Y., \& {Gecam Team}. 2023, GRB Coordinates Network, 33406, 1

\bibitem[{{Xu} {et~al.}(2016){Xu}, {Malesani}, {de Ugarte Postigo}, {Gafton}, \& {Rivero Losada}}]{Xu+16_GCN}
{Xu}, D., {Malesani}, D., {de Ugarte Postigo}, A., {Gafton}, E., \& {Rivero Losada}, I. 2016, GRB Coordinates Network, 19834, 1

\bibitem[{{Xu} {et~al.}(2009){Xu}, {Starling}, {Fynbo}, {Sollerman}, {Yost}, {Watson}, {Foley}, {O'Brien}, \& {Hjorth}}]{Xu+09}
{Xu}, D., {Starling}, R.~L.~C., {Fynbo}, J.~P.~U., {et~al.} 2009, \apj, 696, 971, \dodoi{10.1088/0004-637X/696/1/971}

\bibitem[{{Yang} {et~al.}(2015){Yang}, {Jin}, {Li}, {Covino}, {Zheng}, {Hotokezaka}, {Fan}, {Piran}, \& {Wei}}]{yang+15}
{Yang}, B., {Jin}, Z.-P., {Li}, X., {et~al.} 2015, Nature Communications, 6, 7323, \dodoi{10.1038/ncomms8323}

\bibitem[{{Yang} {et~al.}(2022){Yang}, {Ai}, {Zhang}, {Zhang}, {Liu}, {Wang}, {Yang}, {Yin}, {Li}, \& {L{\"u}}}]{Yang+22}
{Yang}, J., {Ai}, S., {Zhang}, B.-B., {et~al.} 2022, \nat, 612, 232, \dodoi{10.1038/s41586-022-05403-8}

\bibitem[{{Yang} {et~al.}(2024){Yang}, {Troja}, {O'Connor}, {Fryer}, {Im}, {Durbak}, {Paek}, {Ricci}, {Bom}, {Gillanders}, {Castro-Tirado}, {Peng}, {Dichiara}, {Ryan}, {van Eerten}, {Dai}, {Chang}, {Choi}, {De}, {Hu}, {Kilpatrick}, {Kutyrev}, {Jeong}, {Lee}, {Makler}, {Navarete}, \& {P{\'e}rez-Garc{\'\i}a}}]{Yang+24}
{Yang}, Y.-H., {Troja}, E., {O'Connor}, B., {et~al.} 2024, \nat, 626, 742, \dodoi{10.1038/s41586-023-06979-5}

\bibitem[{{Zevin} {et~al.}(2022){Zevin}, {Nugent}, {Adhikari}, {Fong}, {Holz}, \& {Kelley}}]{Zevin+22}
{Zevin}, M., {Nugent}, A.~E., {Adhikari}, S., {et~al.} 2022, \apjl, 940, L18, \dodoi{10.3847/2041-8213/ac91cd}

\bibitem[{{Zheng} {et~al.}(2021){Zheng}, {Filippenko}, \& {KAIT GRB Team}}]{GCN_211211A_KAITag}
{Zheng}, W., {Filippenko}, A.~V., \& {KAIT GRB Team}. 2021, GRB Coordinates Network, 31203, 1

\end{thebibliography}

\section{Appendix}

\startlongtable
\begin{deluxetable*}{cCcCcCCcc}
\savetablenum{5}
\tabletypesize{\footnotesize}
\centering
\tablecolumns{9}
\tabcolsep0.04in
\tablecaption{Multi-wavelength Observations of GRB Afterglows and Kilonovae
\label{tab:obs}}
\tablehead {
\colhead {GRB}		&
\colhead {$z$}		&
\colhead {Tel./Instum.}		&
\colhead {$\delta t$}	&
\colhead {Band}		& 
\colhead {Magnitude$^{\dagger}$}	&
\colhead {Flux} &
\colhead {f$_{\rm AG}^{\dagger}$} &
\colhead {Ref.}  \\
\colhead {} &
\colhead {} &
\colhead {} &
\colhead {(days)}		& 
\colhead {} &
\colhead {(AB mag)}	&
\colhead {($\mu$Jy)} &
\colhead {} &
\colhead {} 
}
\startdata
050709 & 0.161 & Swift/XRT & 1.62 & 0.3-10 keV &  & 0.00019 \pm 0.00014 & * & 1 \\
 &  & Swift/XRT & 2.43 & 0.3-10 keV &  & < 0.00015 & * & 1 \\
 &  & Chandra/ACIS & 2.55 & 0.3-10 keV &  & 0.00011 \pm 0.00002 & * & 1 \\
 &  & Swift/XRT & 3.24 & 0.3-10 keV &  & < 0.00073 & * & 1 \\
 &  & Swift/XRT & 4.28 & 0.3-10 keV &  & < 0.00059 & * & 1 \\
 &  & Chandra/ACIS & 16.1 & 0.3-10 keV &  & 0.000011 \pm 0.00002 & * & 1 \\
 &  & Swope-40/Swope & 1.24 & $i'$ & > 20.5 & < 23 &  & 1 \\
 &  & Danish/DFOSC & 1.42 & $R$ & 23.2 \pm 0.1 & 1.9 \pm 0.099 & * & 2 \\
\enddata
\tablecomments{
Observations are not corrected for Galactic nor local extinction. \\
$\dagger$ Starred observations were employed in our afterglow analysis. \\
\textbf{References}: (1) \citealt{Fox+05}, (2) \citealt{Hjorth+05}, (3) \citealt{Covino+06}, (4) UKSSDC \citep{Evans+07,Evans+09}, (5) \citealt{Mangano+07}, (6) \citealt{Schmidt+06_GCN}, (7) \citealt{Xu+09}, (8) \citealt{Cobb+06_GCN}, (9) \citealt{Fynbo+06}, (10) \citealt{DellaValle+06}, (11) \citealt{yang+15}, (12) \citealt{Londish+06}, (13) \citealt{Fong+14}, (14) \citealt{deUP+13}, (15) \citealt{Cucchiara+13}, (16) \citealt{berger+13}, (17) \citealt{Tanvir+13}, (18) \citealt{troja+19}, (19) \citealt{lamb+19}, (20) \citealt{Kasliwal+17}, (21) \citealt{Fong+21}, (22) \citealt{O'Connor+21}, (23) \citealt{Mei+22}, (24) \citealt{GCN31217}, (25) \citealt{Troja+22}, (26) \citealt{Xiao+24}, (27) \citealt{GCN31227}, (28) \citealt{Rastinejad+22}, (29) \citealt{GCN31214}, (30) \citealt{GCN31232}, (31) \citealt{GCN31234}, (32) \citealt{GCN31299}, (33) \citealt{Yang+24}, (34) \citealt{Levan+24}. }
\end{deluxetable*}

\end{document}